\documentclass[aps,prb,twocolumn,groupedaddress,showpacs,showkeys]{revtex4}

\usepackage{amsmath}
\usepackage{amssymb}
\usepackage{graphicx}

\renewcommand{\d}{\ensuremath{\mathrm{d}}}

\begin{document}

\title{Theory of thermal spin-charge coupling in electronic systems.}
\author{B. Scharf,$^1$ A. Matos-Abiague,$^1$ I. \v{Z}uti\'c,$^2$ and J. Fabian$^1$ }
\affiliation{$^1$Institute for Theoretical Physics, University of Regensburg, 93040 Regensburg, Germany \\
$^2$Department of Physics, State University of New York at Buffalo, NY 14260, USA}
\date{\today}

\begin{abstract}
The interplay between spin transport and thermoelectricity offers several novel ways of generating, manipulating, and detecting nonequilibrium spin in a wide range of materials. Here we formulate a phenomenological model in the spirit of the standard model of electrical spin injection to describe the electronic mechanism coupling charge, spin, and heat transport and employ the model to analyze several different geometries containing ferromagnetic (F) and nonmagnetic (N) regions: F, F/N, and F/N/F junctions which are subject to thermal gradients. We present analytical formulas for the spin accumulation and spin current profiles in those junctions that are valid for both tunnel and transparent (as well as intermediate) contacts. For F/N junctions we calculate the thermal spin injection efficiency and the spin accumulation induced nonequilibrium thermopower. We find conditions for countering thermal spin effects in the N region with electrical spin injection. This compensating effect should be particularly useful for distinguishing electronic from other mechanisms of spin injection by thermal gradients. For F/N/F junctions we analyze the differences in the nonequilibrium thermopower (and chemical potentials) for parallel and antiparallel orientations of the F magnetizations, as evidence and a quantitative measure of the spin accumulation in N. Furthermore, we study the Peltier and spin Peltier effects in F/N and F/N/F junctions and present analytical formulas for the heat evolution at the interfaces of isothermal junctions.
\end{abstract}

\pacs{72.15.Jf, 72.25.-b, 85.75.-d}
\keywords{spintronics, spin caloritronics, spin Seebeck effect, spin Peltier effect, spin-charge coupling}

\maketitle

\section{Introduction}
The central theme in spintronics is the generation and control of nonequilibrium electron spin in solids.\cite{Zutic2004:RMP,Fabian2007:APS,DasSarma2000:Superlattice,DasSarma2001:SolidStateCommunications} Until recently the spin generation has been done by optical, magnetic, and, most important for device prospects, electrical means.\cite{Fabian2007:APS,Fabian2009:JuelichWorkshop} In a typical device spin-polarized electrons from a ferromagnetic conductor are driven by electromagnetic force to a nonmagnetic conductor. There the spin accumulates, with the steady state facilitated by spin relaxation. (There are also novel ways to generate pure spin currents, without accompanying charge currents.\cite{Stevens2003:PRL,Kimura2007:PRL,Zutic2011:NatureMat,Ando2011:NatureMat,Kurebayashi2011:NatureMat}) The concept of electrical spin injection was first proposed by Aronov,\cite{Aronov1976:JETP} and experimentally confirmed by Johnson and Silsbee,\cite{JohnsonSilsbee1985:PRL} who also formulated the problem from a nonequilibrium thermodynamics and drift-diffusion view.\cite{JohnsonSilsbee1987:PRB,JohnsonSilsbee1988:PRB} An equivalent description in terms of quasichemical potentials, convenient to treat discrete (junction) systems was formulated systematically by Rashba.\cite{Rashba2002:EPJB} This model, which we call the standard model of spin injection, is widely used to describe electrical spin injection into metals and semiconductors\cite{Zutic2004:RMP,Fabian2007:APS,Fabian2009:JuelichWorkshop} and can also be extended to ac currents.\cite{Kochan2011:PRL}

Until recently one particularly interesting possibility of generating spin, by spin-heat coupling, has been largely neglected. The generation of nonequilibrium spin by heat currents and the opposite process of generating heat currents by spin accumulation has already been proposed by Johnson and Silsbee\cite{JohnsonSilsbee1987:PRB} based on nonequilibrium thermodynamics concepts (see also Ref.~\onlinecite{Wegrowe2000:PRB}). The spin-heat coupling is now the central point of spin caloritronics (or spin calorics).\cite{Bauer2010:SolidStateCommunications,Johnson2010:SolidStateCommunications}
Although the theory of thermoelectricity has long been known,\cite{Callen1960,AshcroftMermin2006} only experimental improvements over the past few years have made its application in the context of generating and transporting spin appear possible.\cite{Fukushima2005:JJAP,Fukushima2005:IEEE,Gravier2006:PRB,Gravier2006:PRB2,Costache2011:NatureMat}

\begin{figure}[h]
\centering
\includegraphics*[width=8cm]{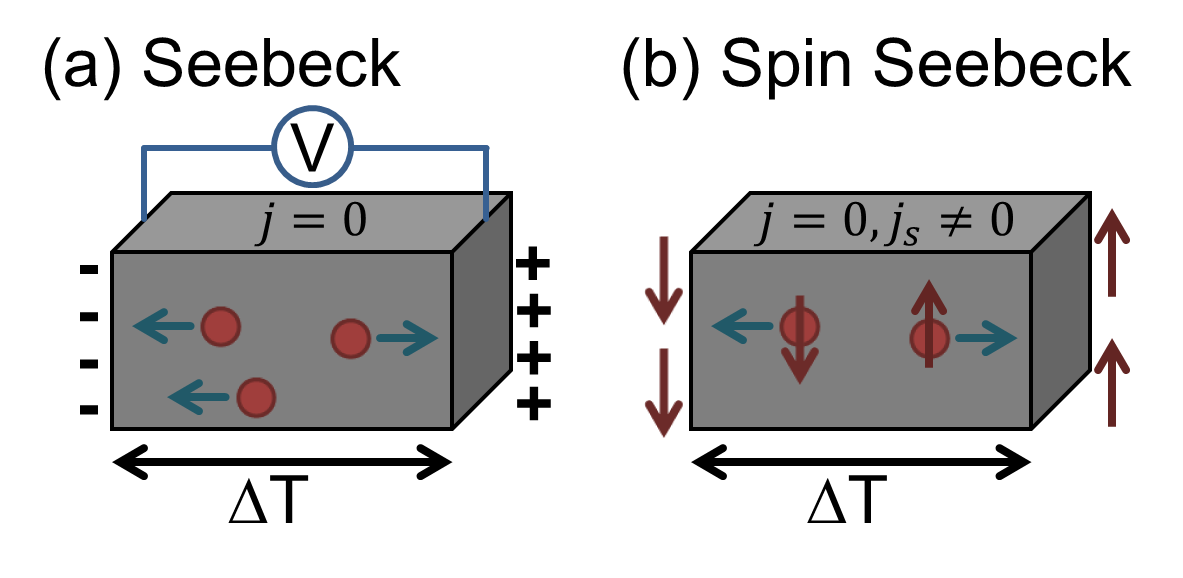}
\caption{(Color online) Schematic illustrations of the Seebeck (a) and spin Seebeck (b) effects. Here $\Delta T$ is the temperature difference, $V$ the voltage, $j$ the charge current, $j_s$ the spin current, and vertical arrows denote up/down spin projections.}\label{fig:Seebeck}
\end{figure}

At the heart of spin caloritronics is the spin Seebeck effect (see Fig.~\ref{fig:Seebeck}).\cite{Uchida2008:Nature,Uchida2010:JoAP,Uchida2010:NatureMat} The conventional Seebeck effect, also called thermopower,\cite{AshcroftMermin2006} describes the generation of an electric voltage if a thermal gradient is applied to a conductor. In analogy, the spin Seebeck effect describes the generation of spin accumulation in ferromagnets by thermal gradients. The effect was originally observed in the ferromagnetic conductor NiFe,\cite{Uchida2008:Nature,Uchida2009:JoAP} where indication of spin accumulation over large length scales (millimeters), independent of the spin relaxation scales in the ferromagnet, was found. Since it also exists at room temperature, the spin Seebeck phenomenon may have some technological applications.\cite{Bader2010:ARCMP}

However, the spin Seebeck effect is not limited to metals. It has also been observed in ferromagnetic insulators\cite{Uchida2010:NatureMat} as well as in the ferromagnetic semiconductor (Ga,Mn)As.\cite{Jaworski2010:NatureMat} This suggests that the spin Seebeck effect does not need to be connected with charge flow. In (Ga,Mn)As the sample was even cut preventing charge redistribution over the whole slab; the spin Seebeck signals were unaffected and in both cases, of compact and disconnected samples, the Pt stripes pick up the same inverse spin Hall signals.\cite{Valenzuela2006:Nature,Saitoh2006:APL} The evidence points to a mechanism of magnon-assisted spin pumping from the ferromagnet into the Pt, producing spin currents there. A theory for this spin pumping from a ferromagnetic insulator was suggested in Ref.~\onlinecite{Xiao2010:PRB}. It was predicted that phonons can play an important role in the spin Seebeck effect, leading to its huge enhancement.\cite{Adachi2010:APL} Recent measurements of the spin Seebeck effect in multiple (Ga,Mn)As samples also suggest that the spin Seebeck effect can be driven by phonons.\cite{Jaworski2011:PRL} In order to explain the main trends of the observed temperature and spatial dependence of the spin Seebeck effect in (Ga,Mn)As, a phenomenological model involving phonon-magnon coupling was introduced.\cite{Jaworski2011:PRL}

\begin{figure}[t]
\centering
\includegraphics*[width=8cm]{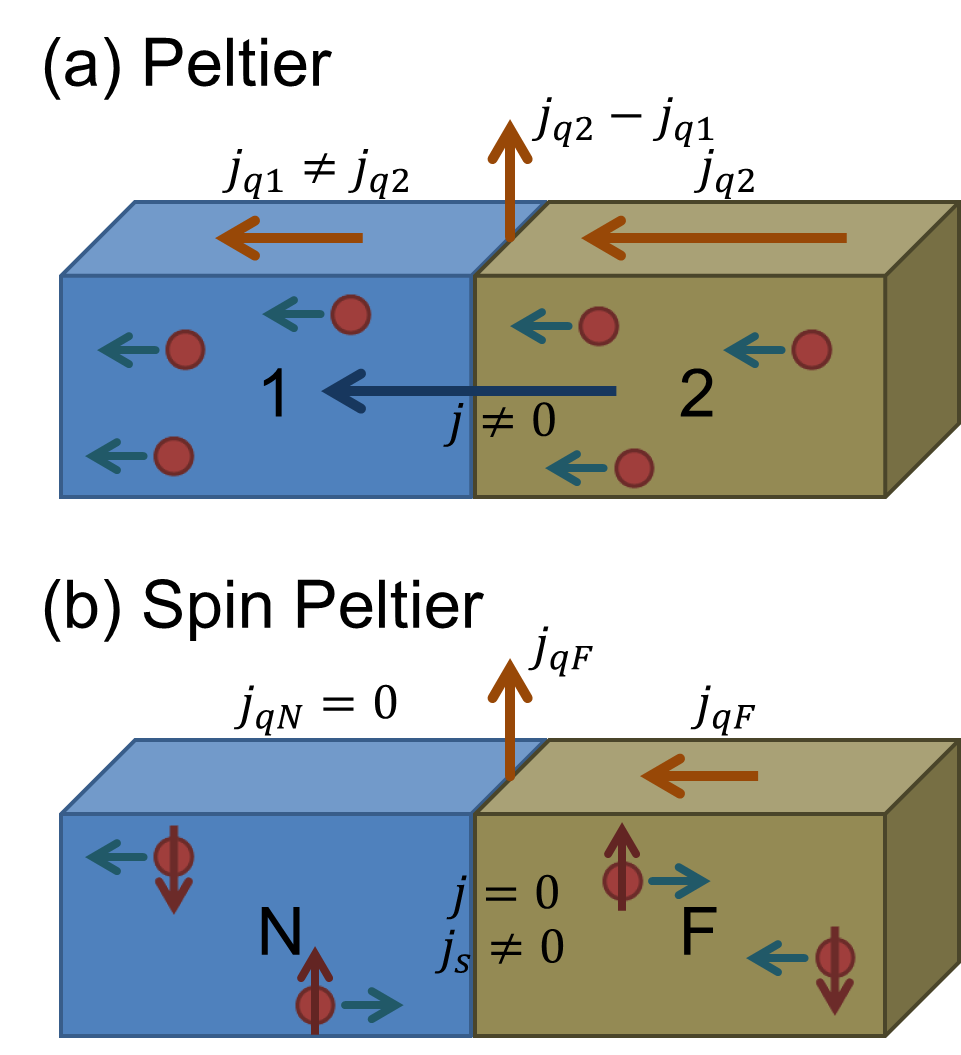}
\caption{(Color online) Schematic illustrations of the Peltier (a) and spin Peltier (b) effects, where $j$ and $j_s$ denote the charge and spin currents. The thermal current $j_q$ is different in each region. Small vertical arrows denote up/down spin projections.}\label{fig:Peltier}
\end{figure}

In addition to the Seebeck effect, there is also another thermoelectric effect, the Peltier effect, which refers to the evolution of heat across an isothermal junction of two different materials due to an electric current being passed through the junction.\cite{Callen1960,AshcroftMermin2006} Recently a spin caloritronics analog to the Peltier effect, termed spin Peltier effect, has been predicted and experimentally observed in a permalloy (Ni$_{80}$Fe$_{20}$)(PY)/copper/PY valve stack.\cite{Slachter2011:PRB,Flipse2011:preprint} The spin Peltier effect describes the heating or cooling at the interface between a ferromagnetic and normal conductor driven by a spin current (see Fig.~\ref{fig:Peltier}).

Another fascinating discovery is that of the thermally driven spin injection from a ferromagnet to a normal conductor.\cite{Slachter2010:Nature} In this experiment thermal currents in permalloy drive spin accumulation into copper, detected in a non-local geometry.\cite{Fabian2007:APS,Fabian2009:JuelichWorkshop} The structures were of submicron sizes, so it is plausible that the effects are electronic in nature, although magnon contributions to such thermal spin injection setups could also be sizable. A practical model was introduced  in Refs.~\onlinecite{Bakker2010:PRL,Slachter2011:PRB} to find, with a finite elements numerical scheme, the profiles of temperature and spin accumulation in the experimental devices. Recently, yet another form of thermal spin flow, coined Seebeck spin tunneling, has been demonstrated in ferromagnet-oxide-silicon tunnel junctions.\cite{LeBreton2011:Nature} Here a temperature difference between the ferromagnet and silicon causes a transfer of spin angular momentum across the interface between both materials.

An important goal for both theory and experiment of the spin Seebeck phenomena is to decipher the roles of the electronic and non-electronic contributions. It is yet unclear under which circumstances the electronic contribution may dominate. It seems likely that when going to smaller, submicron structures in which the spin accumulation will be a bulk effect, the spin phenomena carried by electrons will become important. Similarly, in materials with strong magnon damping, such that magnons are in local equilibrium with the given temperature profile, electrons may ultimately carry the entire spin Seebeck effect. It is thus important to set the benchmarks for the electronic contributions
in useful device geometries. This is what this paper does: we explore the role of the electronic contributions in F/N and F/N/F junctions which are subjected to thermal gradients and derive useful analytical formulas for various spin injection efficiencies.

Our purpose is twofold: First, we use the drift-diffusion framework of the standard model of spin injection presented in Refs.~\onlinecite{Zutic2004:RMP,Fabian2007:APS,Fabian2009:JuelichWorkshop} and generalize it to include electronic heat transport and thereby derive a theory for charge, spin, and heat transport in electronic materials. Secondly, we apply this theory to describe F/N and F/N/F junctions placed in thermal gradients. While the Peltier and Seebeck effects in such structures have been investigated in Ref.~\onlinecite{Hatami2009:PRB}, we focus here on the description of thermal spin injection and the investigation of the corresponding spin accumulation. We also look at the spin injection in the presence of both electric and thermal currents, and find the conditions under which the resulting spin current in N vanishes. In all junctions studied we present, as general as possible, analytical formulas for the spin accumulation and spin current profiles, as well as for the thermal spin injection efficiency and the nonequilibrium (spin accumulation driven) spin Seebeck coefficient. Moreover, we look at several different setups of the Peltier and spin Peltier effects and calculate their respective contributions to the heating/cooling at the interfaces in F/N and F/N/F junctions.

The manuscript is organized as follows: Following the introduction of the formalism and the basic equations in Sec.~\ref{spin-polarized_transport_concepts}, the electronic contribution to the spin Seebeck effect in a ferromagnetic metal is discussed within the framework of this formalism in Sec.~\ref{ferromagnet}, while Secs.~\ref{FN_junction} and~\ref{FNF_junction} are devoted to the discussion of thermal spin injection and related thermoelectric effects in F/N and F/N/F junctions respectively. A short summary concludes the manuscript.

\section{Spin-polarized transport in the presence of thermal fluctuations: concepts and definitions}\label{spin-polarized_transport_concepts}
\subsection{Spin-unpolarized transport equations}\label{spin-unpolarized_equations}
As a first step we will restrict ourselves to the description of transport in an electronic system which consists only of electrons of one species, that is, either of spin up or spin down electrons (denoted by the subscript $\lambda=\uparrow/\downarrow$ throughout this manuscript). The derivation presented here is a textbook
matter \cite{AshcroftMermin2006,Ziman2007} and is given here to introduce the terminology needed for the spin-polarized case and to match the concepts from the standard spin injection model of Ref.~\onlinecite{Fabian2007:APS}.

If this system is in thermodynamic equilibrium, the temperature $T$ and the chemical potential $\eta(T)$ are uniform throughout the system. Knowing the chemical potential,\footnote{The chemical potential is not only a function of the temperature but also of the total electron density.} one can calculate the density of the respective electron species under consideration from
\begin{equation}\label{subsystem_equilibrium_particle_density}
n^0_{\lambda}\left[\eta(T),T\right]=\int\d\varepsilon\; g_\lambda\left(\varepsilon\right)f_0\left[\frac{\varepsilon-\eta(T)}{k_BT}\right],
\end{equation}
where $k_B$ denotes the Boltzmann constant, $g_\lambda\left(\varepsilon\right)$ the electronic density of states at the energy $\varepsilon$, and $f_0$ the equilibrium Fermi-Dirac distribution function. Similarly, the equilibrium energy density is given by
\begin{equation}\label{subsystem_equilibrium_energy_density}
e^0_{\lambda}\left[\eta(T),T\right]=\int\d\varepsilon\;\varepsilon g_\lambda\left(\varepsilon\right) f_0\left[\frac{\varepsilon-\eta(T)}{k_BT}\right].
\end{equation}

The system is not in equilibrium if an electric field $-\nabla\varphi(x)$ is present in its bulk. In this case the chemical potential becomes space dependent. This is taken into account by replacing $\eta(T)$ with $\eta(T)+e\mu_\lambda(x)$, where the quasichemical potential $\mu_\lambda(x)$ now contains the space dependence.\footnote{In general, $\mu_\lambda$ also depends on the temperature $T$. If we consider different, space dependent local equilibrium temperatures $T(x)$, the gradient of the quasichemical potential reads $\nabla\mu_\lambda\left[x,T(x)\right]=\frac{\partial\mu_\lambda}{\partial x}+\frac{\partial\mu_\lambda}{\partial T}\nabla T$. Since we are only interested in first order effects, the temperature dependence of $\mu_\lambda$, which leads to a second order contribution (in the nonequilibrium quantities $\mu_\lambda\left[x,T(x)\right]$, $\varphi(x)$, and $\nabla T(x)$), can be omitted.} Since we want to incorporate the effects of thermal gradients into our formalism, we furthermore allow for different local equilibrium temperatures by replacing the constant temperature $T$ by a space dependent temperature $T(x)$. As a consequence there is an additional position dependence of the chemical potential due to the temperature, that is, $\eta(T)$ has to be replaced by $\eta\left[T(x)\right]$. Thus, the total chemical potential is given by $\eta\left[T(x)\right]+e\mu_\lambda(x)$.\\
Assuming the local nonequilibrium distribution function to be only energy dependent because momentum relaxation happens on length scales much smaller compared to the variation of the electric potential $\varphi(x)$, one obtains
\begin{equation}\label{subsystem_nonequilibrium_distribution}
f_\lambda(\varepsilon,x)=f_0\left\{\frac{\varepsilon-\eta\left[T(x)\right]-e\mu_\lambda(x)-e\varphi(x)}{k_BT(x)}\right\}.
\end{equation}
Therefore, the nonequilibrium electron and energy densities read
\begin{equation}\label{subsystem_nonequilibrium_local_density}
\begin{aligned}
n_\lambda(x)&=\int\d\varepsilon\;g_\lambda\left(\varepsilon\right)f_\lambda(\varepsilon,x)\\
&=n^0_{\lambda}\left\{\eta\left[T(x)\right]+e\mu_\lambda(x)+e\varphi(x),T(x)\right\},
\end{aligned}
\end{equation}
\begin{equation}\label{subsystem_nonequilibrium_energy_density}
\begin{aligned}
e_\lambda(x)&=\int\d\varepsilon\;\varepsilon g_\lambda\left(\varepsilon\right)f_\lambda(\varepsilon,x)\\
&=e^0_{\lambda}\left\{\eta\left[T(x)\right]+e\mu_\lambda(x)+e\varphi(x),T(x)\right\}.
\end{aligned}
\end{equation}

The electrostatic field gives rise to an electric current. This charge current consists of two parts: the drift current, proportional to the electric field $E(x)=-\nabla\varphi(x)$ and the diffusion current, proportional to the gradient of the local electron density.\\
Since the proportionality factor of the diffusion current, the diffusivity $D_\lambda(\varepsilon)$, is energy dependent, it is convenient to treat electrons with different energies separately. The spectral diffusion current density reads
\begin{equation}\label{subsystem_differential_diffusion_current}
j_{D\lambda}(x,\varepsilon)\d\varepsilon=eD_\lambda(\varepsilon)\nabla\left[g_\lambda(\varepsilon)f_\lambda(\varepsilon,x)\right]\d\varepsilon,
\end{equation}
from which the complete diffusion current can be obtained by integrating over the entire energy spectrum. The total charge current for electrons of spin $\lambda$ is given by
\begin{equation}\label{subsystem_complete_charge_current}
j_\lambda(x)=-\sigma_\lambda\nabla\varphi(x)+e\int\d\varepsilon\;D_\lambda(\varepsilon)g_\lambda(\varepsilon)\nabla f_\lambda(\varepsilon,x),
\end{equation}
where $\sigma_\lambda$ is the conductivity.
Inserting Eq.~(\ref{subsystem_nonequilibrium_distribution}) into Eq.~(\ref{subsystem_complete_charge_current}), using the Einstein relation,\footnote{The Einstein relation is obtained by requiring that $j_\lambda=0$ if $\nabla\left\{\frac{\eta\left[T(x)\right]}{e}+\mu_\lambda(x)\right\}=0$ and $\nabla T(x)=0$.} and keeping only terms linear in the nonequilibrium quantities $\mu_\lambda(x)$ and $\varphi(x)$, we find
\begin{equation}\label{subsystem_charge_current}
j_\lambda(x)=\sigma_\lambda\nabla\left\{\frac{\eta\left[T(x)\right]}{e}+\mu_\lambda(x)\right\}-S_\lambda\sigma_\lambda\nabla T(x).
\end{equation}
Here the conductivity is given by the Einstein relation
\begin{equation}\label{subsystem_conductivity}
\sigma_\lambda=e^2\int\d\varepsilon\;D_\lambda(\varepsilon)g_\lambda(\varepsilon)\left(-\frac{\partial f_0}{\partial\varepsilon}\right)\approx e^2D_\lambda(\varepsilon_F)g_\lambda(\varepsilon_F)
\end{equation}
and the Seebeck coefficient by
\begin{equation}\label{subsystem_thermopower}
\begin{aligned}
S_\lambda&=-\frac{e}{\sigma_\lambda}\int\d\varepsilon\;D_\lambda(\varepsilon)g_\lambda(\varepsilon)\left(-\frac{\partial f_0}{\partial\varepsilon}\right)\frac{\varepsilon-\eta\left[T(x)\right]}{T(x)}\\
&\approx-\mathcal{L}eT(x)\left[\frac{g_\lambda'(\varepsilon_F)}{g_\lambda(\varepsilon_F)}+\frac{D_\lambda'(\varepsilon_F)}{D_\lambda(\varepsilon_F)}\right].
\end{aligned}
\end{equation}
In both cases the integrals are calculated to the first non-vanishing order in the Sommerfeld expansion.\cite{AshcroftMermin2006} The Lorenz number is $\mathcal{L}=(\pi^2/3)(k_B/e)^2$ and $g'_\lambda(\varepsilon_F)$ and $D'_\lambda(\varepsilon_F)$ are the derivatives of the density of states and the diffusivity with respect to the energy evaluated at the Fermi level $\varepsilon_F$.

In addition to the charge current, there is a heat current in nonequilibrium. A treatment similar to that of the charge current above yields
\begin{equation}\label{subsystem_heat_current}
\begin{aligned}
j_{q,\lambda}(x)&=S_\lambda\sigma_\lambda T(x)\nabla\left\{\frac{\eta\left[T(x)\right]}{e}+\mu_\lambda(x)\right\}\\
&-\mathcal{L}\sigma_\lambda T(x)\nabla T(x).
\end{aligned}
\end{equation}
If the charge and heat currents are defined as in Eqs.~(\ref{subsystem_charge_current}) and~(\ref{subsystem_heat_current}), currents $j_\lambda(x)>0$ and $j_{q,\lambda}(x)>0$ flow parallel to the x direction.

At sharp contacts the chemical potential and the temperature are generally not continuous. Thus, instead of Eqs.~(\ref{subsystem_charge_current}) and~(\ref{subsystem_heat_current}), discretized versions of these equations are used. The charge current at the contact (C) is given by
\begin{equation}\label{subsystem_charge_current_contact}
j_{\lambda c}=\Sigma_{\lambda c}\left(\frac{1}{e}\Delta\eta_c+\Delta\mu_{\lambda c}\right)-S_{\lambda c}\Sigma_{\lambda c}\Delta T_c
\end{equation}
and the heat current by
\begin{equation}\label{subsystem_heat_current_contact}
j_{q\lambda c}=TS_{\lambda c}\Sigma_{\lambda c}\left(\frac{1}{e}\Delta\eta_c+\Delta\mu_{\lambda c}\right)-\mathcal{L}T\Sigma_{\lambda c}\Delta T_c,
\end{equation}
where $\Delta\eta_c+e\Delta\mu_{\lambda c}$ and $\Delta T_c$ denote the drops of the total chemical potential and the temperature at the contact respectively. The (effective) contact conductance and the contact thermopower are given by $\Sigma_{\lambda c}$ and $S_{\lambda c}$ respectively, while $T$ is the average temperature of the system.

\subsection{Spin-polarized transport equations}\label{spin-polarized_equations}
We now consider spin-polarized systems, which we treat as consisting of two subsystems, one of spin up and one of spin down electrons;
each subsystem is described by the equations from Sec.~\ref{spin-unpolarized_equations}.

Energy as well as particles can be exchanged between the two spin pools (by collisions and spin-flip processes respectively).
As energy relaxation (tens of femtoseconds) happens usually on much shorter time scales than spin relaxation (picoseconds to nanoseconds),
we assume that a local equilibrium exists at each position $x$. Consequently, both subsystems share a common {\it local} equilibrium chemical potential $\eta\left[T(x)\right]$ and temperature $T(x)$. On the other hand, the local nonequilibrium quasichemical potentials $\mu_\lambda(x)$ can be different for each spin subsystem.

From Eq.~(\ref{subsystem_nonequilibrium_local_density}) we obtain
\begin{equation}\label{local_density_definition}
\begin{aligned}
n(x)=&n^0_{\uparrow}\left\{\eta\left[T(x)\right]+e\mu_\uparrow(x)+e\varphi(x),T(x)\right\}\\
&+n^0_{\downarrow}\left\{\eta\left[T(x)\right]+e\mu_\downarrow(x)+e\varphi(x),T(x)\right\}
\end{aligned}
\end{equation}
for the complete local electron density of the system. Expanding the electron density up to the first order in the local nonequilibrium quantities, $\mu_\uparrow(x)$, $\mu_\downarrow(x)$, and $\varphi(x)$, and using the Sommerfeld expansion subsequently to calculate the integrals which enter via Eq.~(\ref{subsystem_equilibrium_particle_density}), we can write the electron density as
\begin{equation}\label{local_density}
n(x)=n_0+\delta n(x).
\end{equation}
Here we have introduced the local equilibrium electron density, $n_0=n^0_{\uparrow}\{\eta\left[T(x)\right],T(x)\}+n^0_{\downarrow}\{\eta\left[T(x)\right],T(x)\}$, and the local nonequilibrium electron density fluctuations,
\begin{equation}\label{local_density_fluctuations}
\delta n(x)=eg\left[\mu(x)+\varphi(x)\right]+eg_s\mu_s(x).
\end{equation}
Additionally, we have introduced the quasichemical potential, $\mu=(\mu_\uparrow+\mu_\downarrow)/2$, the spin accumulation, $\mu_s=(\mu_\uparrow-\mu_\downarrow)/2$, as well as the densities of states $g=g_\uparrow(\varepsilon_F)+g_\downarrow(\varepsilon_F)$ and $g_s=g_\uparrow(\varepsilon_F)-g_\downarrow(\varepsilon_F)$ at the Fermi level. We further assume that there is no accumulation of charge inside the conductor under bias $\varphi(x)$. This assumption of local charge neutrality is valid for metals and highly doped semiconductors and requires $n(x)=n_0$.\footnote{In non-degenerate semiconductors one can relax this condition and obtain the resulting nonlinear current-voltage characteristics and bias-dependent spin injection efficiency [I. \v{Z}uti\'c, J. Fabian, and S. Das Sarma, Phys. Rev. Lett. 88, 066603 (2002);  J. Fabian, I. \v{Z}uti\'c, and S. Das Sarma, Phys. Rev. B 66, 165301 (2002); I. \v{Z}uti\'c, J. Fabian, and S. Das Sarma, Appl. Phys. Lett. 82, 221 (2003)].} Hence, Eq.~(\ref{local_density}) yields the condition
\begin{equation}\label{local_charge_neutrality}
\delta n(x)=0.
\end{equation}

The local spin density,
\begin{equation}\label{local_spin_density_definition}
\begin{aligned}
s(x)=&n^0_{\uparrow}\left\{\eta\left[T(x)\right]+e\mu_\uparrow(x)+e\varphi(x),T(x)\right\}\\
&-n^0_{\downarrow}\left\{\eta\left[T(x)\right]+e\mu_\downarrow(x)+e\varphi(x),T(x)\right\},
\end{aligned}
\end{equation}
can be evaluated analogously to the local electron density: First, Eq.~(\ref{local_spin_density_definition}) is expanded in the local nonequilibrium quantities up to the first order. The resulting integrals are performed employing the Sommerfeld expansion up to the first non-vanishing order and, as a final step, the charge neutrality condition, Eq.~(\ref{local_charge_neutrality}), is used to simplify the result. This procedure yields
\begin{equation}\label{local_spin_density}
s(x)=s_0(x)+\delta s(x),
\end{equation}
with the local equilibrium spin density, $s_0(x)=n^0_{\uparrow}\{\eta\left[T(x)\right],T(x)\}-n^0_{\downarrow}\{\eta\left[T(x)\right],T(x)\}$ and the local nonequilibrium spin density,
\begin{equation}
\delta s(x)=e\frac{g^2-g_s^2}{g}\mu_s(x).\label{local_spin_density_fluctuations}
\end{equation}
It is important to note that $s_0(x)$ is determined by the local temperature $T(x)$, as a result of the rapid energy relaxation as compared to the spin relaxation.

The same procedure can be applied to calculate the energy density from Eq.~(\ref{subsystem_nonequilibrium_energy_density}),
\begin{equation}\label{local_energy_density}
\begin{aligned}
e(x)=&e^0_{\uparrow}\left\{\eta\left[T(x)\right]+e\mu_\uparrow(x)+e\varphi(x),T(x)\right\}\\
&+e^0_{\downarrow}\left\{\eta\left[T(x)\right]+e\mu_\downarrow(x)+e\varphi(x),T(x)\right\},
\end{aligned}
\end{equation}
which can be split in a local equilibrium energy density, $e_0(x)=e^0_{\uparrow}\{\eta\left[T(x)\right],T(x)\}+e^0_{\downarrow}\{\eta\left[T(x)\right],T(x)\}$, and local energy density fluctuations $\delta e(x)$, that is,
\begin{equation}\label{local_energy_density_split}
e(x)=e_0(x)+\delta e(x).
\end{equation}
Calculating $\delta e(x)$ in the same way as $\delta s(x)$, we find that
\begin{equation}\label{local_energy_density_fluctuations}
\delta e(x)=0,
\end{equation}
consistent with our assumption of fast energy relaxation to the local quasiequilibrium.

Next, we consider the currents flowing through the system. Since our goal is to calculate the quasichemical and spin quasichemical potentials, as well as the temperature profile, we not only derive transport equations based on Eqs.~(\ref{subsystem_charge_current}) and~(\ref{subsystem_heat_current}), but also continuity equations for each of the currents considered, that is, charge, spin, and heat currents.

The charge current consists of the electric currents carried by spin up and spin down electrons,
\begin{equation}\label{charge_current}
\begin{aligned}
j(x)&=j_\uparrow(x)+j_\downarrow(x)\\
&=\sigma\nabla\left\{\frac{\eta\left[T(x)\right]}{e}+\mu(x)\right\}+\sigma_s\nabla\mu_s(x)\\
&-\frac{1}{2}\left(S\sigma+S_s\sigma_s\right)\nabla T(x),
\end{aligned}
\end{equation}
where the conductivities are given by $\sigma=\sigma_\uparrow+\sigma_\downarrow$ and $\sigma_s=\sigma_\uparrow-\sigma_\downarrow$, and the Seebeck coefficients by $S=S_\uparrow+S_\downarrow$ and $S_s=S_\uparrow-S_\downarrow$. In nonmagnetic materials $\sigma_s=0$ and $S_s=0$. In our model we consider a steady state, which requires
\begin{equation}\label{charge_current_continuity}
\nabla j(x)=0,
\end{equation}
that is, a uniform electric current, $j(x)=j$.

The spin current is the difference between the electric currents of spin up and spin down electrons,
\begin{equation}\label{spin_current}
\begin{aligned}
j_s(x)&=j_\uparrow(x)-j_\downarrow(x)\\
&=\sigma_s\nabla\left\{\frac{\eta\left[T(x)\right]}{e}+\mu(x)\right\}+\sigma\nabla\mu_s(x)\\
&-\frac{1}{2}\left(S_s\sigma+S\sigma_s\right)\nabla T(x).
\end{aligned}
\end{equation}
As we have seen, the spin density $s(x)$ deviates from its local equilibrium value $s_0(x)$. Unlike charge, spin is not conserved and spin relaxation processes lead to a decrease of the local nonequilibrium spin to $s_0(x)$. Therefore, the continuity equation for the spin current is given by
\begin{equation}\label{spin_current_continuity}
\nabla j_s(x)=e\frac{\delta s(x)}{\tau_s},
\end{equation}
where $\tau_s$ is the spin relaxation time. We will not distinguish between different spin relaxation mechanisms in our model. Instead, we treat $\tau_s$ as an effective spin relaxation time which incorporates all the different spin relaxation mechanisms. We stress that spin relaxation processes bring the nonequilibrium spin $s(x)$ to the (quasi)equilibrium value $s_0(x)$, defined {\it locally} by $T(x)$. Here we deviate from the treatment given in Ref.~\onlinecite{Uchida2009:JoAP}.

The heat current,
\begin{equation}\label{heat_current}
\begin{aligned}
j_q(x)&=j_{q,\uparrow}(x)+j_{q,\downarrow}(x)\\
&=\frac{T\left(S\sigma+S_s\sigma_s\right)}{2}\nabla\left\{\frac{\eta\left[T(x)\right]}{e}+\mu(x)\right\}\\
&+\frac{T\left(S_s\sigma+S\sigma_s\right)}{2}\nabla\mu_s(x)-\mathcal{L}T\sigma\nabla T(x),
\end{aligned}
\end{equation}
is the heat carried through the system by the electrons of both spin species. Closely related is the energy current,
\begin{equation}\label{energy_current}
j_u(x)=j_q(x)-\left\{\frac{\eta\left[T(x)\right]}{e}+\mu(x)\right\}j-\mu_s(x)j_s(x).
\end{equation}
Inserting Eqs.~(\ref{charge_current}), (\ref{spin_current}), and~(\ref{heat_current}) and using that the divergence of the charge current vanishes in a steady state, that is, Eq.~(\ref{charge_current_continuity}), we find
\begin{equation}\label{energy_current_divergence}
\begin{aligned}
\nabla j_u(x)&=\frac{T(x)}{2}\nabla\left[Sj+S_sj_s(x)\right]-\mu_s(x)\nabla j_s(x)\\
&-\nabla\left[\mathcal{L}\sigma T(x)\left(1-\frac{S^2+S_s^2+2SS_sP_\sigma}{4\mathcal{L}}\right)\nabla T(x)\right]\\
&-\frac{j^2_\uparrow(x)}{\sigma_\uparrow}-\frac{j^2_\downarrow(x)}{\sigma_\downarrow},
\end{aligned}
\end{equation}
where $P_\sigma=\sigma_s/\sigma$ is the conductivity spin polarization. The above formula contains Thomson (first term) as well as Joule heating (final two terms). Equation~(\ref{local_energy_density_fluctuations}) can be used to formulate the continuity equation for the energy current by enforcing the energy conservation,
\begin{equation}\label{energy_current_continuity}
\nabla j_u(x)=0.
\end{equation}

Thus, if $j$ is treated as an external parameter, the transport equation for the charge current, Eq.~(\ref{charge_current}), as well as the transport and continuity equations for the spin and heat currents, Eqs.~(\ref{spin_current}), (\ref{spin_current_continuity}), (\ref{heat_current}), and (\ref{energy_current_continuity}), form a complete set of inhomogeneous differential equations to determine the quasichemical potentials $\mu(x)$ and $\mu_s(x)$, the temperature profile $T(x)$, as well as the currents $j_s(x)$ and $j_q(x)$. The solution to this set of differential equations, that couple charge, spin, and heat transport, will be discussed in the next section.

\subsection{Spin diffusion equation and its general solution}\label{spin_diffusion}
In the following the general solutions to the equations introduced in Sec.~\ref{spin-polarized_equations} will be discussed. Inserting Eq.~(\ref{spin_current}) into the spin current continuity equation, Eq.~(\ref{spin_current_continuity}), and using Eqs.~(\ref{local_spin_density_fluctuations}), (\ref{charge_current}), and (\ref{charge_current_continuity}) generalizes the standard\cite{vanSon1987:PRL,ValetFert1993:PRB} spin diffusion equation,
\begin{equation}\label{generalized_spin_diffusion_equation}
\nabla^2\mu_s(x)=\frac{\mu_s(x)}{\lambda_s^2}+\frac{1}{2}\nabla\cdot\left[S_s\nabla T(x)\right].
\end{equation}
Here we have introduced the spin diffusion length\cite{Zutic2004:RMP,Fabian2007:APS}
\begin{equation}\label{spin_diffusion_length}
\lambda_s=\sqrt{\tau_sg\sigma(1-P_\sigma^2)/\left[e^2\left(g^2-g_s^2\right)\right]}.
\end{equation}
As we are primarily interested in linear effects, we neglect the position-dependence of the spin Seebeck coefficient $S_s$, which enters via $T(x)$, and arrive at a simplified diffusion equation for the spin accumulation,
\begin{equation}\label{simplified_spin_diffusion_equation}
\nabla^2\mu_s(x)=\frac{\mu_s(x)}{\lambda_s^2}+\frac{S_s}{2}\nabla^2T(x),
\end{equation}
where $S_s$ is evaluated at the mean temperature $T$. In order to solve this equation, we need the temperature profile which can be determined from Eq.~(\ref{energy_current_continuity}). If only first order effects are taken into account, Eq.~(\ref{energy_current_continuity}) gives the differential equation
\begin{equation}\label{temperature_differential_equation}
\nabla^2T(x)=\frac{2S_s(1-P_\sigma^2)}{\lambda_s^2\left(4\mathcal{L}-S^2-S_s^2-2SS_sP_\sigma\right)}\mu_s(x),
\end{equation}
deforming the typically linear profile of $T(x)$. The solution to the coupled differential Eqs.~(\ref{simplified_spin_diffusion_equation}) and (\ref{temperature_differential_equation}) reads
\begin{equation}\label{solution_spinpotential}
\mu_s(x)=A\exp\left(\frac{x}{\tilde{\lambda}_s}\right)+B\exp\left(-\frac{x}{\tilde{\lambda}_s}\right),
\end{equation}
\begin{equation}\label{solution_temperature}
T(x)=\frac{2S_s(1-P_\sigma^2)}{4\mathcal{L}-(S+S_sP_\sigma)^2}\mu_s(x)+Cx+D,
\end{equation}
with the modified spin diffusion length
\begin{equation}\label{modified_spin_diffusion_length}
\tilde{\lambda}_s=\lambda_s\sqrt{\frac{4\mathcal{L}-S^2-S_s^2-2SS_sP_\sigma}{4\mathcal{L}-(S+S_sP_\sigma)^2}}.
\end{equation}
Integration of Eq.~(\ref{charge_current}) yields the total chemical potential,
\begin{equation}\label{solution_chemicalpotential}
\frac{\eta\left[T(x)\right]}{e}+\mu(x)=\frac{j}{\sigma}x-P_\sigma\mu_s(x)+\frac{S+S_sP_\sigma}{2}T(x)+E.
\end{equation}
The integration constants $A$, $B$, $C$, $D$, and $E$ have to be determined by including the respective boundary conditions of the system under consideration.

If $S_\lambda\ll\sqrt{\mathcal{L}}$ (see the next section), it is often possible to assume a uniform temperature gradient, that is,
\begin{equation}\label{simple_solution_temperature}
T(x)=Cx+D.
\end{equation}
Then Eq.~(\ref{simplified_spin_diffusion_equation}) reduces to the standard spin diffusion equation and its solution is given by
\begin{equation}\label{simple_solution_spinpotential}
\mu_s(x)=A\exp\left(\frac{x}{\lambda_s}\right)+B\exp\left(-\frac{x}{\lambda_s}\right),
\end{equation}
while integration of Eq.~(\ref{charge_current}) yields the total chemical potential,
\begin{equation}\label{simple_solution_chemicalpotential}
\frac{\eta\left[T(x)\right]}{e}+\mu(x)=\left(\frac{j}{\sigma}+\frac{S+S_sP_\sigma}{2}C\right)x-P_\sigma\mu_s(x)+E.
\end{equation}
As before, $A$, $B$, $C$, $D$, and $E$ are integration constants to be specified by boundary conditions. However, assuming a constant temperature gradient in ferromagnets is not consistent with Eq.~(\ref{energy_current_continuity}) and therefore this approximation cannot be used in situations which depend crucially on the heat current profile (see next section).

The spin and heat currents can be obtained by inserting the solutions found above into Eqs.~(\ref{spin_current}) and~(\ref{heat_current}).

\subsection{Contact properties}\label{contact_properties}
To find the specific solution for a system consisting of different materials, such as a F/N junction, we have to know the behavior of the currents at the interfaces between two different materials. The currents at a contact can be obtained by applying Eqs.~(\ref{subsystem_charge_current_contact}) and~(\ref{subsystem_heat_current_contact}), giving
\begin{equation}\label{charge_current_contact}
\begin{aligned}
j_c=&j_{\uparrow c}+j_{\downarrow c}=\Sigma_{c}\left(\frac{1}{e}\Delta\eta_{c}+\Delta\mu_{c}\right)+\Sigma_{sc}\Delta\mu_{sc}\\
&-\frac{1}{2}\left(S_{c}\Sigma_{c}+S_{sc}\Sigma_{sc}\right)\Delta T_{c},
\end{aligned}
\end{equation}
\begin{equation}\label{spin_current_contact}
\begin{aligned}
j_{sc}=&j_{\uparrow c}-j_{\downarrow c}=\Sigma_{sc}\left(\frac{1}{e}\Delta\eta_{c}+\Delta\mu_{c}\right)+\Sigma_{c}\Delta\mu_{sc}\\
&-\frac{1}{2}\left(S_{sc}\Sigma_{c}+S_{c}\Sigma_{sc}\right)\Delta T_{c},
\end{aligned}
\end{equation}
\begin{equation}\label{heat_current_contact}
\begin{aligned}
j_{qc}&=j_{q\uparrow c}+j_{q\downarrow c}=\frac{T}{2}\left(S_{c}\Sigma_{c}+S_{sc}\Sigma_{sc}\right)\left(\frac{1}{e}\Delta\eta_{c}+\Delta\mu_{c}\right)\\
&+\frac{T}{2}\left(S_{sc}\Sigma_{c}+S_{c}\Sigma_{sc}\right)\Delta\mu_{sc}-\mathcal{L}T\Sigma_{c}\Delta T_{c},
\end{aligned}
\end{equation}
where $\Delta T_{c}$ is the temperature drop at the contact, and $\Delta\eta_{c}$, $\Delta\mu_{c}$, and $\Delta\mu_{sc}$ are the drops of the local equilibrium chemical, quasichemical and spin quasichemical potentials. Moreover, the contact conductances $\Sigma_c=\Sigma_{\uparrow c}+\Sigma_{\downarrow c}$ and $\Sigma_s^c=\Sigma_{\uparrow c}-\Sigma_{\downarrow c}$ as well as the contact thermopowers $S_c=S_{\uparrow c}+S_{\downarrow c}$ and $S_{sc}=S_{\uparrow c}-S_{\downarrow c}$ have been introduced.\\
Equations~(\ref{charge_current_contact})-(\ref{heat_current_contact}) will be used in Secs.~\ref{FN_junction} and~\ref{FNF_junction} to fix the integration constants of the general solutions, Eqs.~(\ref{simple_solution_temperature})-(\ref{simple_solution_chemicalpotential}) and Eqs.~(\ref{solution_spinpotential})-(\ref{simple_solution_chemicalpotential}) found in Sec.~\ref{spin_diffusion}.

\section{Ferromagnet placed in a thermal gradient}\label{ferromagnet}
\begin{figure}[t]
\centering
\includegraphics*[width=8cm]{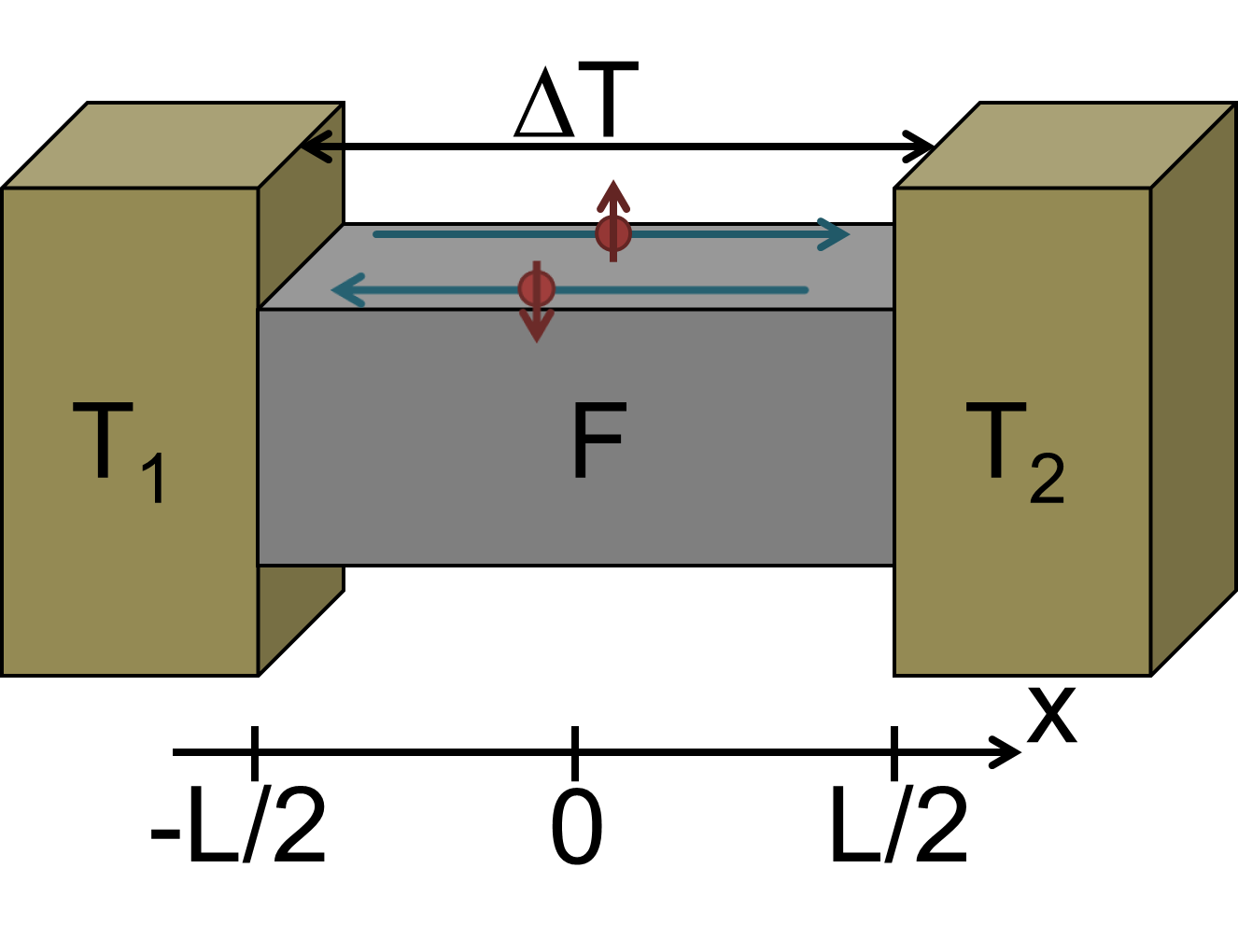}
\caption{(Color online) A schematic illustration of a ferromagnet metal placed in a thermal gradient which leads to the generation of a spin current.}\label{fig:ferromagnet}
\end{figure}

As a first example we consider a ferromagnetic metal F of length $L$ ($-L/2<x<L/2$) subject to a thermal gradient under open-circuit conditions, that is, $j=0$. The gradient is applied by creating a temperature difference $\Delta T=T_2-T_1$ between both ends of the metal which are held at temperatures $T_1$ and $T_2$ respectively, as shown in Fig.~\ref{fig:ferromagnet}.

At the ends of the ferromagnet we impose the boundary conditions $T(-L/2)=T_1$, $T(L/2)=T_2$, and set $j_s(\pm L/2)=0$. Since we consider only first order effects, the Seebeck coefficients are assumed to be constant over the length of the ferromagnet and are evaluated at the mean temperature $T=(T_1+T_2)/2$. Using the above boundary conditions and Eqs.~(\ref{solution_spinpotential})-(\ref{solution_chemicalpotential}) yields the spin accumulation
\begin{equation}\label{ferromagnet_spin_accumulation}
\mu_s(x)=\frac{S_{s}}{2}\;\tilde{\lambda}_s\;\frac{\Delta T}{L}\;\frac{\sinh(x/\tilde{\lambda}_s)}{\cosh(L/2\tilde{\lambda}_s)}\frac{4\mathcal{L}-(S+S_{s}P_{\sigma})^2}{N(L)},
\end{equation}
and the spin current
\begin{equation}\label{ferromagnet_spin_current}
\begin{aligned}
j_s(x)=&-\frac{S_{s}}{2}\;\frac{\tilde{\lambda}_s}{\tilde{R}}\;\frac{\Delta T}{L}\;\left[1-\frac{\cosh(x/\tilde{\lambda}_s)}{\cosh(L/2\tilde{\lambda}_s)}\right]\\
&\times\frac{4\mathcal{L}-S^2-S_{s}^2-2SS_{s}P_{\sigma}}{N(L)},
\end{aligned}
\end{equation}
where $\tilde{R}=\tilde{\lambda}_s/\left[\sigma(1-P_{\sigma}^2)\right]$ and
\begin{equation}
\begin{aligned}
N(L)=&4\mathcal{L}-S^2-S_{s}^2-2SS_{s}P_{\sigma}\\
&+S_{s}^2\left(1-P_{\sigma}^2\right)\frac{\tanh(L/2\tilde{\lambda}_s)}{L/2\tilde{\lambda}_s}.
\end{aligned}
\end{equation}

If a constant temperature gradient is assumed and the reduced model given by Eqs.~(\ref{simple_solution_temperature})-(\ref{simple_solution_chemicalpotential}) is used, the spin accumulation reads
\begin{equation}\label{ferromagnet_spin_accumulation_simple}
\mu_s(x)=\frac{S_{s}}{2}\;\lambda_{s}\;\frac{\Delta T}{L}\;\frac{\sinh(x/\lambda_{s})}{\cosh(L/2\lambda_{s})},
\end{equation}
and the spin current
\begin{equation}\label{ferromagnet_spin_current_simple}
j_s(x)=-\frac{S_{s}}{2}\;\frac{\lambda_{s}}{R}\;\frac{\Delta T}{L}\;\left[1-\frac{\cosh(x/\lambda_{s})}{\cosh(L/2\lambda_{s})}\right],
\end{equation}
where $R=\lambda_{s}/\left[\sigma(1-P_{\sigma}^2)\right]$ is the effective resistance of the ferromagnet.

For metals $S_{\lambda}\ll\sqrt{\mathcal{L}}$ and Eqs.~(\ref{ferromagnet_spin_accumulation}) and~(\ref{ferromagnet_spin_current}) reduce to Eqs.~(\ref{ferromagnet_spin_accumulation_simple}) and~(\ref{ferromagnet_spin_current_simple}), that is, the assumption of a uniform temperature gradient $\nabla T=\Delta T/L$ is justified. Only at the boundaries of the sample both temperature profiles differ (insignificantly) as there is a small exponential decay within the spin diffusion length $\tilde{\lambda}_s\approx\lambda_{s}$ if the full model is used compared to a perfectly linear temperature profile of the reduced model.

Equations~(\ref{ferromagnet_spin_accumulation_simple}) and~(\ref{ferromagnet_spin_current_simple}) from the
reduced model correspond to the profiles of the spin accumulation and spin current found in Ref.~\onlinecite{Hatami2010:SolidStateCommunications}, where a Boltzmann equation approach has been used to describe thermoelectric spin diffusion in a ferromagnetic metal.

\begin{figure}[t]
\includegraphics*[clip,trim=1cm 1cm 1cm 15cm,width=8cm]{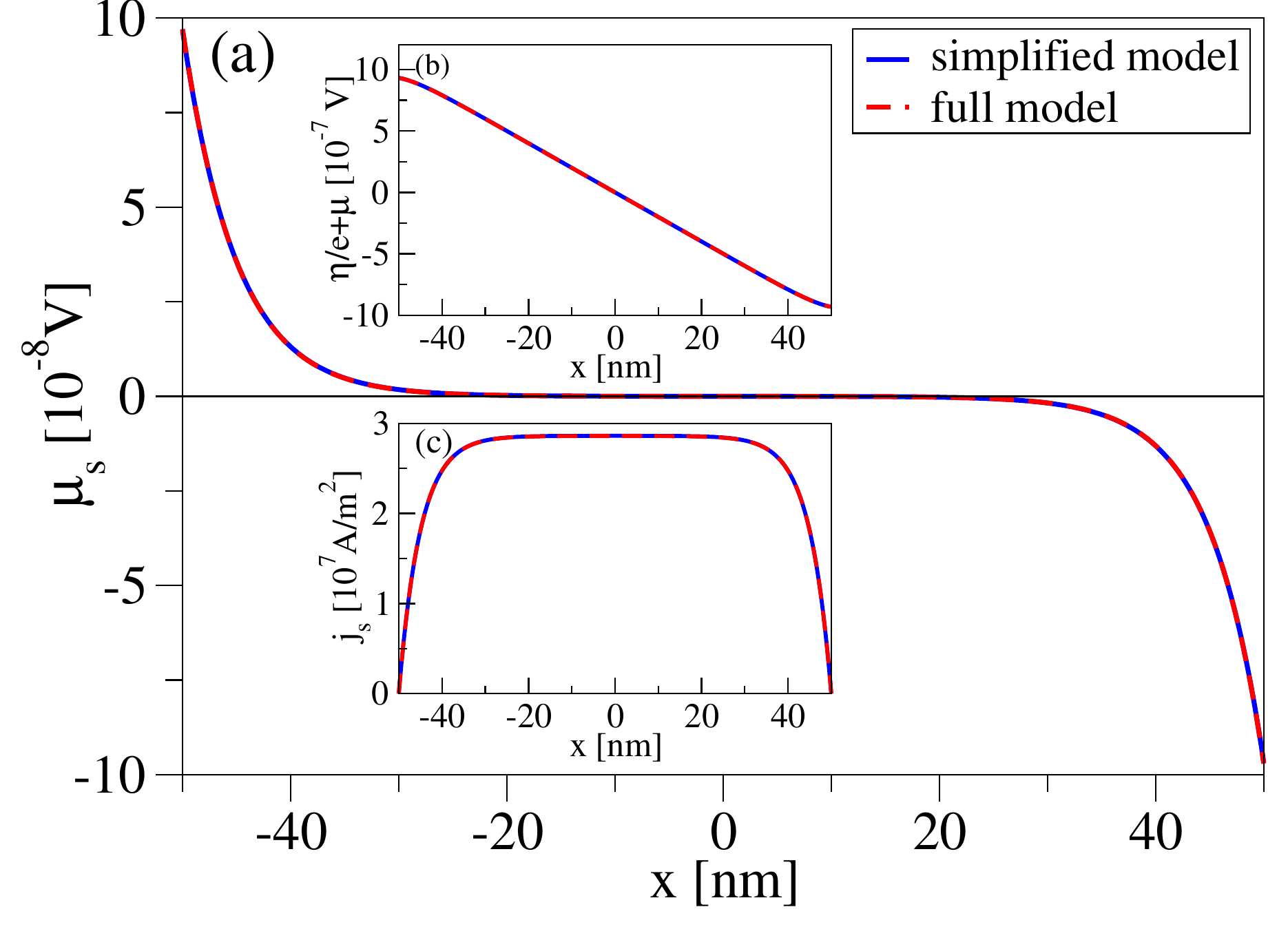}
\caption{(Color online) Profiles of the spin accumulation (a), the total chemical potential (b), and the spin current (c) for Ni$_{81}$Fe$_{19}$ at $T=300$ K with $L=100$ nm and $\Delta T=100$ mK. The solid lines show the results obtained if a constant temperature gradient  $\nabla T=\Delta T/L_F$ is assumed, while the dashed lines (fully overlapping
with the solid ones) show the results obtained if the temperature profile is determined by $\nabla j_u=0$.}\label{fig:ferromagnet_graphs}
\end{figure}

In Fig.~\ref{fig:ferromagnet_graphs} the results calculated for a model Ni$_{81}$Fe$_{19}$ film with realistic parameters\cite{Uchida2009:JoAP} [$\lambda_{s}=5$ nm, $\sigma=2.9\times10^6$ 1/$\Omega$m, $S_0=(S_{\uparrow}\sigma_{\uparrow}+S_{\downarrow}\sigma_{\downarrow})/(\sigma_{\uparrow}+\sigma_{\downarrow})=-2.0\times10^{-5}$ V/K with $P_{\sigma}=0.7$ and $P_{S}=(S_{\uparrow}-S_{\downarrow})/(S_{\uparrow}+S_{\downarrow})=3.0$] at a mean temperature $T=300$ K are displayed. The length of the sample is $L=100$ nm and the temperature difference is $\Delta T=100$ mK. As can be seen in Fig.~\ref{fig:ferromagnet_graphs}, the agreement between both solutions is very good.

Figure~\ref{fig:ferromagnet_graphs}~(b) shows an almost linear drop of the total chemical potential between both ends of the ferromagnet. Only at the contacts this linear drop is superimposed by an exponential decay. It is also at the contacts that nonequilibrium spin accumulates and decays within the spin diffusion length [see Figs.~\ref{fig:ferromagnet_graphs}~(a) and ~\ref{fig:ferromagnet_graphs}~(c)]. Thus, only near the contacts there is an electronic contribution to the spin voltage and our electronic model does not reproduce the linear inverse spin Hall voltage observed in this system,\cite{Uchida2008:Nature} which suggests that a mechanism different from electronic spin diffusion is responsible for the detected spin Hall voltage.\cite{Hatami2010:SolidStateCommunications} Also, the {\lq\lq entropic\rq\rq} terms in the spin accumulation as introduced in Ref.~\onlinecite{Uchida2009:JoAP}, which would lead to a uniform decay of the spin accumulation across the whole sample, not just at the distances of the spin diffusion lengths off of the edges, do not arise in our theory. 

\section{F/N junctions}\label{FN_junction}
\subsection{F/N junctions placed in thermal gradients}\label{FN_junction_thermal_gradient}
\begin{figure}[t]
\centering
\includegraphics*[width=8cm]{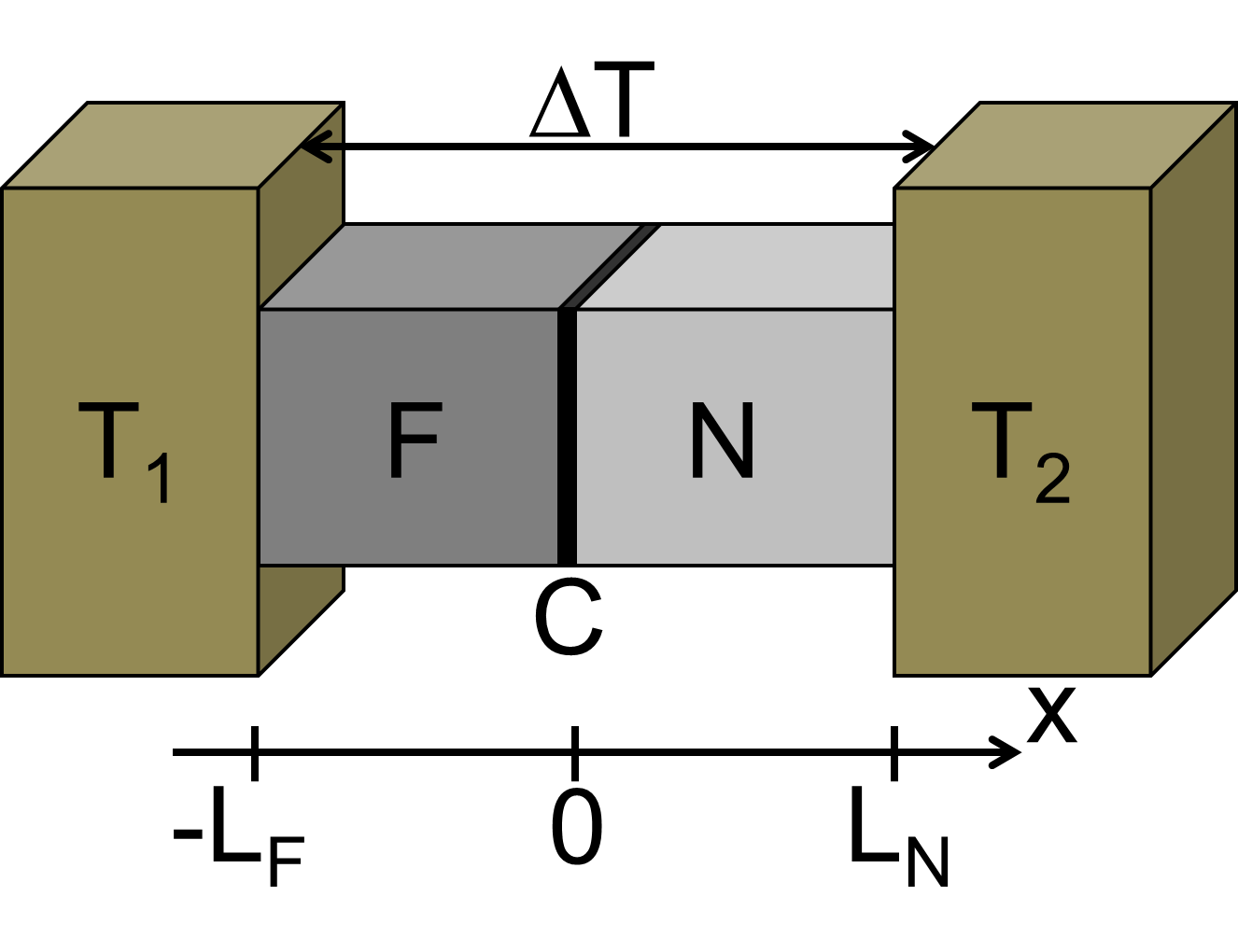}
\caption{(Color online) A schematic illustration of a F/N junction placed in a thermal gradient.}\label{fig:FNjunction}
\end{figure}

In this section we investigate an open ($j=0$) F/N junction under a thermal gradient. The F/N junction consists of a ferromagnet and a nonmagnetic conductor, denoted by the additional subscripts F and N in the quantities defined in the previous sections. The extension of the ferromagnet is given by $-L_F<x<0$, whereas the nonmagnetic conductor is described by values $0<x<L_N$. We also assume that the properties of the contact region C, located at $x=0$, are known. By coupling the F and N regions to reservoirs with different temperatures, $T_2$ and $T_1$ respectively, a temperature gradient is created across the junction. The model investigated in the following is summarized in Fig.~\ref{fig:FNjunction}.

Like in the previous section, we can assume uniform (but for each region different) temperature gradients $\nabla T_F$ and $\nabla T_N$ and use the simplified spin diffusion equation, Eq.~(\ref{simplified_spin_diffusion_equation}), and the corresponding solutions, Eqs.~(\ref{simple_solution_temperature})-(\ref{simple_solution_chemicalpotential}), to describe the total chemical potential, the spin accumulation, and the temperature profile in each region separately. The integration constants are solved invoking the following boundary conditions: $T(-L_F)=T_1$, $T(L_N)=T_2$, and $j_s(-L_F)=j_s(L_N)=0$. Furthermore, we use Eqs.~(\ref{charge_current_contact})-(\ref{heat_current_contact}) and assume, as in the standard spin injection model,\cite{Fabian2007:APS} that the charge, spin, and heat currents are continuous at the interface, giving us five additional equations for the integration constants. From this set of equations the integration constants, including the gradients $\nabla T_F$ and $\nabla T_N$, can be obtained. Depending on the choice of the direction of the gradient, one finds that spin is either injected from the F region into the N region or extracted from the N region by a pure spin current, that is, a spin current without accompanying charge current.

In order to measure the efficiency of the thermal spin injection [$j_s(0)<0$] and extraction [$j_s(0)>0$] at the interface, we calculate the \emph{thermal spin injection efficiency} $\kappa=j_s(x=0)/\nabla T_N$, which corresponds to a spin thermal conductivity. Our model gives
\begin{widetext}
\begin{equation}\label{FN_junction_spin_injection_efficiency}
\kappa=-\frac{\sigma_N}{2}\;\frac{\tanh\left(L_N/\lambda_{sN}\right)\left\{\tanh\left(L_F/\lambda_{sF}\right)S_{sc}R_c\left(1-P_{\Sigma}^2\right)+\left[1-\cosh^{-1}\left(L_F/\lambda_{sF}\right)\right]S_{sF}R_F\left(1-P_{\sigma F}^2\right)\right\}}{R_F\tanh\left(L_N/\lambda_{sN}\right)+R_c\tanh\left(L_N/\lambda_{sN}\right)\tanh\left(L_F/\lambda_{sF}\right)+R_N\tanh\left(L_F/\lambda_{sF}\right)},
\end{equation}
\end{widetext}
with the effective resistances for the F, N, and contact regions,
\begin{eqnarray} 
R_N & = & \lambda_{sN}/\sigma_N, \\
R_F & = & \lambda_{sF}/\left[\sigma_F(1-P_{\sigma F}^2)\right], \\
R_c & = & 1/\left[\Sigma_{c}(1-P_{\Sigma}^2)\right],
\end{eqnarray}  
and the contact conductance spin polarization 
\begin{equation}
P_{\Sigma}=\Sigma_{sc}/\Sigma_c.
\end{equation} 
Equation~(\ref{FN_junction_spin_injection_efficiency}) has been derived in the limit of $S_{\lambda F/N/c}\ll\sqrt{\mathcal{L}}$, in which the temperature gradients are given by
\begin{equation}\label{FN_junction_F_region_gradient}
\nabla T_F=\frac{\Delta T}{\sigma_F\mathcal{R}_{FN}},
\end{equation}
\begin{equation}\label{FN_junction_N_region_gradient}
\nabla T_N=\frac{\Delta T}{\sigma_N\mathcal{R}_{FN}},
\end{equation}
where
\begin{equation}\label{FN_junction_resistance}
\mathcal{R}_{FN}=\frac{L_F}{\sigma_F}+\frac{1}{\Sigma_c}+\frac{L_N}{\sigma_N}.
\end{equation}

\begin{figure}[t]
\includegraphics*[clip,trim=1cm 1cm 1cm 15cm,width=8cm]{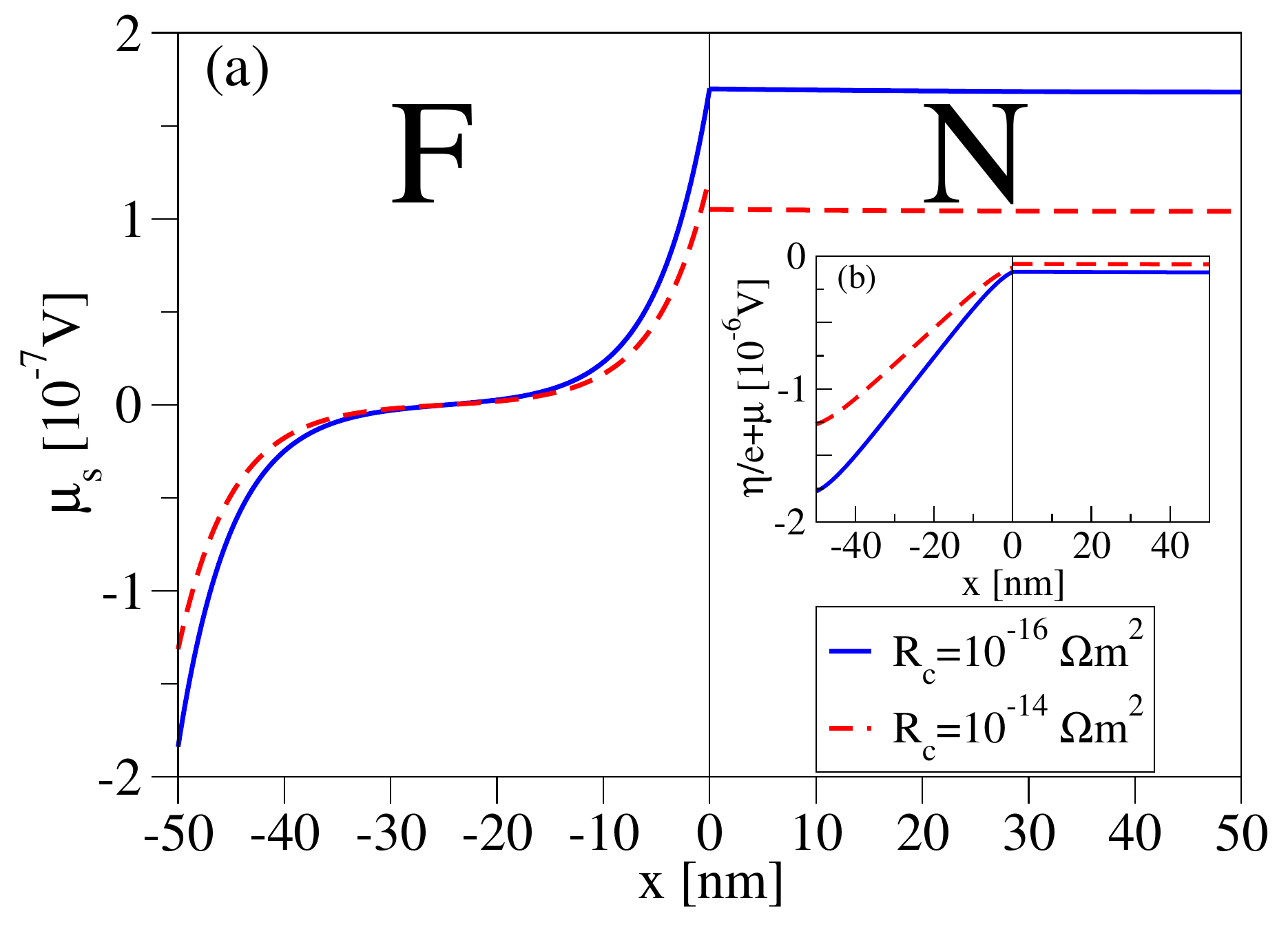}
\caption{(Color online) Profiles of the spin accumulation (a) and the total chemical potential (b) for a Ni$_{81}$Fe$_{19}$/Cu junction at $T=300$ K with $L_F=L_N=50$ nm and $\Delta T=-100$ mK. The solid lines show the results for $R_c=1\times10^{-16}\;\Omega\mathrm{m}^2$, the dashed lines for $R_c=1\times10^{-14}\;\Omega\mathrm{m}^2$.}\label{fig:FN_junction_potentials_graphs}
\end{figure}

\begin{figure}[t]
\includegraphics*[clip,trim=1cm 1cm 1cm 15cm,width=8cm]{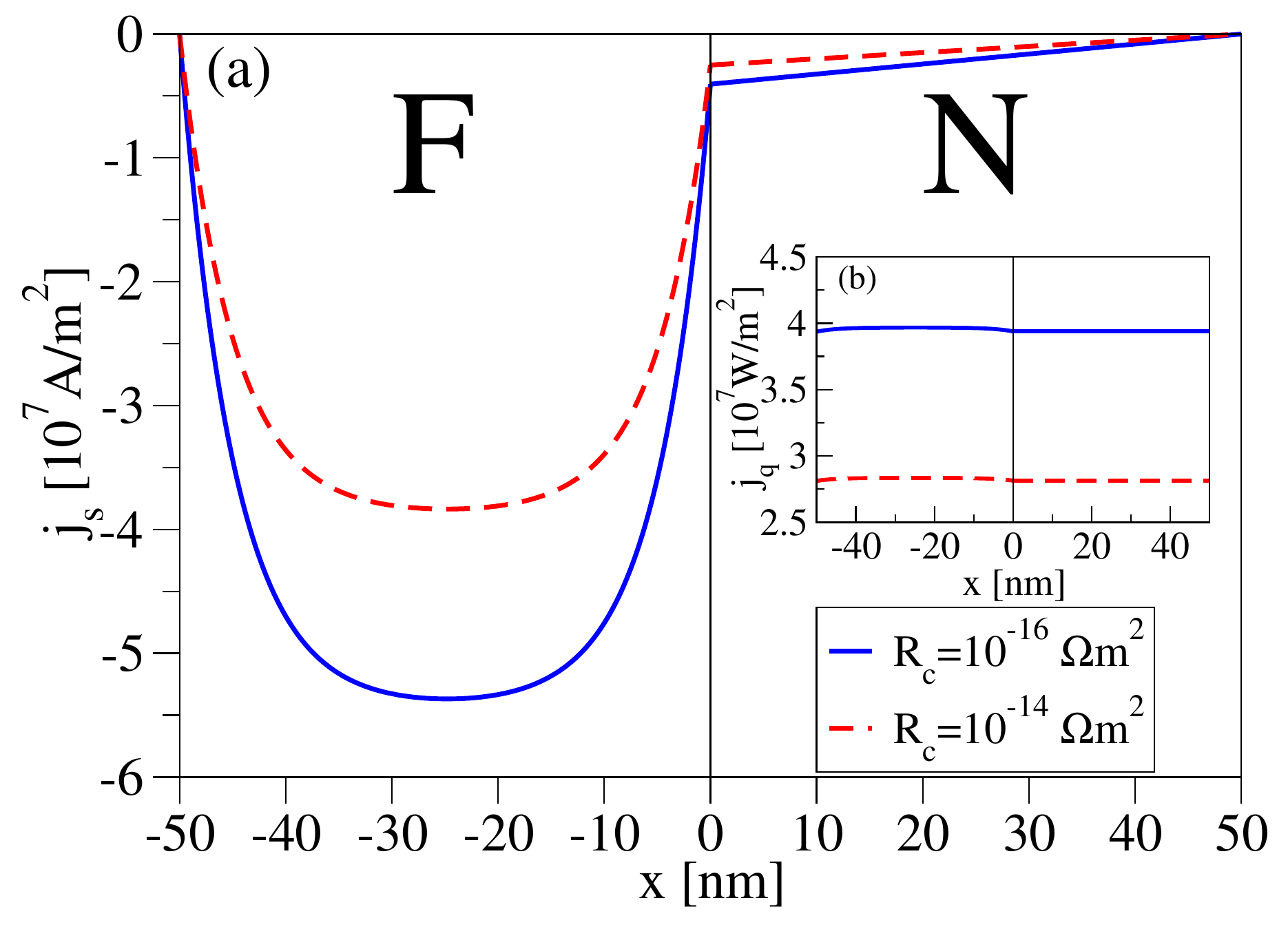}
\caption{(Color online) Profiles of the spin current (a) and the heat current (b) for a Ni$_{81}$Fe$_{19}$/Cu junction at $T=300$ K with $L_F=L_N=50$ nm and $\Delta T=-100$ mK. The solid lines show the results for $R_c=1\times10^{-16}\;\Omega\mathrm{m}^2$, the dashed lines for $R_c=1\times10^{-14}\;\Omega\mathrm{m}^2$.}\label{fig:FN_junction_currents_graphs}
\end{figure}

If the sample sizes are large, that is, if $L_F\gg\lambda_{sF}$ and $L_N\gg\lambda_{sN}$, as is usually the case (but not in Figs.~\ref{fig:FN_junction_potentials_graphs} and~\ref{fig:FN_junction_currents_graphs} where $L_N<\lambda_{sN}$), the situation at the interface is not sensitive to the boundary conditions far away from the interface and Eq.~(\ref{FN_junction_spin_injection_efficiency}) reduces to
\begin{equation}\label{FN_junction_spin_injection_efficiency_large_device}
\begin{aligned}
\kappa&=-\frac{\sigma_N}{2}\;\frac{S_{sc}R_c\left(1-P_{\Sigma}^2\right)+S_{sF}R_F\left(1-P_{\sigma F}^2\right)}{R_F+R_c+R_N}\\
&=-\frac{\sigma_N}{2}\langle S_s(1-P_\sigma^2)\rangle_R,
\end{aligned}
\end{equation}
where $\langle ...\rangle_R$ denotes an average over the effective resistances. The above expressions for the spin injection efficiency and the gradients, Eqs.~(\ref{FN_junction_spin_injection_efficiency})-(\ref{FN_junction_spin_injection_efficiency_large_device}), could have also been obtained by using Eqs.~(\ref{solution_spinpotential})-(\ref{solution_chemicalpotential}) to calculate the profiles and taking the limit $S_{\lambda F/N/c}\ll\sqrt{\mathcal{L}}$. Equation~(\ref{FN_junction_spin_injection_efficiency_large_device}) is the spin-heat coupling equivalent of the well-known formula for the electrical spin injection efficiency.\cite{Zutic2004:RMP, Fabian2007:APS}

Using the spin injection efficiency, Eq.~(\ref{FN_junction_spin_injection_efficiency}) [or Eq.~(\ref{FN_junction_spin_injection_efficiency_large_device}) for large devices], the profiles of the spin current and accumulation in the N region ($0<x<L_N$) can be written compactly as
\begin{equation}\label{FN_junction_spin_current}
j_s(x)=-\kappa\nabla T_N\frac{\sinh\left[(x-L_N)/\lambda_{sN}\right]}{\sinh(L_N/\lambda_{sN})}
\end{equation}
and
\begin{equation}\label{FN_junction_spin_accumulation}
\mu_s(x)=-R_N\kappa\nabla T_N\frac{\cosh\left[(x-L_N)/\lambda_{sN}\right]}{\sinh(L_N/\lambda_{sN})},
\end{equation}
which reduce to
\begin{equation}\label{FN_junction_spin_current_simplified}
j_s(x)=\kappa\nabla T_N\exp\left(-x/\lambda_{sN}\right)
\end{equation}
and
\begin{equation}\label{FN_junction_spin_accumulation_simplified}
\mu_s(x)=-R_N\kappa\nabla T_N\exp\left(-x/\lambda_{sN}\right)
\end{equation}
for $L_N\gg\lambda_{sN}$. In particular, at the contact the spin accumulation in the nonmagnetic material can be calculated as
\begin{equation}\label{FN_junction_spin_accumulation_simplified_contact_N}
\mu_s(0^+)=-R_N\kappa\nabla T_N\coth\left(L_N/\lambda_{sN}\right).
\end{equation}
Equation~(\ref{FN_junction_spin_injection_efficiency}) also makes it clear that whether there is spin injection or extraction depends not only on the direction of the temperature gradient, but also on the specific materials chosen.

Another quantity of interest is the total drop of the chemical potential across the F/N junction,
\begin{equation}\label{FN_junction_chemical_potential_drop}
\Delta\left(\eta/e+\mu\right)=\left[\eta(T_2)-\eta(T_1)\right]/e+\mu(L_N)-\mu(-L_F),
\end{equation}
because---in analogy to the calculation of the total resistance of the F/N junction in the case of the electrical spin injection\cite{Fabian2007:APS}---it allows us to define the total Seebeck coefficient $S$ of the device, which can be separated into an equilibrium and a nonequilibrium contribution:
\begin{equation}\label{FN_junction_total_Seebeck_coefficient}
\Delta\left(\eta/e+\mu\right)\equiv S\Delta T\equiv\left(S_0+\delta S\right)\Delta T.
\end{equation}
Here
\begin{equation}\label{FN_junction_Seebeck_equilibrium}
S_0=\frac{\left(S_F+S_{sF}P_{\sigma F}\right)\frac{L_F}{\sigma_F}+\left(S_c+S_{sc}P_{\Sigma}\right)\frac{1}{\Sigma_c}+S_N\frac{L_N}{\sigma_N}}{2\mathcal{R}_{FN}}
\end{equation}
denotes the Seebeck coefficient of the F/N junction in the absence of spin accumulation, whereas
\begin{equation}\label{FN_junction_Seebeck_nonequilibrium}
\delta S=\frac{P_{\sigma F}\left[\mu_{s}(-L_F)-\mu_{s}\left(0^-\right)\right]+P_{\Sigma}\left[\mu_s\left(0^-\right)-\mu_s\left(0^+\right)\right]}{\Delta T}
\end{equation}
is the nonequilibrium contribution to the Seebeck coefficient due to spin accumulation. If the extensions of the F/N junction are much larger than the spin diffusion lengths, the nonequilibrium Seebeck coefficient can be expressed as
\begin{equation}\label{FN_junction_Seebeck_nonequilibrium_large_device}
\delta S=\frac{\frac{S_{sF}\lambda_{sF}\left(P_{\Sigma}-2P_{\sigma F}\right)}{2\sigma_F}+\frac{\kappa\left[\left(P_{\Sigma}-P_{\sigma F}\right)R_F+P_{\Sigma}R_N\right]}{\sigma_N}}{\mathcal{R}_{FN}}.
\end{equation}

For illustration, the profiles of the total chemical potential and the spin accumulation are displayed in Fig.~\ref{fig:FN_junction_potentials_graphs} for a junction consisting of Ni$_{81}$Fe$_{19}$ (see Sec.~\ref{ferromagnet} for the corresponding parameters) and Cu ($\lambda_{sN}=350$ nm, $\sigma_N=5.88\times10^7$ 1/$\Omega$m, $S_N=1.84\times10^{-6}$ V/K) with a temperature difference $\Delta T=T_2-T_1=-100$ mK between both ends of the junction and the mean temperature $T=300$ K.\cite{Slachter2010:Nature,FertJaffres2001:PRB,Uchida2009:JoAP} Figure~\ref{fig:FN_junction_currents_graphs} shows the spin and heat currents for the same system. In Figs.~\ref{fig:FN_junction_potentials_graphs} and~\ref{fig:FN_junction_currents_graphs} we have chosen $R_c=1\times10^{-16}\;\Omega\mathrm{m}^2$ and $R_c=1\times10^{-14}\;\Omega\mathrm{m}^2$, as well as $P_{\Sigma}=0.5$, $S_c=-1.0\times10^{-6}$ V/K, and $S_{sc}=0.5S_c$.\cite{FertJaffres2001:PRB} There is a drop of the total chemical potential across the junction [see Fig.~\ref{fig:FN_junction_potentials_graphs}~(b)]. For the chosen parameters spin is injected from the F region into the N region, where nonequilibrium spin accumulates at the F/N interface and decays within the spin diffusion length [see Figs.~\ref{fig:FN_junction_potentials_graphs}~(a) and~\ref{fig:FN_junction_currents_graphs}~(a) where $L_N<\lambda_{sN}$]. By applying the temperature difference $\Delta T$ into the opposite direction, that is, by choosing $T_1<T_2$, the situation reverses and spin would be extracted from the N region. Figure~\ref{fig:FN_junction_potentials_graphs}~(a) also illustrates that the spin accumulation in the N region decreases with increasing contact resistance. The heat current flows from the hot to the cold end of the junction [$j_q(x)>0$], as can be seen in Fig.~(\ref{fig:FN_junction_currents_graphs})~(b). Furthermore, one can observe that in the F region the heat current is not perfectly constant and decreases at $x=-L_F$ as well as at the contact,\footnote{This is due to the assumption of constant temperature gradients and would not be the case if the full model was used.} while in the N region the heat current remains constant.

We now discuss two important cases: transparent and tunnel contacts in large F/N junctions where $L_F\gg\lambda_{sF}$ and $L_N\gg\lambda_{sN}$. For transparent contacts $R_c\ll R_F,R_N$ and the spin injection efficiency reduces to
\begin{equation}\label{FN_junction_spin_injection_efficiency_transparent_contact}
\kappa=-\frac{\sigma_N}{2}\;\frac{S_{sF}R_F\left(1-P_{\sigma F}^2\right)}{R_F+R_N}.
\end{equation}
Thermal electronic spin injection from a ferromagnetic metal to a semiconductor, that is, the case of $R_N \gg R_F$, would suffer from the same "conductivity/resistance mismatch problem"\cite{Zutic2004:RMP,JohnsonSilsbee1987:PRB,Schmidt2000:PRB,Rashba2000:PRB} as the usual electrical spin injection does. The nonequilibrium Seebeck coefficient can then be written as
\begin{equation}\label{FN_junction_Seebeck_nonequilibrium_transparent_contact}
\delta S=-\frac{S_{sF}\lambda_{sF}P_{\sigma F}}{2\sigma_F\left(L_F/\sigma_F+L_N/\sigma_N\right)}\left(1+\frac{R_N}{R_F+R_N}\right).
\end{equation}
In this case $\kappa$ and $\delta S$ are restricted only by the individual effective resistances $R_F$ and $R_N$ of the F and N regions. Moreover, the spin accumulation $\mu_s$ is continuous at transparent contacts, that is, $\mu_s(0^+)=\mu_s(0^-)$ and Eq.~(\ref{FN_junction_spin_accumulation_simplified_contact_N}) yields the expression found in Ref.~\onlinecite{Slachter2010:Nature} for $\mu_s(0)/\nabla T_F$.\footnote{Here we use that $\nabla T_N=(\sigma_F/\sigma_N)\nabla T_F$.} 

Tunnel contacts, on the other hand, have very large effective resistances $R_c\gg R_F,R_N$ for which Eqs.~(\ref{FN_junction_spin_injection_efficiency_large_device}) and~(\ref{FN_junction_Seebeck_nonequilibrium_large_device}) reduce to
\begin{equation}\label{FN_junction_spin_injection_efficiency_tunnel_contact}
\kappa=-\frac{\sigma_N}{2}\;S_{sc}\left(1-P_{\Sigma}^2\right)
\end{equation}
and
\begin{equation}\label{FN_junction_Seebeck_nonequilibrium_tunnel_contact}
\delta S=\frac{\frac{S_{sF}\lambda_{sF}\left(P_\Sigma-2P_{\sigma F}\right)}{2\sigma_F}+\frac{S_{sc}\left(1-P_{\Sigma}^2\right)\left[P_{\sigma F}R_F-P_\Sigma\left(R_F+R_N\right)\right]}{2}}{\mathcal{R}_{FN}}.
\end{equation}
The thermal spin injection efficiency for the tunnel junction is determined by the spin-polarization properties of the contact and the conductivity mismatch issue does not arise in this case. A similar result has also been obtained recently in Ref.~\onlinecite{Jansen2011:preprint}.

\subsection{Interplay between thermal gradients and simultaneous charge currents}\label{FN_junction_gradient_and_current}
\begin{figure}[t]
\centering
\includegraphics*[width=8cm]{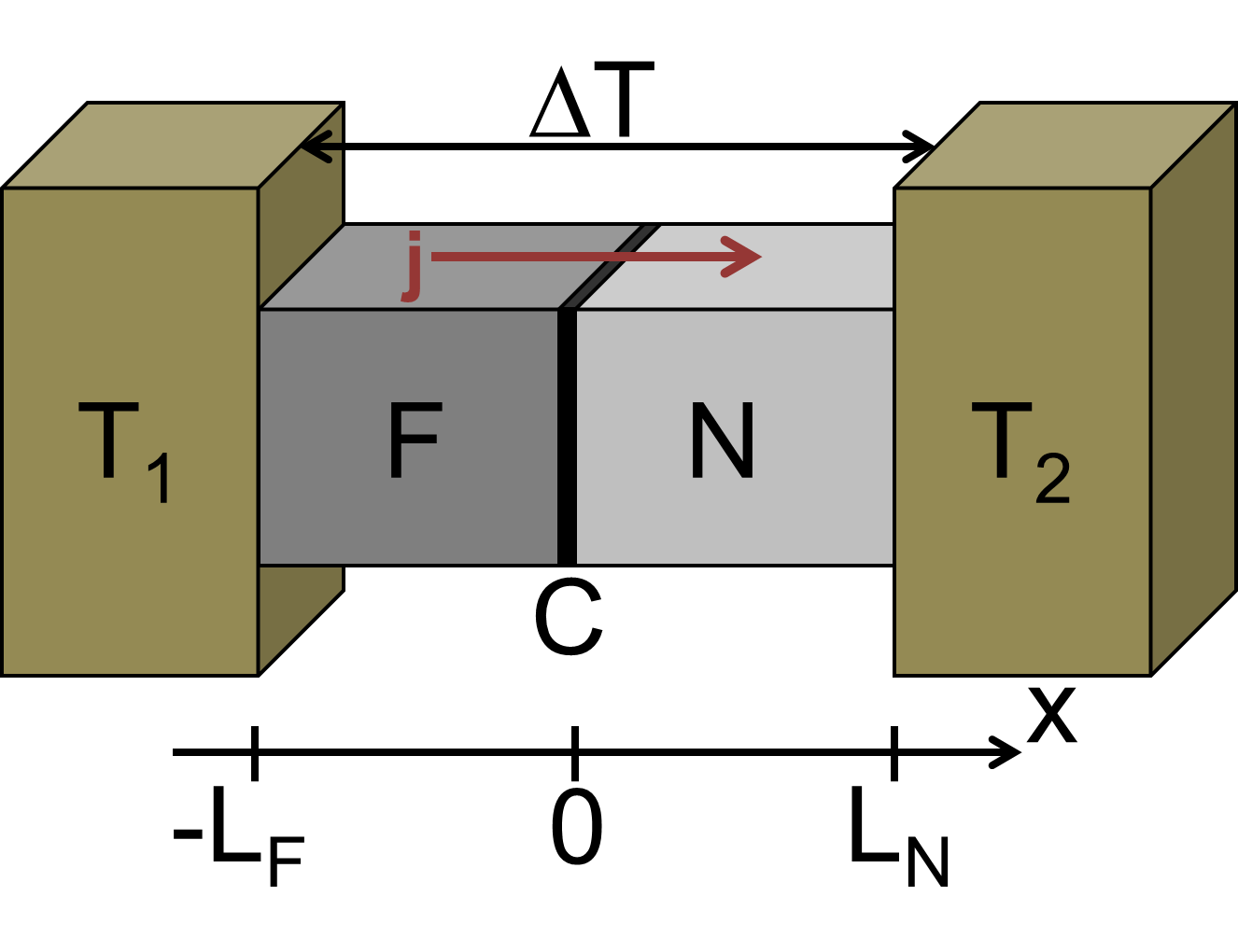}
\caption{(Color online) A schematic illustration of a F/N junction placed in a thermal gradient with a charge current being simultaneously driven through the junction.}\label{fig:FNjunction_gradient_and_current}
\end{figure}

Another interesting effect is the interplay between a thermal gradient across the F/N junction and a simultaneous charge current (see Fig.~\ref{fig:FNjunction_gradient_and_current}). To analyze this process, we take Eqs.~(\ref{simple_solution_temperature})-(\ref{simple_solution_chemicalpotential}), this time with a finite charge current $j$, and replace the boundary condition for the spin current at $x=-L_F$ by $j_s(-L_F)=P_{\sigma F}j$ while leaving the boundary conditions for the temperature unchanged and also taking $j_s(L_N)=0$ as before. By choosing the charge current $j=j_{\rm{com}}$ appropriately, the effects of the charge current and the thermal gradient, each by itself applicable for injecting spin into the N region or extracting spin from it, can cancel each other out. As a result we find that for $L_F\gg\lambda_{sF}$ a charge current
\begin{equation}\label{FN_junction_compensating_current_large_device}
j_{\rm{com}}=\frac{R_F\left(1-P_{\sigma F}^2\right)S_{sF}+R_c\left(1-P_{\Sigma}^2\right)S_s^c}{2\mathcal{R}_{FN}\left(R_FP_{\sigma F}+R_cP_{\Sigma}\right)}\Delta T
\end{equation}
extracts (injects) the spin injected (extracted) through a given temperature difference $\Delta T$ with no net spin current in the N region.

\begin{figure}[t]
\includegraphics*[clip,trim=1cm 1cm 1cm 15cm,width=8cm]{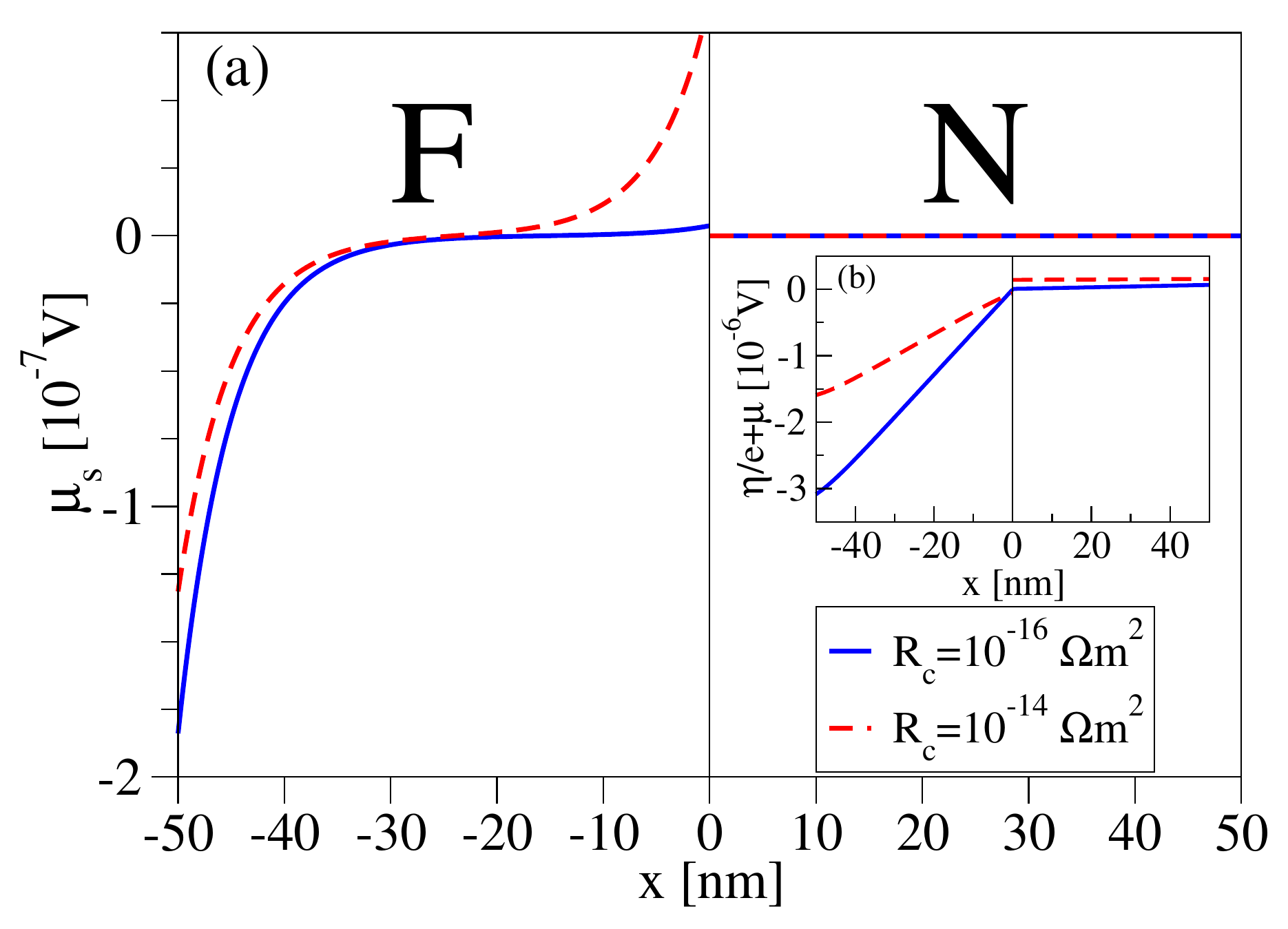}
\caption{(Color online) Profiles of the spin accumulation (a) and the total chemical potential (b) for a Ni$_{81}$Fe$_{19}$/Cu junction at $T=300$ K with $L_F=L_N=50$ nm and $\Delta T=-100$ mK if an electric current compensates the spin accumulation due to the thermal gradient. The solid lines show the results for $R_c=1\times10^{-16}\;\Omega\mathrm{m}^2$, the dashed lines for $R_c=1\times10^{-14}\;\Omega\mathrm{m}^2$.}\label{fig:FN_junction_current_potentials_graphs}
\end{figure}

\begin{figure}[t]
\includegraphics*[clip,trim=1cm 1cm 1cm 15cm,width=8cm]{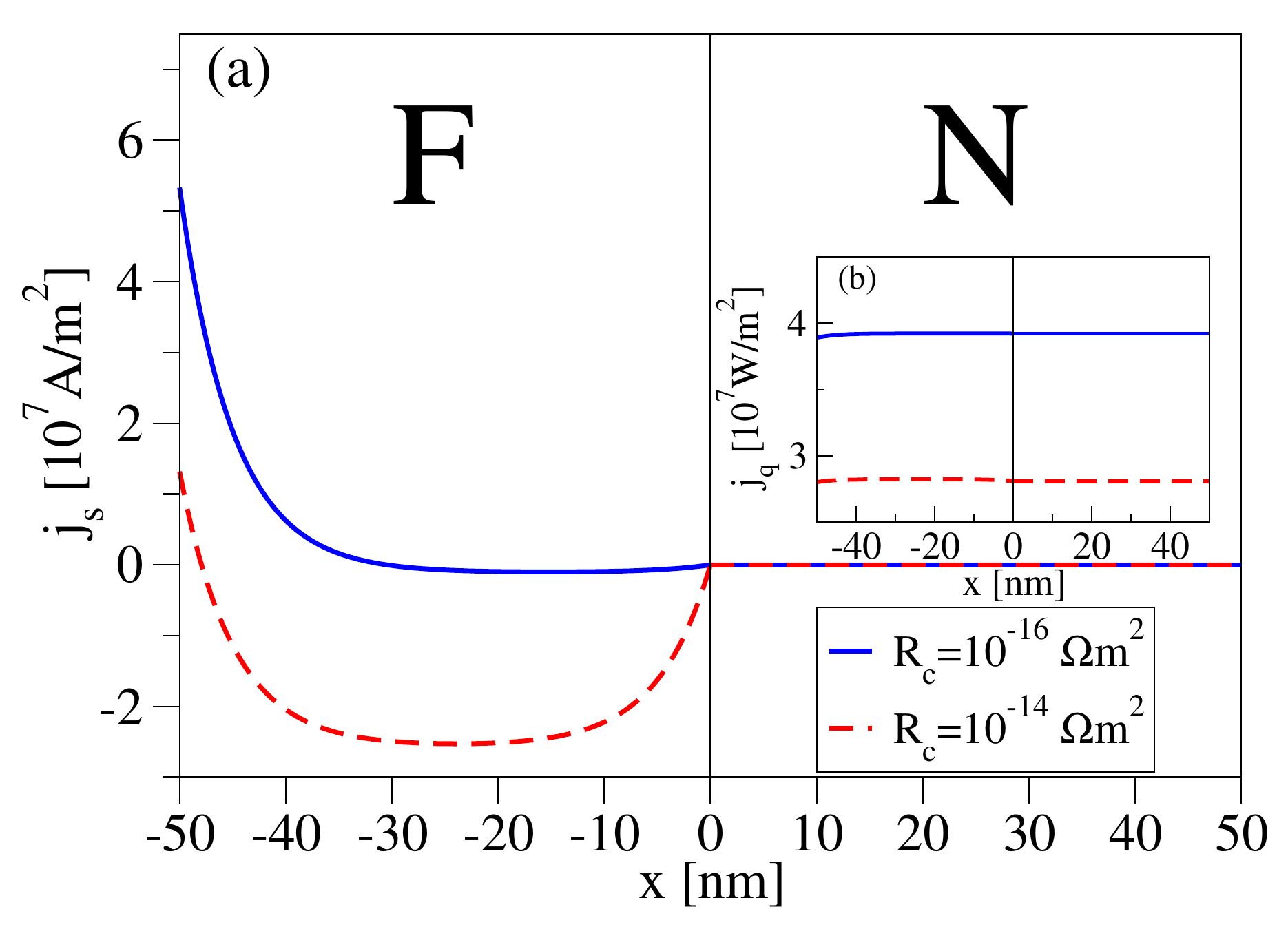}
\caption{(Color online) Profiles of the spin current (a) and the heat current (b) for a Ni$_{81}$Fe$_{19}$/Cu junction at $T=300$ K with $L_F=L_N=50$ nm and $\Delta T=-100$ mK if an electric current compensates the spin accumulation due to the thermal gradient. The solid lines show the results for $R_c=1\times10^{-16}\;\Omega\mathrm{m}^2$, the dashed lines for $R_c=1\times10^{-14}\;\Omega\mathrm{m}^2$.}\label{fig:FN_junction_current_currents_graphs}
\end{figure}

This effect is shown in Figs.~\ref{fig:FN_junction_current_potentials_graphs} and~\ref{fig:FN_junction_current_currents_graphs} for the Ni$_{81}$Fe$_{19}$/Cu junction investigated in this section (see above). We find that a current density of $j_{\rm{com}}=7.6\times10^{7}$ A/m$^2$ ($j_{\rm{com}}=1.9\times10^{7}$ A/m$^2$) is needed to compensate a temperature difference of $\Delta T=-100$ mK if $R_c=1\times10^{-16}\;\Omega\mathrm{m}^2$ ($R_c=1\times10^{-14}\;\Omega\mathrm{m}^2$). Figures~\ref{fig:FN_junction_current_potentials_graphs}~(a) and~\ref{fig:FN_junction_current_currents_graphs}~(a) show that there is no spin accumulation and no spin current in the nonmagnetic material under the compensating electric current condition. The drop of the chemical potential across the F/N junction is shown in Fig.~\ref{fig:FN_junction_current_potentials_graphs}~(b) and the heat current flowing from the hot to the cold end of the junction in Fig.~\ref{fig:FN_junction_current_currents_graphs}~(b). The spin injection compensation should be useful for experimental investigation of the purely electronic contribution to the spin Seebeck effect.

Moreover, we remark that $j_{\rm{com}}$ can be used to describe the efficiency of thermal spin injection if one investigates an open-circuit F/N junction ($j=0$) placed in a thermal gradient as above. In this case the spin current at the interface, $j_s(x=0)$, is described by Eqs.~(\ref{FN_junction_spin_injection_efficiency}) or~(\ref{FN_junction_spin_injection_efficiency_large_device}) respectively. We can then define the ratio between the spin current at the interface and the charge current one would have to drive through the junction to cancel the thermal spin injection, $P=j_s(x=0)/j_{\rm{com}}$. For large devices and $S_{\lambda F/N/c}\ll\sqrt{\mathcal{L}}$ this ratio can be calculated as
\begin{equation}
P=-\langle P_\sigma\rangle_R,
\end{equation}
which represents the negative spin injection efficiency of the electrical spin injection.\cite{Fabian2007:APS}

\subsection{Peltier effects in F/N junctions}\label{FN_junction_Peltier}
As mentioned above, the spin Peltier effect describes the heating or cooling at the interface between a ferromagnetic and normal conductor driven by a spin current.\cite{Flipse2011:preprint} In the following we study several different setups in which a spin current passes through the interface of an isothermal (or nearly isothermal) F/N junction and which therefore give rise to the spin Peltier effect.

\begin{figure}[t]
\centering
\includegraphics*[width=8cm]{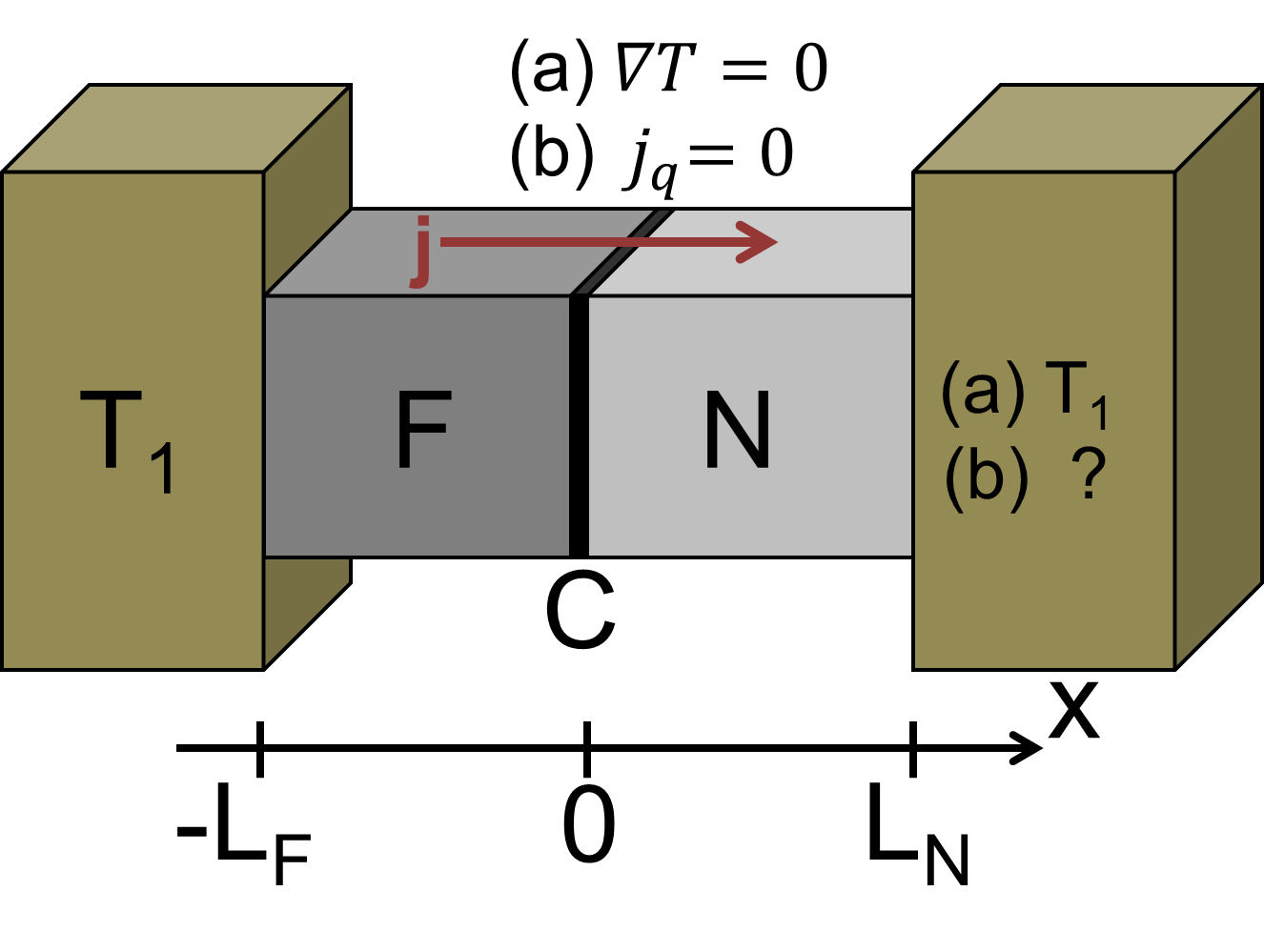}
\caption{(Color online) A schematic illustration of a F/N junction in the electrical spin injection setup, where (a) refers to an isothermal junction and (b) to the situation where $j_q(x)=0$. The fact that in (b) the temperature at one end of the junction is not given as an external boundary condition, but has to be calculated from the model is implied by {\lq\lq ?\rq\rq}.}\label{fig:FNjunction_electrical_spin_injection}
\end{figure}

For every setup investigated in this subsection we assume $L_{N/F}\gg\lambda_{sN/F}$. The first setup considered is the electrical spin injection in a F/N junction: An electric current is driven across an isothermal F/N junction, that is, $\nabla T=0$ [see Fig.~\ref{fig:FNjunction_electrical_spin_injection}~(a)]. Since the entire junction is kept at constant temperature, the continuity of the heat/energy current, Eq.~(\ref{energy_current_continuity}), does not apply and it is sufficient to solve just Eqs.~(\ref{charge_current}), (\ref{spin_current}), and~(\ref{spin_current_continuity}), that is, the formulas obtained for the electrical spin injection can be used. The spin current at the interface is given by\cite{Fabian2007:APS}
\begin{equation}\label{FNjunction_electrical_spin_injection_spin_current}
j_s(0)=\frac{P_{\sigma F}R_F+P_{\Sigma}R_c}{R_F+R_c+R_N}j=\langle P_\sigma\rangle_R j.
\end{equation}
For constant temperature profiles the heat current, Eq.~(\ref{heat_current}), is not continuous at the interface and reads
\begin{equation}\label{FNjunction_electrical_spin_injection_heat_current_F}
j_{q}(x)=\frac{TS_{F}}{2}j+\frac{TS_{sF}}{2}j_s(x),\quad x<0
\end{equation}
and
\begin{equation}\label{FNjunction_electrical_spin_injection_heat_current_N}
j_{q}(x)=\frac{TS_{N}}{2}j,\quad x>0.
\end{equation}
Therefore, the total heat produced/dissipated per time at the interface is given by
\begin{equation}\label{FNjunction_electrical_spin_injection_spin_current_total_heat_production}
\Gamma^{\mathrm{tot}}_q=j_{q}(0^-)-j_{q}(0^+)=\Gamma_q+\Gamma_q^s,
\end{equation}
where
\begin{equation}\label{FNjunction_electrical_spin_injection_spin_current_heat_production_charge}
\Gamma_q=\frac{T\left(S_{F}-S_{N}\right)j}{2}
\end{equation}
and
\begin{equation}\label{FNjunction_electrical_spin_injection_spin_current_heat_production_spin}
\Gamma_q^s=\frac{TS_{sF}\langle P_\sigma\rangle_Rj}{2}
\end{equation}
denote the rates of heat production/dissipation due to the conventional (charge) Peltier and spin Peltier effects.

If the temperature is fixed at just one end of the junction, a temperature drop arises across the F/N junction due to the heat evolution at the interface. In order to estimate this temperature drop, we follow the approach used in Ref.~\onlinecite{Flipse2011:preprint} and investigate the hypothetical situation where no heat enters or leaves the F/N junction and no heat is generated inside the junction, that is, $j_q(x)=0$ [see Fig.~\ref{fig:FNjunction_electrical_spin_injection}~(b)]. For $S_{\lambda}\ll\sqrt{\mathcal{L}}$ the profiles of the chemical potential, the spin accumulation, and the spin current are nearly identical in the cases of an isothermal F/N junction and a F/N junction with $j_q(x)=0$ (see below) and Eq.~(\ref{FNjunction_electrical_spin_injection_spin_current_total_temperature_drop}) should give a good estimate for the temperature difference arising across the junction due to the heating/cooling at the interface.

Thus, instead of $\nabla T=0$, we apply the condition $j_q(x)=0$ for any $x$. This situation requires us to solve the full system of differential equations given by Eqs.~(\ref{charge_current}), (\ref{spin_current}), (\ref{spin_current_continuity}), (\ref{heat_current}), and~(\ref{energy_current_continuity}). Since this situation depends crucially on the heat current [via $j_q(x)=0$], the full solution given by Eqs.~(\ref{solution_spinpotential})-(\ref{solution_chemicalpotential}) has to be used, which---in contrast to the assumption of constant gradients in each region---ensures constant heat currents. The temperature far away from the interface is fixed at a given value for one region [for example, at $T_1$ in the F region as shown in Fig.~\ref{fig:FNjunction_electrical_spin_injection}~(b)]. At the interface we impose the boundary conditions that the charge, spin, and heat currents given by Eqs.~(\ref{charge_current_contact})-(\ref{heat_current_contact}) have to be continuous. As before, we assume that $L_{N/F}\gg\lambda_{sN/F}$, in which case the situation at the interface is not sensitive to the boundary conditions far away from the interface. Thus, we choose $\lim\limits_{x\to\pm\infty}\mu_s(x)=0$ as boundary conditions for convenience.

The quantity we are interested in is the temperature drop across the entire junction, which can be obtained as
\begin{equation}\label{FNjunction_electrical_spin_injection_spin_current_total_temperature_drop}
\Delta T=\Delta T_\mathrm{ch}+\Delta T_s.
\end{equation}
As usual, $S_{\lambda F/N/c}\ll\sqrt{\mathcal{L}}$ and the conventional contribution to the temperature drop then reads
\begin{equation}\label{FNjunction_electrical_spin_injection_spin_current_temperature_drop_charge}
\begin{aligned}
\Delta T_\mathrm{ch}=&\left[\frac{\left(S_{F}+S_{sF}P_{\sigma F}\right)L_F}{2\mathcal{L}\sigma_F}+\frac{S_{c}+S_{sc}P_{\Sigma}}{2\mathcal{L}\Sigma_c}\right.\\
&+\left.\frac{S_{N}L_N}{2\mathcal{L}\sigma_N}\right]j,
\end{aligned}
\end{equation}
while the contribution due to the spin accumulation in the region around the interface can be obtained from
\begin{equation}\label{FNjunction_electrical_spin_injection_spin_current_temperature_drop_spin}
\begin{aligned}
\Delta T_{s}=&\frac{S_{sF}\left(1-P_{\sigma F}^2\right)}{2\mathcal{L}}\mu_s\left(0^-\right)\\
&+\frac{S_{sc}\left(1-P_{\Sigma}^2\right)}{2\mathcal{L}}\left[\mu_s\left(0^+\right)-\mu_s\left(0^-\right)\right].
\end{aligned}
\end{equation}
In this limit the spin current at the interface is given by the same expression as in Eq.~(\ref{FNjunction_electrical_spin_injection_spin_current}) and we find
\begin{equation}\label{FNjunction_electrical_spin_injection_spin_current_temperature_drop_spin_final}
\begin{aligned}
\Delta T_{s}=&\frac{S_{sF}\left(1-P_{\sigma F}^2\right)}{2\mathcal{L}}R_F\left(\langle P_\sigma\rangle_R-P_{\sigma F}\right)j\\
&+\frac{S_{sc}\left(1-P_{\Sigma}^2\right)}{2\mathcal{L}}\left[R_FP_{\sigma F}-\left(R_F+R_N\right)\langle P_\sigma\rangle_R\right]j.
\end{aligned}
\end{equation}

\begin{figure}[t]
\centering
\includegraphics*[clip,trim=1cm 1cm 1cm 15cm,width=8cm]{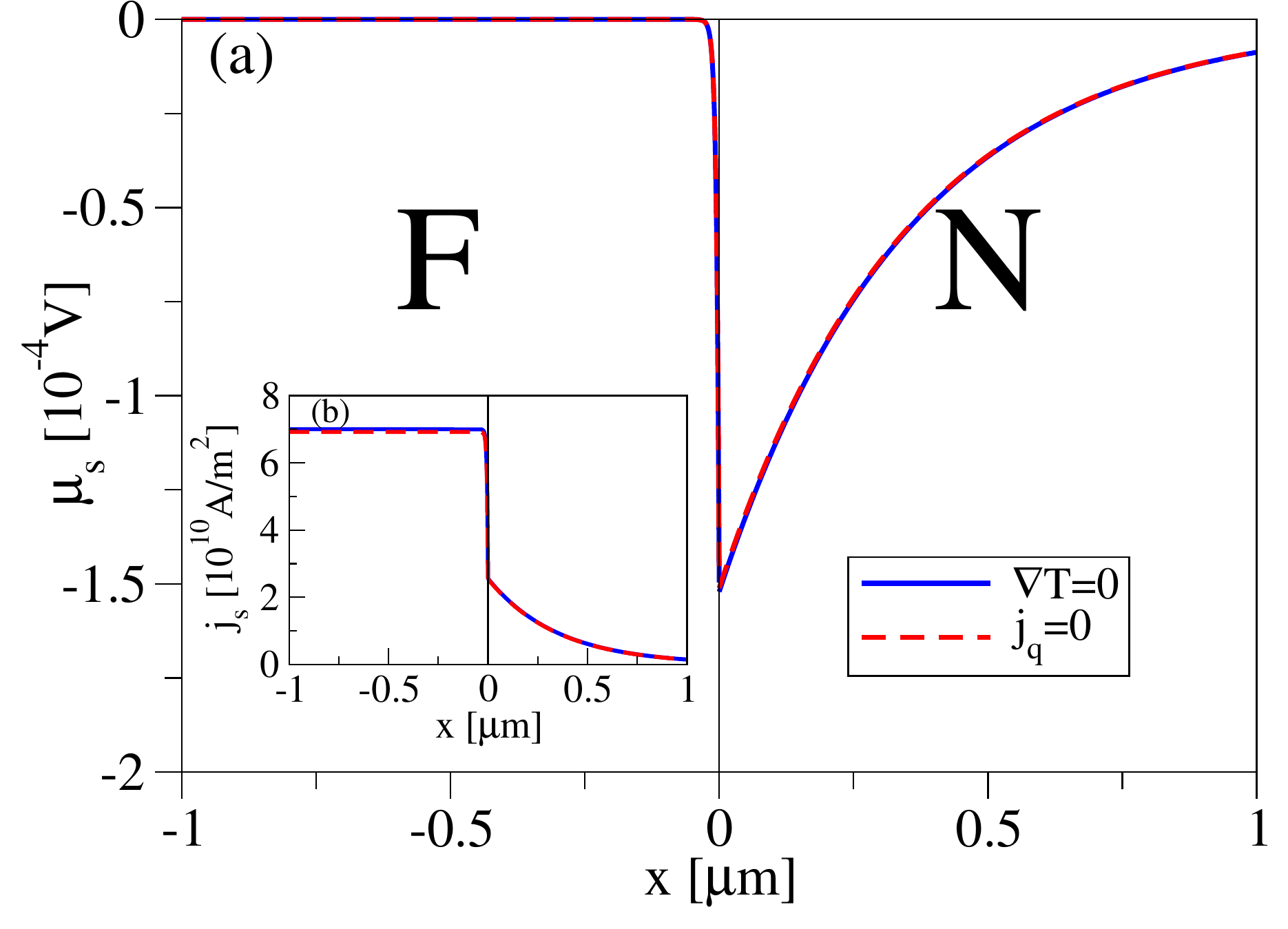}
\caption{(Color online) Profiles of the spin accumulation (a) and the spin current (b) for a Ni$_{81}$Fe$_{19}$/Cu junction with $L_F=L_N=1\;\mu\mathrm{m}$, $R_c=1\times10^{-16}\;\Omega\mathrm{m}^2$, and $j=10^{11}$ A/m$^2$. The solid lines show the results obtained for an isothermal junction at $T=300$ K, whereas the dashed lines show the results obtained for a junction with $j_q(x)=0$ and $T(-L_1)=300$ K.}\label{fig:FNjunction_spin_accumulation_and_current}
\end{figure}

In Fig.~\ref{fig:FNjunction_spin_accumulation_and_current} we display the profiles of the spin accumulation [Fig.~\ref{fig:FNjunction_spin_accumulation_and_current}~(a)] and the spin current [Fig.~\ref{fig:FNjunction_spin_accumulation_and_current}~(b)] in Ni$_{81}$Fe$_{19}$/Cu junctions ($L_F=L_N=1\;\mu\mathrm{m}$, and $R_c=1\times10^{-16}\;\Omega\mathrm{m}^2$) across which a current $j=10^{11}$ A/m$^2$ is driven. As can be seen in Fig~\ref{fig:FNjunction_spin_accumulation_and_current}, the agreement between the solutions of an isothermal junction at $T=300$ K and those of a junction where $j_q(x)=0$ and $T(-L_F)=300$ K is very good, that is, for $S_{\lambda}\ll\sqrt{\mathcal{L}}$ the behavior of the spin accumulation and current is relatively insensitive in these cases. 

\begin{figure}[t]
\centering
\includegraphics*[width=8cm]{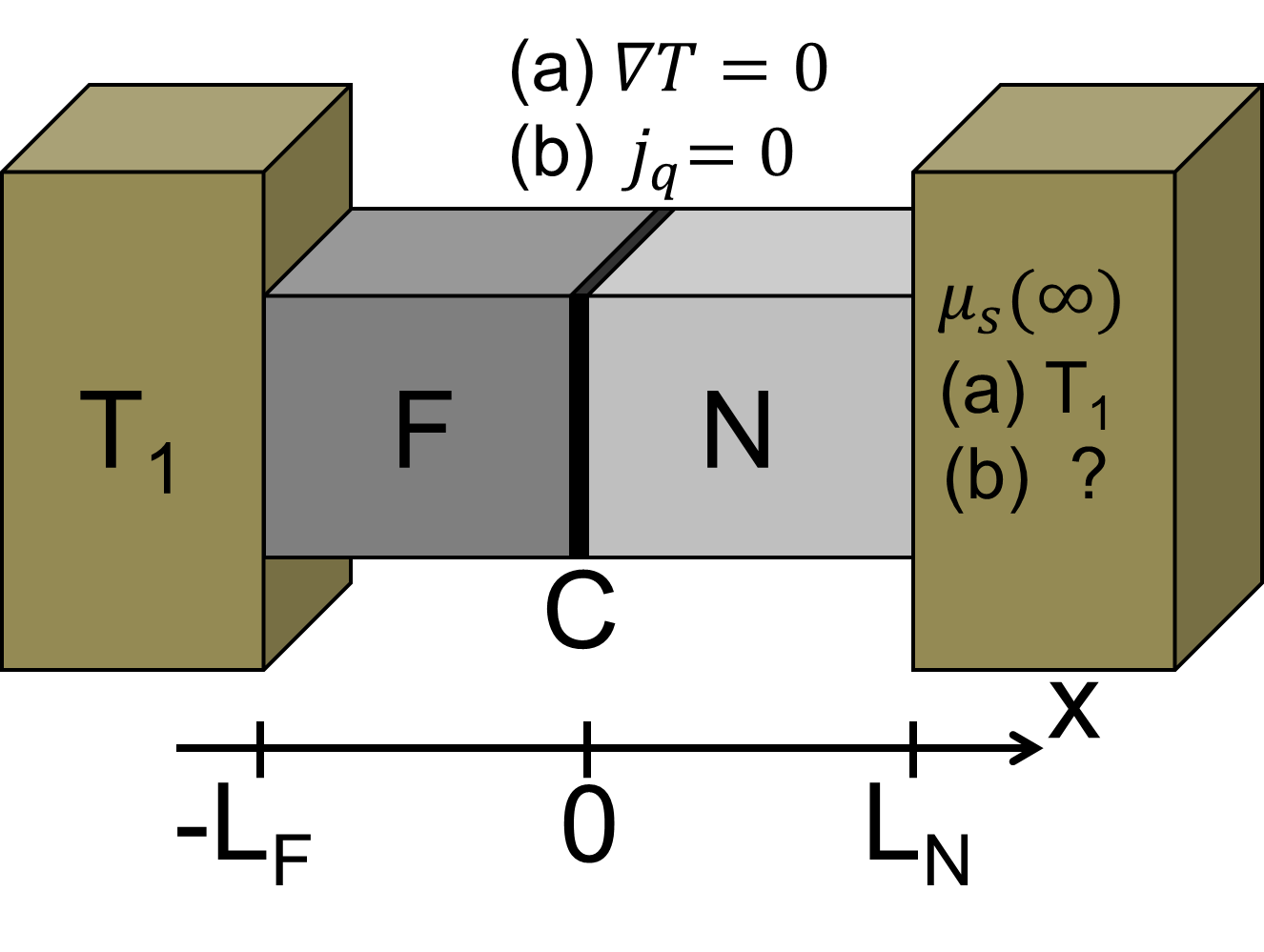}
\caption{(Color online) A schematic illustration of a F/N junction in the Silsbee-Johnson spin-charge coupling setup, where (a) refers to an isothermal junction and (b) to the situation where $j_q(x)=0$. The fact that in (b) the temperature at one end of the junction is not given as an external boundary condition, but has to be calculated from the model is implied by {\lq\lq ?\rq\rq}.}\label{fig:FNjunction_Silsbee_Johnson}
\end{figure}

Having studied the spin Peltier effect in situations where the spin current is driven by an accompanying charge current, we now turn to a different scenario in which we are dealing with a pure spin current ($j=0$) and there consequently is no contribution from the conventional Peltier effect. First, we study heating/cooling effects at the interface of a F/N junction in the Silsbee-Johnson spin-charge coupling setup,\cite{JohnsonSilsbee1985:PRL,Silsbee1980:BMR} that is, we investigate the heat generated at the F/N interface while keeping the temperature constant across the entire structure, $\nabla T=0$ [see Fig.~\ref{fig:FNjunction_Silsbee_Johnson}~(a)]. The inverse process of spin injection, the Silsbee-Johnson spin-charge coupling describes the generation of an electromotive force across the junction due to the presence of nonequilibrium spin in the proximity of the ferromagnet for $j=0$. This nonequilibrium spin in the N region generates a spin current which then drives the spin Peltier effect. For $\nabla T=0$ and the boundary conditions $\mu_{s}\left(-\infty\right)=0$ and $\mu_{s}\left(\infty\right)\neq0$ (modeling the spin accumulation in the N region) the standard model of electrical spin injection yields
\begin{equation}\label{Silsbee_Johnson_spin_current}
j_s(0)=\frac{\mu_{s}\left(\infty\right)}{R_F+R_c+R_N}
\end{equation}
for the spin current at the interface.\cite{Fabian2007:APS,Fabian2009:JuelichWorkshop} Equations~(\ref{FNjunction_electrical_spin_injection_heat_current_F}) and~(\ref{FNjunction_electrical_spin_injection_heat_current_N}), which apply to any case of $\nabla T=0$, show that the heat current vanishes in the N region and the rate of heat flowing to or away from the interface is given by
\begin{equation}\label{Silsbee_Johnson_spin_current_heat_production}
\Gamma_q^{\mathrm{tot}}=\Gamma_q^{s}=\frac{TS_{sF}}{2}\frac{\mu_{s}\left(\infty\right)}{R_F+R_c+R_N}.
\end{equation}

Finally, we look at the Silsbee-Johnson spin-charge coupling setup, but instead of keeping the junction at a constant temperature, we impose the condition $j_q(x)=0$ while keeping one end at a fixed temperature and calculate the temperature drop across the junction [see Fig.~\ref{fig:FNjunction_Silsbee_Johnson}~(b)]. Applying the additional boundary conditions $\lim\limits_{x\to\pm\infty}\mu_s(x)=0$ and requiring the currents to be continuous at the interface, we can use Eqs.~(\ref{FNjunction_electrical_spin_injection_spin_current_total_temperature_drop})-(\ref{FNjunction_electrical_spin_injection_spin_current_temperature_drop_spin}) with $j=0$. Thus, $\Delta T_\mathrm{ch}=0$ and the temperature drop across the junction is entirely due to the spin current/accumulation, $\Delta T=\Delta T_{s}$. We find that the spin current at the interface is given by Eq.~(\ref{Silsbee_Johnson_spin_current}) for $S_{\lambda F/N/c}\ll\sqrt{\mathcal{L}}$ and thus the temperature drop across the junction is given by
\begin{equation}\label{Silsbee_Johnson_temperature_drop_final}
\Delta T=\frac{\langle S_s(1-P_\sigma^2)\rangle_R}{2\mathcal{L}}\mu_{s}\left(\infty\right)=-\frac{\kappa}{\mathcal{L}\sigma_N}\mu_{s}\left(\infty\right),
\end{equation}
where $\kappa$ is the thermal spin injection efficiency of the F/N junction defined in Eq.~(\ref{FN_junction_spin_injection_efficiency_large_device}). Equation~(\ref{Silsbee_Johnson_temperature_drop_final}) is the thermal analog of the Silsbee-Johnson spin-charge coupling. The sign of the temperature drop changes when changing the spin accumulation $\mu_s(\infty)$ from parallel to antiparallel to $\kappa$.

\section{F/N/F junctions}\label{FNF_junction}
\subsection{F/N/F junctions placed in thermal gradients}\label{FNF_junction_thermal_gradient}
\begin{figure}[t]
\centering
\includegraphics*[width=8cm]{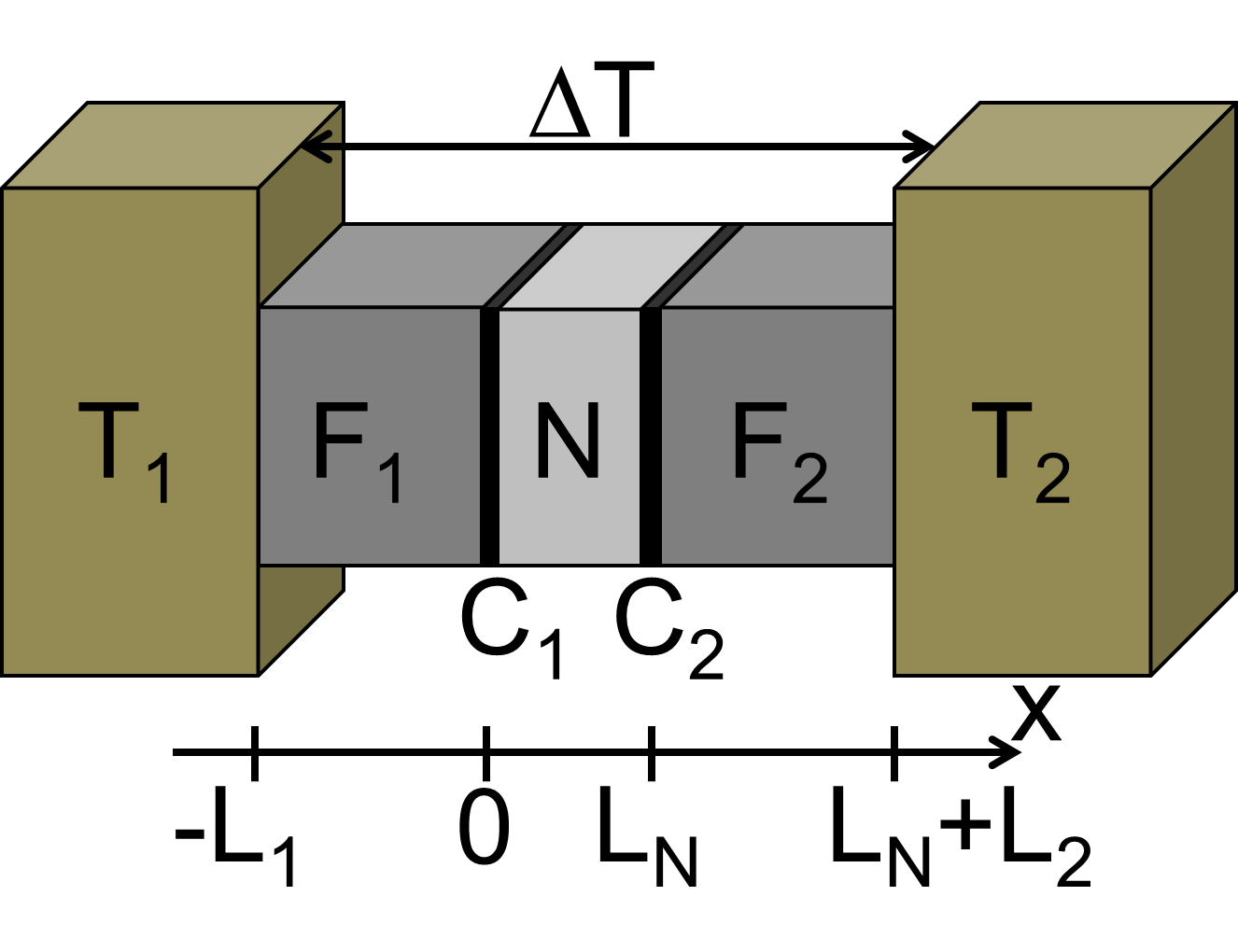}
\caption{(Color online) A schematic illustration of a F/N/F junction placed in a thermal gradient.}\label{fig:FNFjunction}
\end{figure}

The procedure which we used in the previous section to describe spin injection in a F/N junction can also be applied to more complex structures. Here we will discuss spin injection in a F/N/F junction consisting of two ferromagnets F$_1$ and F$_2$ (denoted by the additional subscripts 1 and 2) of lengths $L_{1}$ and $L_{2}$ and a nonmagnetic conductor N (denoted by the additional subscript $N$) of length $L_{N}$ between the ferromagnets. By adjusting the orientations of the magnetization in each ferromagnet independently, the junction can be either in a parallel ($\uparrow\uparrow$) or antiparallel ($\uparrow\downarrow$) configuration, that is, we restrict ourselves to collinear configurations. The interfaces C$_1$ and C$_2$ between the ferromagnets and the nonmagnetic material are located at $x=0$ and $x=L_N$. In Ref.~\onlinecite{Hatami2009:PRB} the influence of electric currents on the temperature profile in such structures has been investigated if both ends of the device were held at the same temperature. Here we consider a different situation: We investigate an open curcuit geometry ($j=0$) in which both ends of the device are coupled to different temperature reservoirs. Holding the opposite ends of the device at different temperatures, $T_2$ and $T_1$, gives rise to temperature gradients across the junction. Figure~\ref{fig:FNFjunction} gives a schematic overview of this geometry.

The chemical potential, the spin accumulation, and the spin current are calculated as in the previous section: Assuming uniform temperature gradients $\nabla T_{1}$, $\nabla T_{2}$, and $\nabla T_N$, we use the simplified spin diffusion equation, Eq.~(\ref{simplified_spin_diffusion_equation}), and fix the integration constants by the boundary conditions $T(-L_{1})=T_1$, $T(L_N+L_{2})=T_2$, and $j_s(-L_1)=j_s(L_N+L_{2})=0$. Each of the contact regions C$_1$ and C$_2$ is characterized by Eqs.~(\ref{charge_current_contact})-(\ref{heat_current_contact}) and we require that the currents are continuous at each interface. This allows us to obtain the profiles of the chemical potential, the spin accumulation, and the spin current.

As in the case of the F/N junction, spin is either injected or extracted at the interfaces between the ferromagnets and the nonmagnetic material. We investigate the spin injection efficiencies, $\kappa_1=j_s(0)/\nabla T_N$ and $\kappa_2=j_s(L_N)/\nabla T_N$, at the contacts C$_1$ and C$_2$. In general, the expressions for $\kappa_1$ and $\kappa_2$ are quite unwieldy, but can be simplified somewhat if we assume the case of $L_{1}\gg\lambda_{s1}$ and $L_{2}\gg\lambda_{s2}$:
\begin{equation}\label{FNF_junction_spin_injection_efficiency}
\begin{aligned}
\kappa_i=&\kappa_i^0R_{FN}^i\frac{R_N\coth\left(L_N/\lambda_{sN}\right)+R_{cj}+R_j}{D_0}\\
&+\kappa_j^0R_{FN}^j\frac{R_N}{D_0\sinh\left(L_N/\lambda_{sN}\right)},
\end{aligned}
\end{equation}
with $i,j=1,2$ and $i\neq j$, the thermal spin injection efficiencies of the individual F/N junctions,
\begin{equation}\label{FNF_junction_spin_injection_efficiency_FN}
\kappa_i^0=-\frac{\sigma_N}{2}\;\frac{S_{sci}R_{ci}\left(1-P_{\Sigma ci}^2\right)+S_{si}R_i\left(1-P_{\sigma i}^2\right)}{R_i+R_{ci}+R_N},
\end{equation}
as defined in Eq.~(\ref{FN_junction_spin_injection_efficiency_large_device}), their effective resistances,
\begin{equation}\label{FNF_junction_effective_resistance_FN}
R_{FN}^i=R_i+R_{ci}+R_N,
\end{equation}
and
\begin{equation}\label{D0}
\begin{aligned}
D_0=&R_N^2+(R_{c1}+R_1)(R_{c2}+R_2)\\
&+\coth\left(L_N/\lambda_{sN}\right)\left(R_1+R_{c1}+R_{c2}+R_2\right)R_N.
\end{aligned}
\end{equation}
Comparing the thermal and electrical\cite{Fabian2009:JuelichWorkshop} spin injection efficiencies of the F/N/F junction, we find that the structure of Eq.~(\ref{FNF_junction_spin_injection_efficiency}) is similar to the structure of the electrical spin injection efficiency. Here the temperature gradient in the N region reads
\begin{equation}\label{FNF_junction_N_region_gradient}
\nabla T_N=\frac{\Delta T}{\sigma_N\mathcal{R}_{FNF}},
\end{equation}
where
\begin{equation}\label{FNF_junction_resistance}
\mathcal{R}_{FNF}=\frac{L_1}{\sigma_1}+\frac{1}{\Sigma_{c1}}+\frac{L_N}{\sigma_N}+\frac{1}{\Sigma_{c2}}+\frac{L_2}{\sigma_2}.
\end{equation}
For a given temperature gradient Eq.~(\ref{FNF_junction_spin_injection_efficiency}) can be used to determine whether there is spin injection [$j_s(0)<0$ or $j_s(L_N)>0$] or extraction [$j_s(0)>0$ or $j_s(L_N)<0$] at the interface C$_i$. The profiles of the spin current and the spin accumulation in the N region ($0<x<L_N$) are
\begin{equation}\label{FNF_junction_spin_current}
\frac{j_s(x)}{\nabla T_N}=\frac{\kappa_2\sinh\left(x/\lambda_{sN}\right)-\kappa_1\sinh\left[\left(x-L_N\right)/\lambda_{sN}\right]}{\sinh\left(L_N/\lambda_{sN}\right)}
\end{equation}
and
\begin{equation}\label{FNF_junction_spin_accumulation}
\frac{\mu_s(x)}{R_N\nabla T_N}=\frac{\kappa_2\cosh\left(x/\lambda_{sN}\right)-\kappa_1\cosh\left[\left(x-L_N\right)/\lambda_{sN}\right]}{\sinh\left(L_N/\lambda_{sN}\right)}.
\end{equation}
If $L_N\gg\lambda_{sN}$, Eq.~(\ref{FNF_junction_spin_injection_efficiency}) reduces to Eq.~(\ref{FN_junction_spin_injection_efficiency_large_device}), that is, the spin injection efficiency of a simple F/N junction.

In analogy to the procedure employed in Sec.~\ref{FN_junction} we can calculate the drop of the chemical potential across the F/N/F junction,
\begin{equation}\label{FNF_junction_chemical_potential_drop}
\Delta\left(\eta/e+\mu\right)=\left[\eta(T_2)-\eta(T_1)\right]/e+\mu(L_N+L_2)-\mu(-L_1),
\end{equation}
and relate this drop to the Seebeck coefficient $S$ of the entire device,
\begin{equation}\label{FNF_junction_total_Seebeck_coefficient}
\Delta\left(\eta/e+\mu\right)\equiv S\Delta T\equiv\left(S_0+\delta S\right)\Delta T,
\end{equation}
which we split into the equilibrium contribution $S_0$ and a nonequilibrium contribution $\delta S$ due to spin accumulation. By investigating the chemical potential drops in the different regions and at the contacts we obtain the equilibrium and nonequilibrium Seebeck coefficients,
\begin{widetext}
\begin{equation}\label{FNF_junction_Seebeck_equilibrium}
S_0=\frac{\left(S_1+S_{s1}P_{\sigma 1}\right)L_1/\sigma_1+\left(S_{c1}+S_{sc1}P_{\Sigma 1}\right)/\Sigma_{c1}+S_NL_N/\sigma_N+\left(S_{c2}+S_{sc2}P_{\Sigma 2}\right)/\Sigma_{c2}+\left(S_2+S_{s2}P_{\sigma 2}\right)L_2/\sigma_2}{2\mathcal{R}_{FNF}}
\end{equation}
and
\begin{equation}\label{FNF_junction_Seebeck_nonequilibrium}
\begin{aligned}
\delta S=&\left\{P_{\Sigma1}\left[\frac{R_1\kappa_1}{\sigma_N}+\frac{S_{s1}\lambda_{s1}}{2\sigma_1}-\lambda_{sN}\frac{\kappa_2-\kappa_1\cosh(L_N/\lambda_{sN})}{\sinh(L_N/\lambda_{sN})}\right]+P_{\Sigma2}\left[\frac{R_2\kappa_2}{\sigma_N}+\frac{S_{s2}\lambda_{s2}}{2\sigma_2}-\lambda_{sN}\frac{\kappa_1-\kappa_2\cosh(L_N/\lambda_{sN})}{\sinh(L_N/\lambda_{sN})}\right]\right.\\
&\left.-P_{\sigma1}\left(\frac{\kappa_1R_1}{\sigma_N}+\frac{S_{s1}\lambda_{s1}}{\sigma_1}\right)-P_{\sigma2}\left(\frac{\kappa_2R_2}{\sigma_N}+\frac{S_{s2}\lambda_{s2}}{\sigma_2}\right)\right\}\mathcal{R}_{FNF}^{-1}.
\end{aligned}
\end{equation}
\end{widetext}
Once more, Eq.~(\ref{FNF_junction_Seebeck_nonequilibrium}) has been derived in the limit of $L_{1}\gg\lambda_{s1}$ and $L_{2}\gg\lambda_{s2}$, which usually applies to most devices.

\begin{figure}[t]
\includegraphics*[clip,trim=1cm 1cm 1cm 15cm,width=8cm]{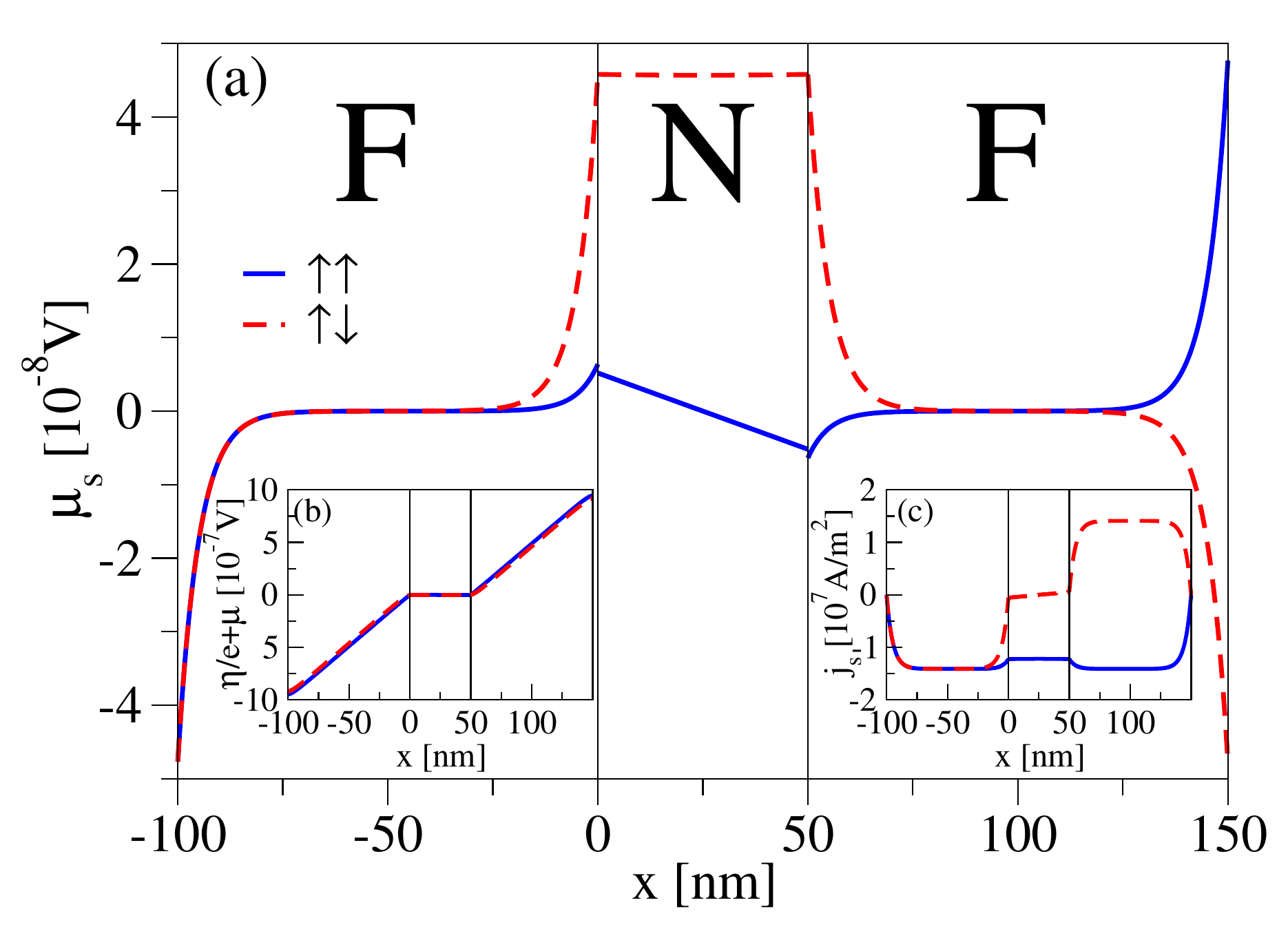}
\caption{(Color online) Profiles of the spin potential (a), the total chemical potential (b), and the spin current (c) for a Ni$_{81}$Fe$_{19}$/Cu/Ni$_{81}$Fe$_{19}$ junction at $T=300$ K with $L_{1}=L_{2}=100$ nm, $L_N=50$ nm, and $\Delta T=-100$ mK. The solid lines show the profiles for the parallel configuration, the dashed lines for the antiparallel configuration.}\label{fig:FNF_junction_graphs}
\end{figure}

Figure~\ref{fig:FNF_junction_graphs} shows the profiles for a symmetric F/N/F junction consisting of Ni$_{81}$Fe$_{19}$ as ferromagnets and Cu as the nonmagnetic material (for the corresponding parameters see Secs.~\ref{ferromagnet} and ~\ref{FN_junction}) for $T=(T_1+T_2)/2=300$ K and $\Delta T=T_2-T_1=-100$ mK. Here the lengths of the individual constituents are chosen to be $L_{1}=L_{2}=100$ nm and $L_N=50$ nm. The contact parameters are $R_{c1}=R_{c2}=1\times10^{-16}\;\Omega\mathrm{m}^2$, $S_{c1}=S_{c2}=-1.0\times10^{-6}$ V/K, and $P_{\Sigma1}=\pm P_{\Sigma2}=0.5$ and $S_{sc1}=\pm S_{sc2}=0.5S_{c1}$ depending on whether the parallel $(+)$ or antiparallel $(-)$ configuration is investigated. As shown in Figs.~\ref{fig:FNF_junction_graphs}~(a) and (c), spin is injected into the N region from both F regions in the antiparallel configuration. If the F/N/F junction is in the parallel configuration spin is injected into the N region from one F region, while at the opposite interface spin is extracted from the N region. Changing the sign of $\Delta T$ would lead to spin extraction from the N region in the antiparallel configuration, whereas spin would still be injected at one interface and extracted at the other interface.
In Fig.~\ref{fig:FNF_junction_graphs}~(b) one can observe a drop of the total chemical potential across the F/N/F junction for both, the parallel and antiparallel configurations. If an asymmetric F/N/F junction (for example by choosing different lengths $L_1$ and $L_2$ or different materials for F$_1$ and F$_2$) is considered, the qualitative properties of Fig.~\ref{fig:FNF_junction_graphs} will remain the same, although the graphs will be distorted compared to the symmetric case.

Next, we look at the difference between the drops of the chemical potential [given by Eq.~(\ref{FNF_junction_chemical_potential_drop})] in the parallel and antiparallel configurations (denoted by the superscripts $i=\uparrow\uparrow,\uparrow\downarrow$ in the following), as a quantitative measure of the
spin accumulation in the N region (thermal analog of the giant magnetoresistance). If one analyses the temperature profile $T(x)$ and the local equilibrium chemical potential $\eta\left[T(x)\right]$, one finds that within our model they are the same for the parallel and antiparallel configurations (in the limit $S_{\lambda j}\ll\sqrt{\mathcal{L}}$). Hence, the difference between the drops of the chemical potential is just the drop of the quasichemical potentials, that is,
\begin{equation}
\Delta\left(\eta/e+\mu\right)^{\uparrow\uparrow}-\Delta\left(\eta/e+\mu\right)^{\uparrow\downarrow}=\Delta\mu^{\uparrow\uparrow}-\Delta\mu^{\uparrow\downarrow},
\end{equation}
where $\Delta\mu^{i}=\mu^{i}(L_N+L_{2})-\mu^{i}(-L_{1})$. Moreover, the equilibrium Seebeck coefficients given by Eq.~(\ref{FNF_junction_Seebeck_equilibrium}) are the same for both configurations and consequently
\begin{equation}\label{}
\Delta\mu^{\uparrow\uparrow}-\Delta\mu^{\uparrow\downarrow}=\left(\delta S^{\uparrow\uparrow}-\delta S^{\uparrow\downarrow}\right)\Delta T,
\end{equation}
which, in the limit of $L_{1}\gg\lambda_{s1}$ and $L_{2}\gg\lambda_{s2}$, yields
\begin{widetext}
\begin{equation}\label{FNF_junction_chemical_potential_difference}
\begin{aligned}
\Delta\mu^{\uparrow\uparrow}-\Delta\mu^{\uparrow\downarrow}=\frac{\left[\left(S_{s1}\lambda_{s1}/\sigma_1+S_{sc1}/\Sigma_{c1}\right)\left(R_{2}P_{2}+R_{c2}P_{\Sigma2}\right)+\left(S_{s2}\lambda_{s2}/\sigma_2+S_{sc2}/\Sigma_{c2}\right)\left(R_{1}P_{1}+R_{c1}P_{\Sigma1}\right)\right]\lambda_{sN}\nabla T_{N}}{D_0\sinh\left(L_N/\lambda_{sN}\right)},
\end{aligned}
\end{equation}
\end{widetext}
if Eq.~(\ref{FNF_junction_Seebeck_nonequilibrium}) is inserted for each of the nonequilibrium Seebeck coefficients. In Eq.~(\ref{FNF_junction_chemical_potential_difference}) as well as in the following we choose to express the system parameters in terms of the parallel configuration (for example, $P_2=P_2^{\uparrow\uparrow}$ etc.). As mentioned before, in our approximation the temperature gradient in the N region, given by Eq.~(\ref{FNF_junction_N_region_gradient}), does not depend on whether the system is in its parallel or antiparallel configuration.

The charge neutrality condition~(\ref{local_charge_neutrality}) enables us to relate $\Delta\mu^{i}$ to the voltage drop measured across the junction, $\Delta\varphi^{i}=\varphi^{i}(L_N+L_2)-\varphi^{i}(-L_1)$. Using this the difference between the voltage drops in both configurations can be written as
\begin{equation}\label{FNF_junction_voltage_drop_difference_general}
\begin{aligned}
\Delta\varphi^{\uparrow\uparrow}-\Delta\varphi^{\uparrow\downarrow}=&\frac{g_{s1}}{g_1}\left(\mu^{\uparrow\uparrow}_{sL}-\mu^{\uparrow\downarrow}_{sL}\right)-\frac{g_{s2}}{g_2}\left(\mu^{\uparrow\uparrow}_{sR}+\mu^{\uparrow\downarrow}_{sR}\right)\\
&-\left(\Delta\mu^{\uparrow\uparrow}-\Delta\mu^{\uparrow\downarrow}\right),
\end{aligned}
\end{equation}
where the shorthand notations $\mu_{sL}^{i}=\mu_{s}^{i}(-L_1)$ and $\mu_{sR}^{i}=\mu_{s}^{i}(L_N+L_2)$ have been introduced. For $L_{1}\gg\lambda_{s1}$ and $L_{2}\gg\lambda_{s2}$ the contributions to Eq.~(\ref{FNF_junction_voltage_drop_difference_general}) originating from the spin accumulation at $x=-L_1$ and $x=L_N+L_2$, $\mu^{\uparrow\uparrow}_{sL}-\mu^{\uparrow\downarrow}_{sL}$ and $\mu^{\uparrow\uparrow}_{sR}+\mu^{\uparrow\downarrow}_{sR}$, are small compared to $\Delta\mu^{\uparrow\uparrow}-\Delta\mu^{\uparrow\downarrow}$ and consequently
\begin{equation}
\Delta\varphi^{\uparrow\uparrow}-\Delta\varphi^{\uparrow\downarrow}\approx-\left(\Delta\mu^{\uparrow\uparrow}-\Delta\mu^{\uparrow\downarrow}\right).
\end{equation}
Thus, one can also measure the difference between the quasichemical potential drops electrostatically, namely as the difference between the voltage drops across the F/N/F junction.

\begin{figure}[t]
\includegraphics*[clip,trim=1cm 1cm 1cm 14.5cm,width=8cm]{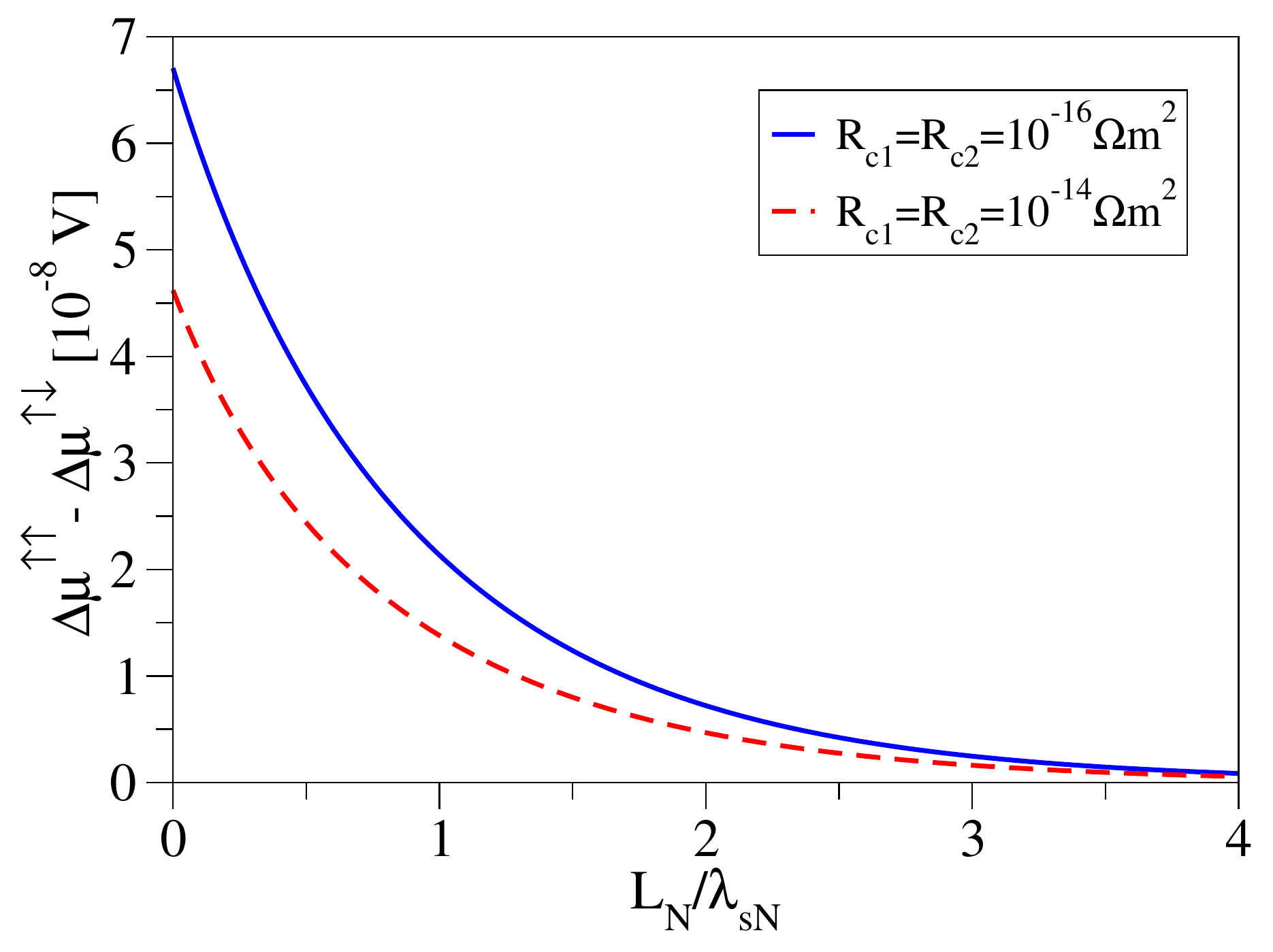}
\caption{(Color online) Difference between the chemical potential drops of the parallel and antiparallel configurations, $\Delta\mu^{\uparrow\uparrow}-\Delta\mu^{\uparrow\downarrow}$, as a function of the length of the N region, $L_N$, for a Ni$_{81}$Fe$_{19}$/Cu/Ni$_{81}$Fe$_{19}$ junction at $T=300$ K with $L_{1}=L_{2}=100$ nm and $\Delta T=-100$ mK.}\label{fig:FNF_junction_voltage_difference}
\end{figure}

Figure~\ref{fig:FNF_junction_voltage_difference} shows the dependence of $\Delta\mu^{\uparrow\uparrow}-\Delta\mu^{\uparrow\downarrow}$ on the length of the N region, $L_N$, for a symmetric Ni$_{81}$Fe$_{19}$/Cu/Ni$_{81}$Fe$_{19}$ junction similar to the one considered above (apart from $L_N$, $R_{c1}$, and $R_{c2}$ the parameters are the same as in Fig.~\ref{fig:FNF_junction_graphs}) for the contact resistances $R_{c1}=R_{c2}=1\times10^{-16}\;\Omega\mathrm{m}^2$ and $R_{c1}=R_{c2}=1\times10^{-14}\;\Omega\mathrm{m}^2$. With increasing length of the N region the amplitude of the voltage difference decreases until, for very large N regions with $L_N\gg\lambda_{sN}$, there is no difference between the voltage drops in the parallel and antiparallel configurations and $\Delta\mu^{\uparrow\uparrow}-\Delta\mu^{\uparrow\downarrow}\to0$. If $L_N$ is comparable or even smaller than the spin diffusion length ($\lambda_{sN}\approx350$ nm in Cu), the voltage drops across the F/N/F junction are different for the different configurations with $\Delta\mu^{\uparrow\uparrow}-\Delta\mu^{\uparrow\downarrow}$ given by Eq.~(\ref{FNF_junction_chemical_potential_difference}).

\subsection{Peltier effects in F/N/F junctions}\label{FNF_junction_Peltier}

\begin{figure}[t]
\centering
\includegraphics*[width=8cm]{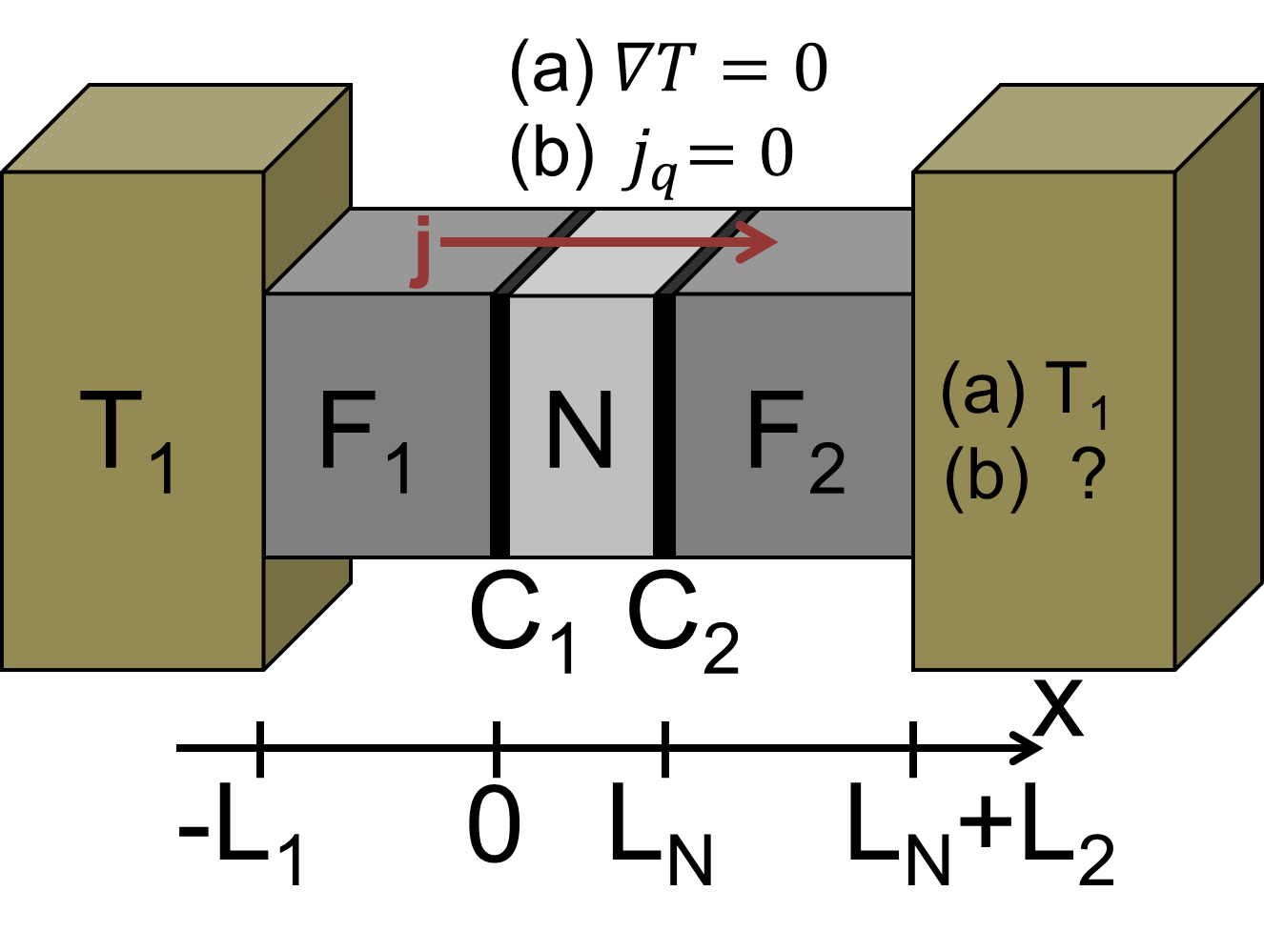}
\caption{(Color online) A schematic illustration of a F/N/F junction in the electrical spin injection setup, where (a) refers to an isothermal junction and (b) to the situation where $j_q(x)=0$. The fact that in (b) the temperature at one end of the junction is not given as an external boundary condition, but has to be calculated from the model is implied by {\lq\lq ?\rq\rq}.}\label{fig:FNFjunction_electrical_spin_injection}
\end{figure}
The section on F/N/F junctions is concluded by a brief discussion of Peltier effects in such structures in the limit of $S_{\lambda j}\ll\sqrt{\mathcal{L}}$ and $L_{1/2}\gg\lambda_{s1/2}$.

Figure~\ref{fig:FNFjunction_electrical_spin_injection}~(a) summarizes the first system considered: A charge current $j$ is driven across an isothermal F/N/F junction and there is heating/cooling the interfaces. Similarly to Sec.~\ref{FN_junction_Peltier}, the electrical spin injection efficiencies at the interfaces, $P_{j1}=j_s(0)/j$ and $P_{j2}=j_s(L_N)/j$, are given by the standard model of electrical spin injection and, as described in detail in Ref.~\onlinecite{Fabian2009:JuelichWorkshop}, read
\begin{equation}\label{FNF_junction_electrical_spin_injection_efficiency}
\begin{aligned}
P_{jk}=&P_{jk}^0R_{FN}^k\frac{R_N\coth\left(L_N/\lambda_{sN}\right)+R_{cl}+R_l}{D_0}\\
&+P_{jl}^0R_{FN}^l\frac{R_N}{D_0\sinh\left(L_N/\lambda_{sN}\right)},
\end{aligned}
\end{equation}
where $D_0$ is given by Eq.~(\ref{D0}) and $k,l=1,2$ and $k\neq l$. The effective resistances of the individual F/N junctions, $R_{FN}^k$, are given by Eq.~(\ref{FNF_junction_effective_resistance_FN}) and their electrical spin injection efficiencies by
\begin{equation}\label{FNF_junction_electrical_spin_injection_efficiency_FN}
P_{jk}^0=\frac{P_{\Sigma k}R_{ck}+P_{\sigma k}R_{k}}{R_k+R_{ck}+R_N}.
\end{equation}
As noted above, the electrical spin injection efficiencies of a F/N/F junction, Eq.~(\ref{FNF_junction_electrical_spin_injection_efficiency}), are composed of the electrical spin injection efficiencies of the individual F/N junctions in the same way the thermal spin injection efficiencies, Eq.~(\ref{FNF_junction_spin_injection_efficiency}), are composed of the thermal spin injection efficiencies of the individual F/N junctions.

\begin{figure}[t]
\includegraphics*[clip,trim=1cm 1cm 1cm 15cm,width=8cm]{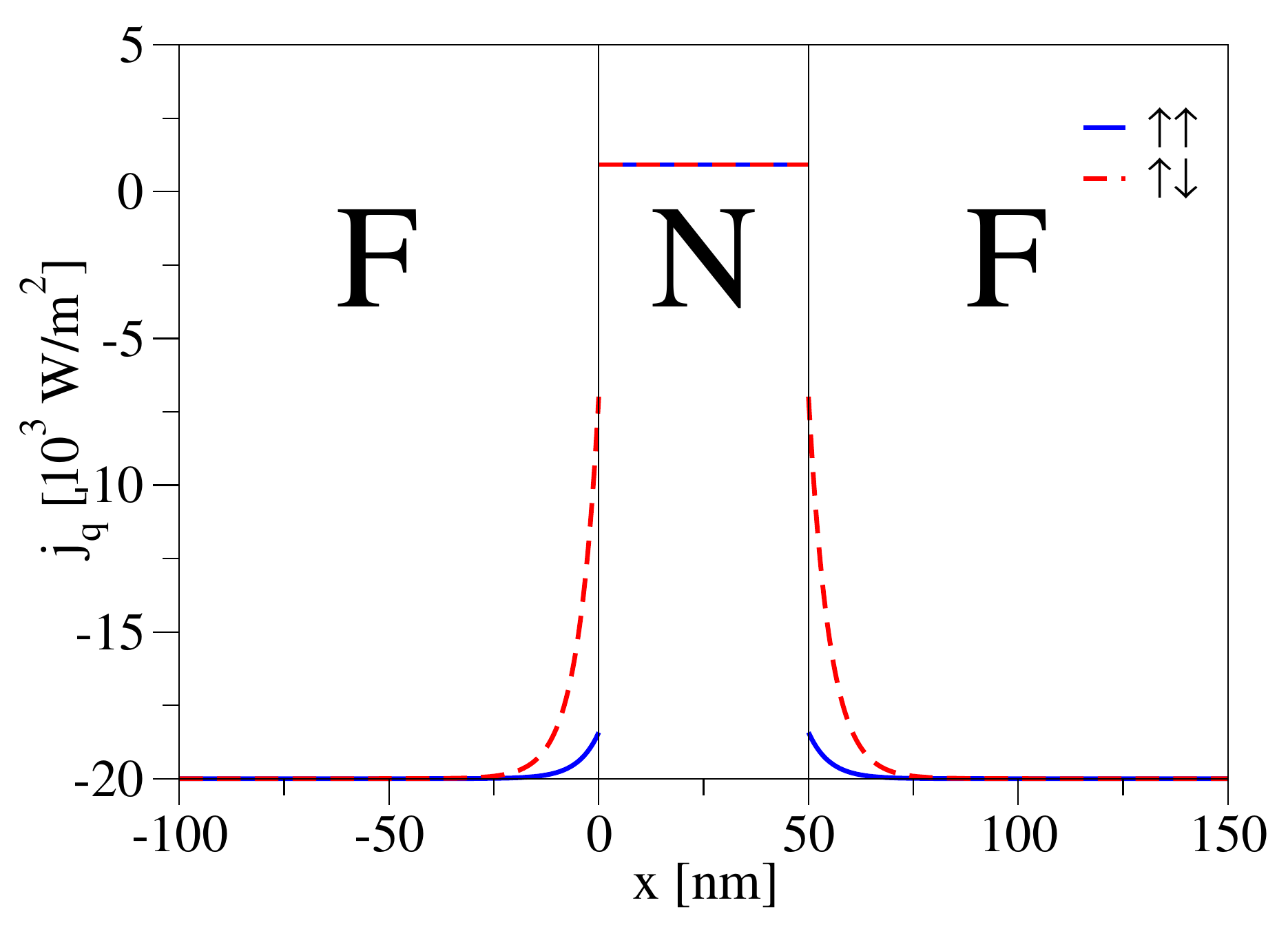}
\caption{(Color online) Profiles of the heat current for an isothermal Ni$_{81}$Fe$_{19}$/Cu/Ni$_{81}$Fe$_{19}$ junction at $T=300$ K with $L_{1}=L_{2}=100$ nm, $L_N=50$ nm, and $j=10^7$ A/m$^2$. The solid line shows the profile for the parallel configuration, the dashed line for the antiparallel configuration.}\label{fig:FNFjunction_electrical_spin_injection_heat}
\end{figure}

Consequently, the rates of heat production/dissipation at contacts C$_1$ and C$_2$ read
\begin{equation}\label{FNFjunction_electrical_spin_injection_spin_current_total_heat_production_c1}
\Gamma^{\mathrm{tot}}_{q1}=j_{q}(0^-)-j_{q}(0^+)=\Gamma_{q1}+\Gamma_{q1}^s,
\end{equation}
\begin{equation}\label{FNFjunction_electrical_spin_injection_spin_current_total_heat_production_c2}
\Gamma^{\mathrm{tot}}_{q2}=j_{q}(L_N^-)-j_{q}(L_N^+)=\Gamma_{q2}+\Gamma_{q2}^s
\end{equation}
and consist of contributions from the conventional Peltier effect,
\begin{equation}\label{FNFjunction_electrical_spin_injection_spin_current_heat_production_charge_c1}
\Gamma_{q1}=\frac{T\left(S_{1}-S_{N}\right)j}{2},
\end{equation}
\begin{equation}\label{FNFjunction_electrical_spin_injection_spin_current_heat_production_charge_c2}
\Gamma_{q2}=\frac{T\left(S_{N}-S_{2}\right)j}{2},
\end{equation}
as well as contributions from the spin Peltier effect,
\begin{equation}\label{FNFjunction_electrical_spin_injection_spin_current_heat_production_spin_c1}
\Gamma_{q1}^s=\frac{TS_{s1}P_{j1}j}{2},
\end{equation}
\begin{equation}\label{FNFjunction_electrical_spin_injection_spin_current_heat_production_spin_c2}
\Gamma_{q2}^s=-\frac{TS_{s2}P_{j2}j}{2}.
\end{equation}
Figure~\ref{fig:FNFjunction_electrical_spin_injection_heat} illustrates this situation for an isothermal Ni$_{81}$Fe$_{19}$/Cu/Ni$_{81}$Fe$_{19}$ junction (in parallel and antiparallel configurations) at $T=300$ K with $L_{1}=L_{2}=100$ nm, $L_N=50$ nm, $R_{c1}=R_{c2}=1\times10^{-16}\;\Omega\mathrm{m}^2$, $S_{c1}=S_{c2}=-1.0\times10^{-6}$ V/K, $P_{\Sigma1}=\pm P_{\Sigma2}=0.5$, $S_{sc1}=\pm S_{sc2}=0.5S_{c1}$, and $j=10^7$ A/m$^2$. The profiles of the heat current in Fig.~\ref{fig:FNFjunction_electrical_spin_injection_heat} show that---for the parameters chosen---there is cooling at C$_1$ ($x=0$) as heat flows away from it, while heat flows to C$_2$ and leads to heating in the region around the C$_2$ ($x=L_N$). The widths of those regions of heating/cooling are given by the individual spin diffusion lengths.

The second system considered is a F/N/F junction where $j_q(x)=0$ and across which an electric current $j$ is driven and one end of which is anchored at a fixed temperature [see Fig.~\ref{fig:FNjunction_electrical_spin_injection}~(b)]. Requiring the charge, spin, and heat currents given by Eqs.~(\ref{charge_current_contact})-(\ref{heat_current_contact}) to be continuous and imposing the additional boundary conditions $\lim\limits_{x\to\pm\infty}\mu_s(x)=0$, we find that the temperature drop across the junction, $\Delta T=\Delta T_\mathrm{ch}+\Delta T_{s}$, is composed of a drop due to the conventional Peltier effect,
\begin{equation}\label{FNFjunction_electrical_spin_injection_temperature_drop_charge}
\begin{aligned}
\Delta T_\mathrm{ch}=&\left[\frac{\left(S_{1}+S_{s1}P_{\sigma 1}\right)L_1}{2\mathcal{L}\sigma_1}+\frac{S_{c1}+S_{sc1}P_{\Sigma 1}}{2\mathcal{L}\Sigma_{c1}}+\frac{S_{N}L_N}{2\mathcal{L}\sigma_N}\right.\\
&+\left.\frac{S_{c2}+S_{sc2}P_{\Sigma 2}}{2\mathcal{L}\Sigma_{c2}}+\frac{\left(S_{2}+S_{s2}P_{\sigma 2}\right)L_2}{2\mathcal{L}\sigma_2}\right]j,
\end{aligned}
\end{equation}
and a contribution due to the spin accumulation in the region around the interfaces,
\begin{equation}\label{FNFjunction_electrical_spin_injection_temperature_drop_spin}
\begin{aligned}
\Delta T_{s}=&\frac{S_{s1}\left(1-P_{\sigma 1}^2\right)}{2\mathcal{L}}\mu_s\left(0^-\right)-\frac{S_{s2}\left(1-P_{\sigma 2}^2\right)}{2\mathcal{L}}\mu_s\left(L_N^+\right)\\
&+\frac{S_{sc1}\left(1-P_{\Sigma 1}^2\right)}{2\mathcal{L}}\left[\mu_s\left(0^+\right)-\mu_s\left(0^-\right)\right]\\
&+\frac{S_{sc2}\left(1-P_{\Sigma 2}^2\right)}{2\mathcal{L}}\left[\mu_s\left(L_N^+\right)-\mu_s\left(L_N^-\right)\right].
\end{aligned}
\end{equation}

Here we are mainly interested in the difference between those temperature drops in configurations of parallel and antiparallel magnetizations of the ferromagnets (denoted by the superscripts $i=\uparrow\uparrow,\uparrow\downarrow$ as in Sec.~\ref{FNF_junction_thermal_gradient}). With the temperature drop due to the conventional Peltier effect being the same for both configurations, this difference is exclusively due to the spin accumulation, that is, $\Delta T^{\uparrow\uparrow}-\Delta T^{\uparrow\downarrow}=\Delta T_s^{\uparrow\uparrow}-\Delta T_s^{\uparrow\downarrow}$, which can be calculated as
\begin{widetext}
\begin{equation}\label{FNF_junction_temperature_drop_difference}
\begin{aligned}
\Delta T^{\uparrow\uparrow}-\Delta T^{\uparrow\downarrow}=\frac{\left[\left(S_{s1}\lambda_{s1}/\sigma_1+S_{sc1}/\Sigma_{c1}\right)\left(R_{2}P_{2}+R_{c2}P_{\Sigma2}\right)+\left(S_{s2}\lambda_{s2}/\sigma_2+S_{sc2}/\Sigma_{c2}\right)\left(R_{1}P_{1}+R_{c1}P_{\Sigma1}\right)\right]R_Nj}{\mathcal{L}D_0\sinh\left(L_N/\lambda_{sN}\right)},
\end{aligned}
\end{equation}
\end{widetext}
where we have expressed the system parameters in terms of the parallel configuration (see Sec.~\ref{FNF_junction_thermal_gradient}).

\begin{figure}[t]
\includegraphics*[clip,trim=1cm 1cm 1cm 15cm,width=8cm]{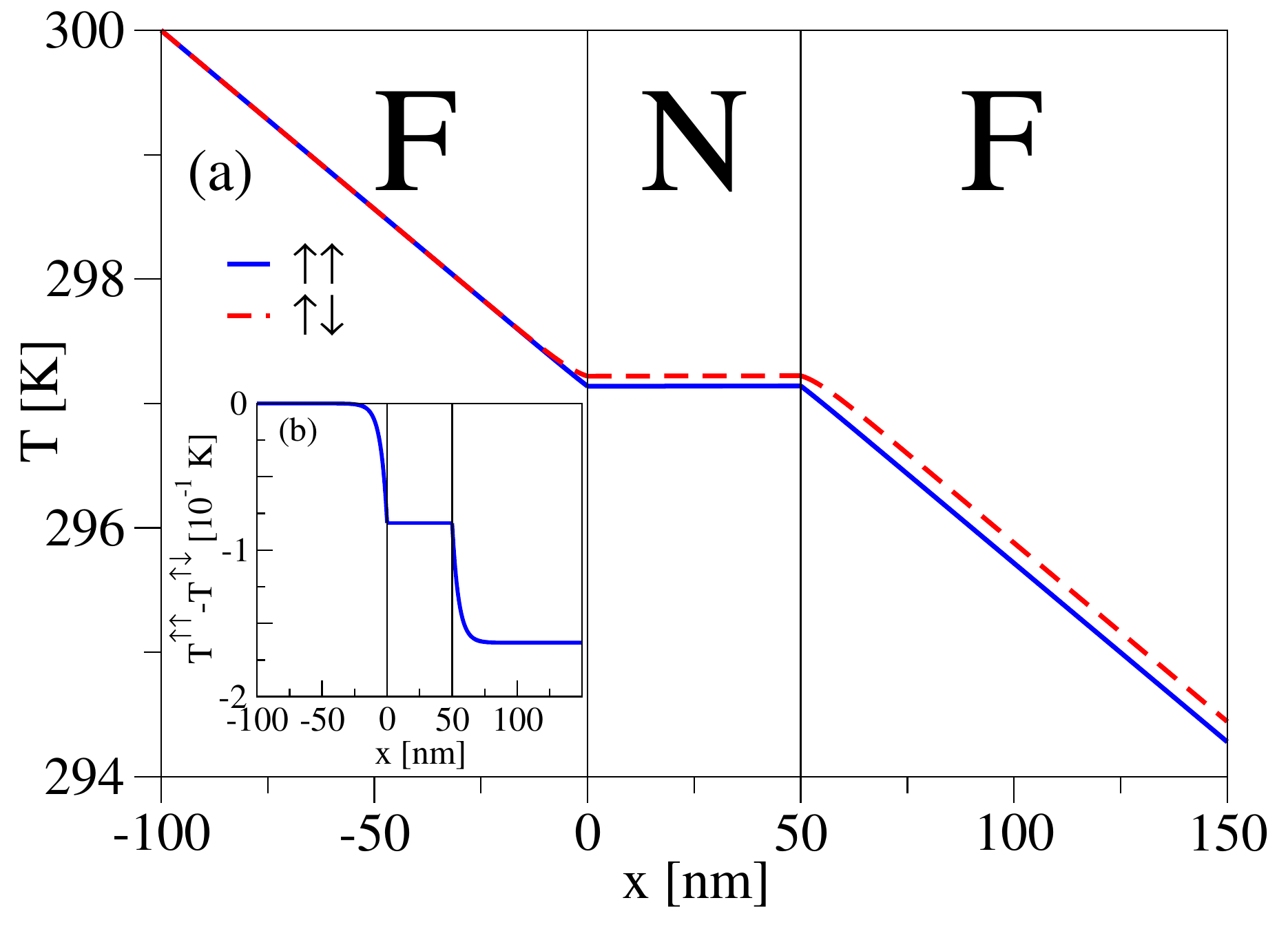}
\caption{(Color online) Temperature profile (a) of a Ni$_{81}$Fe$_{19}$/Cu/Ni$_{81}$Fe$_{19}$ junction with with $j_q(x)=0$, $T_1=300$ K, $L_{1}=L_{2}=100$ nm, $L_N=50$ nm, and $j=10^{11}$ A/m$^2$. The solid line shows the profile for the parallel configuration, the dashed line for the antiparallel configuration. The profile of the temperature difference between the parallel and antiparallel configurations is shown in the inset~(b).}\label{fig:FNFjunction_electrical_spin_injection_temperature}
\end{figure}

For illustration the temperature profiles of a Ni$_{81}$Fe$_{19}$/Cu/Ni$_{81}$Fe$_{19}$ junction at $T=300$ K with $j_q(x)=0$, $L_{1}=L_{2}=100$ nm, $L_N=50$ nm, $R_{c1}=R_{c2}=1\times10^{-16}\;\Omega\mathrm{m}^2$, $S_{c1}=S_{c2}=-1.0\times10^{-6}$ V/K, $P_{\Sigma1}=\pm P_{\Sigma2}=0.5$, $S_{sc1}=\pm S_{sc2}=0.5S_{c1}$, and $j=10^{11}$ A/m$^2$ are shown in Fig.~\ref{fig:FNFjunction_electrical_spin_injection_heat}~(a) for both, parallel and antiparallel magnetizations in the ferromagnets. While the main (linear) contribution to the temperature drop originates from the charge Peltier effect and is the same for both configurations, the spin accumulation near the interfaces is different for each configuration and accounts for different temperature profiles. Figure~\ref{fig:FNFjunction_electrical_spin_injection_heat}~(b), which depicts the difference between the temperature profiles of the parallel and antiparallel configurations, also shows that this difference in the temperature profiles arises in the F regions near the interfaces and within the spin diffusion lengths. Outside these regions the temperature difference remains constant.

\section{Conclusion}\label{conclusion}
We have generalized the standard model of spin injection as explained in Refs.~\onlinecite{Zutic2004:RMP,Fabian2007:APS,Fabian2009:JuelichWorkshop} to describe the coupling between charge, spin, and heat transport in metals. The formalism has then been used to describe the electronic contribution to the spin Seebeck effect in such materials, where we found that only at the boundaries of the ferromagnet there is significant electronic spin accumulation, which, however, decays within the spin diffusion length and can therefore not be responsible for the linear inverse spin Hall voltage measured by Uchida et al.\cite{Uchida2008:Nature} Furthermore, we have analyzed F/N and F/N/F junctions. For F/N junctions we have shown that a temperature difference between both ends of the junction generates pure spin currents which can be used to extract or inject spin at the interface between the F and N regions. We have also derived a formula to measure the efficiency of the spin injection(extraction). In the case of a F/N/F junction a temperature difference can also be used to extract or inject spin into the N region if the junction is in a antiparallel configuration. Moreover, a formula has been derived to calculate the difference between the voltage drops across the junction in the parallel and antiparallel configurations. Finally, we have investigated the Peltier and spin Peltier effects in F/N and F/N/F junctions and derived analytical formulas to describe their respective contributions to the heating/cooling in these systems.

\acknowledgments
This work was supported by the DFG SPP 1538 and GRK 1570. I. \v{Z}. acknowledges support from AFOSR-DCT, U.S. ONR, NSF-NRI NEB 2020, SRC, and DOE-BES.

\bibliographystyle{apsrev}

\begin{thebibliography}{4}
\expandafter\ifx\csname
natexlab\endcsname\relax\def\natexlab#1{#1}\fi
\expandafter\ifx\csname bibnamefont\endcsname\relax
  \def\bibnamefont#1{#1}\fi
\expandafter\ifx\csname bibfnamefont\endcsname\relax
  \def\bibfnamefont#1{#1}\fi
\expandafter\ifx\csname citenamefont\endcsname\relax
  \def\citenamefont#1{#1}\fi
\expandafter\ifx\csname url\endcsname\relax
  \def\url#1{\texttt{#1}}\fi
\expandafter\ifx\csname
urlprefix\endcsname\relax\def\urlprefix{URL }\fi
\providecommand{\bibinfo}[2]{#2}
\providecommand{\eprint}[2][]{\url{#2}}

\bibitem[{\citenamefont{Zutic}(2004)}]{Zutic2004:RMP}
\bibinfo{author}{\bibfnamefont{I.}~\bibnamefont{\v{Z}uti\'c}},
\bibinfo{author}{\bibfnamefont{J.}~\bibnamefont{Fabian}}, \bibnamefont{and}
\bibinfo{author}{\bibfnamefont{S.}~\bibnamefont{Das Sarma}},
\bibinfo{journal}{{Rev. Mod. Phys.}} \textbf{\bibinfo{volume}{76}},
\bibinfo{pages}{323} (\bibinfo{year}{2004}).

\bibitem[{\citenamefont{Fabian}(2007)}]{Fabian2007:APS}
\bibinfo{author}{\bibfnamefont{J.}~\bibnamefont{Fabian}},
\bibinfo{author}{\bibfnamefont{A.}~\bibnamefont{Matos-Abiague}},
\bibinfo{author}{\bibfnamefont{C.}~\bibnamefont{Ertler}},
\bibinfo{author}{\bibfnamefont{P.}~\bibnamefont{Stano}}, \bibnamefont{and}
\bibinfo{author}{\bibfnamefont{I.}~\bibnamefont{\v{Z}uti\'c}},
\bibinfo{journal}{{Acta Phys. Slov.}} \textbf{\bibinfo{volume}{57}},
\bibinfo{pages}{565} (\bibinfo{year}{2007}).

\bibitem[{\citenamefont{Das Sarma}(2000)}]{DasSarma2000:Superlattice}
\bibinfo{author}{\bibfnamefont{S.}~\bibnamefont{Das Sarma}},
\bibinfo{author}{\bibfnamefont{J.}~\bibnamefont{Fabian}},
\bibinfo{author}{\bibfnamefont{X.~D.}~\bibnamefont{Hu}}, \bibnamefont{and}
\bibinfo{author}{\bibfnamefont{I.}~\bibnamefont{\v{Z}uti\'c}},
\bibinfo{journal}{{Superlattice Microst.}} \textbf{\bibinfo{volume}{27}},
\bibinfo{pages}{289} (\bibinfo{year}{2000}).

\bibitem[{\citenamefont{Das Sarma}(2001)}]{DasSarma2001:SolidStateCommunications}
\bibinfo{author}{\bibfnamefont{S.}~\bibnamefont{Das Sarma}},
\bibinfo{author}{\bibfnamefont{J.}~\bibnamefont{Fabian}},
\bibinfo{author}{\bibfnamefont{X.~D.}~\bibnamefont{Hu}}, \bibnamefont{and}
\bibinfo{author}{\bibfnamefont{I.}~\bibnamefont{\v{Z}uti\'c}},
\bibinfo{journal}{{Solid State Commun.}} \textbf{\bibinfo{volume}{119}},
\bibinfo{pages}{207} (\bibinfo{year}{2001}).

\bibitem[{\citenamefont{Fabian}(2009)}]{Fabian2009:JuelichWorkshop}
\bibinfo{author}{\bibfnamefont{J.}~\bibnamefont{Fabian}} \bibnamefont{and}
\bibinfo{author}{\bibfnamefont{I.}~\bibnamefont{\v{Z}uti\'c}},
\emph{\bibinfo{title}{{The standard model of spin injection}}} \bibnamefont{in}
\emph{\bibinfo{title}{{From GMR to Quantum Information}}}
(\bibinfo{publisher}{{Forschungszentrum J\"{u}lich}}, \bibinfo{year}{2009}, \bibnamefont{Eds. S. Bl\"{u}gel et al}).

\bibitem[{\citenamefont{Stevens}(2003)}]{Stevens2003:PRL}
\bibinfo{author}{\bibfnamefont{M.~J.}~\bibnamefont{Stevens}},
\bibinfo{author}{\bibfnamefont{A.~L.}~\bibnamefont{Smirl}},
\bibinfo{author}{\bibfnamefont{R.~D.~R.}~\bibnamefont{Bhat}},
\bibinfo{author}{\bibfnamefont{A.}~\bibnamefont{Najmaie}},
\bibinfo{author}{\bibfnamefont{J.~E.}~\bibnamefont{Sipe}}, \bibnamefont{and}
\bibinfo{author}{\bibfnamefont{H.~M.}~\bibnamefont{van Driel}},
\bibinfo{journal}{{Phys. Rev. Lett.}} \textbf{\bibinfo{volume}{90}},
\bibinfo{pages}{136603} (\bibinfo{year}{2003}).

\bibitem[{\citenamefont{Kimura}(2007)}]{Kimura2007:PRL}
\bibinfo{author}{\bibfnamefont{T.}~\bibnamefont{Kimura}},
\bibinfo{author}{\bibfnamefont{Y.}~\bibnamefont{Otani}},
\bibinfo{author}{\bibfnamefont{T.}~\bibnamefont{Sato}},
\bibinfo{author}{\bibfnamefont{S.}~\bibnamefont{Takahashi}}, \bibnamefont{and}
\bibinfo{author}{\bibfnamefont{S.}~\bibnamefont{Maekawa}},
\bibinfo{journal}{{Phys. Rev. Lett.}} \textbf{\bibinfo{volume}{98}},
\bibinfo{pages}{156601} (\bibinfo{year}{2007}).

\bibitem[{\citenamefont{Zutic}(2011)}]{Zutic2011:NatureMat}
\bibinfo{author}{\bibfnamefont{I.}~\bibnamefont{\v{Z}uti\'c}} \bibnamefont{and}
\bibinfo{author}{\bibfnamefont{H.}~\bibnamefont{Dery}},
\bibinfo{journal}{{Nature Mater.}} \textbf{\bibinfo{volume}{10}},
\bibinfo{pages}{647} (\bibinfo{year}{2011}).

\bibitem[{\citenamefont{Ando}(2011)}]{Ando2011:NatureMat}
\bibinfo{author}{\bibfnamefont{K.}~\bibnamefont{Ando}},
\bibinfo{author}{\bibfnamefont{S.}~\bibnamefont{Takahashi}},
\bibinfo{author}{\bibfnamefont{J.}~\bibnamefont{Ieda}},
\bibinfo{author}{\bibfnamefont{H.}~\bibnamefont{Kurebayashi}},
\bibinfo{author}{\bibfnamefont{T.}~\bibnamefont{Trypiniotis}},
\bibinfo{author}{\bibfnamefont{C.~H.~W.}~\bibnamefont{Barnes}},
\bibinfo{author}{\bibfnamefont{S.}~\bibnamefont{Maekawa}}, \bibnamefont{and}
\bibinfo{author}{\bibfnamefont{E.}~\bibnamefont{Saitoh}},
\bibinfo{journal}{{Nature Mater.}} \textbf{\bibinfo{volume}{10}},
\bibinfo{pages}{655} (\bibinfo{year}{2011}).

\bibitem[{\citenamefont{Kurebayashi}(2011)}]{Kurebayashi2011:NatureMat}
\bibinfo{author}{\bibfnamefont{H.}~\bibnamefont{Kurebayashi}},
\bibinfo{author}{\bibfnamefont{O.}~\bibnamefont{Dzyapko}},
\bibinfo{author}{\bibfnamefont{V.~E.}~\bibnamefont{Demidov}},
\bibinfo{author}{\bibfnamefont{D.}~\bibnamefont{Fang}},
\bibinfo{author}{\bibfnamefont{A.~J.}~\bibnamefont{Ferguson}}, \bibnamefont{and}
\bibinfo{author}{\bibfnamefont{S.~O.}~\bibnamefont{Demokritov}},
\bibinfo{journal}{{Nature Mater.}} \textbf{\bibinfo{volume}{10}},
\bibinfo{pages}{660} (\bibinfo{year}{2011}).

\bibitem[{\citenamefont{Aronov}(1976)}]{Aronov1976:JETP}
\bibinfo{author}{\bibfnamefont{A.~G.}~\bibnamefont{Aronov}},
\bibinfo{journal}{{JETP Lett.}} \textbf{\bibinfo{volume}{24}},
\bibinfo{pages}{32} (\bibinfo{year}{1976}).

\bibitem[{\citenamefont{Johnson and Silsbee}(1985)}]{JohnsonSilsbee1985:PRL}
\bibinfo{author}{\bibfnamefont{M.}~\bibnamefont{Johnson}} \bibnamefont{and}
\bibinfo{author}{\bibfnamefont{R.~H.}~\bibnamefont{Silsbee}},
\bibinfo{journal}{{Phys. Rev. Lett.}} \textbf{\bibinfo{volume}{55}},
\bibinfo{pages}{1790} (\bibinfo{year}{1985}).

\bibitem[{\citenamefont{Johnson and Silsbee}(1987)}]{JohnsonSilsbee1987:PRB}
\bibinfo{author}{\bibfnamefont{M.}~\bibnamefont{Johnson}} \bibnamefont{and}
\bibinfo{author}{\bibfnamefont{R.~H.}~\bibnamefont{Silsbee}},
\bibinfo{journal}{{Phys. Rev. B}} \textbf{\bibinfo{volume}{35}},
\bibinfo{pages}{4959} (\bibinfo{year}{1987}).

\bibitem[{\citenamefont{Johnson and Silsbee}(1988)}]{JohnsonSilsbee1988:PRB}
\bibinfo{author}{\bibfnamefont{M.}~\bibnamefont{Johnson}} \bibnamefont{and}
\bibinfo{author}{\bibfnamefont{R.~H.}~\bibnamefont{Silsbee}},
\bibinfo{journal}{{Phys. Rev. B}} \textbf{\bibinfo{volume}{37}},
\bibinfo{pages}{5312} (\bibinfo{year}{1988}).

\bibitem[{\citenamefont{Rashba}(2002)}]{Rashba2002:EPJB}
\bibinfo{author}{\bibfnamefont{E.~I.}~\bibnamefont{Rashba}},
\bibinfo{journal}{{Eur. Phys. J. B}} \textbf{\bibinfo{volume}{29}},
\bibinfo{pages}{513} (\bibinfo{year}{2002}).

\bibitem[{\citenamefont{Kochan}(2011)}]{Kochan2011:PRL}
\bibinfo{author}{\bibfnamefont{D.}~\bibnamefont{Kochan}},
\bibinfo{author}{\bibfnamefont{M.}~\bibnamefont{Gmitra}}, \bibnamefont{and}
\bibinfo{author}{\bibfnamefont{J.}~\bibnamefont{Fabian}},
\bibinfo{journal}{{Phys. Rev. Lett.}} \textbf{\bibinfo{volume}{107}},
\bibinfo{pages}{176604} (\bibinfo{year}{2011}).

\bibitem[{\citenamefont{Wegrowe}(2000)}]{Wegrowe2000:PRB}
\bibinfo{author}{\bibfnamefont{J.-E.}~\bibnamefont{Wegrowe}},
\bibinfo{journal}{{Phys. Rev. B}} \textbf{\bibinfo{volume}{62}},
\bibinfo{pages}{1067} (\bibinfo{year}{2000}).

\bibitem[{\citenamefont{Bauer}(2010)}]{Bauer2010:SolidStateCommunications}
\bibinfo{author}{\bibfnamefont{G.~E.}~\bibnamefont{Bauer}},
\bibinfo{author}{\bibfnamefont{A.~H.}~\bibnamefont{MacDonald}}, \bibnamefont{and}
\bibinfo{author}{\bibfnamefont{S.}~\bibnamefont{Maekawa}},
\bibinfo{journal}{{Solid State Commun.}} \textbf{\bibinfo{volume}{150}},
\bibinfo{pages}{459} (\bibinfo{year}{2010}).

\bibitem[{\citenamefont{Johnson}(2010)}]{Johnson2010:SolidStateCommunications}
\bibinfo{author}{\bibfnamefont{M.}~\bibnamefont{Johnson}},
\bibinfo{journal}{{Solid State Commun.}} \textbf{\bibinfo{volume}{150}},
\bibinfo{pages}{543} (\bibinfo{year}{2010}).

\bibitem[{\citenamefont{Ashcroft and Mermin}(2006)}]{AshcroftMermin2006}
\bibinfo{author}{\bibfnamefont{N.~W.}~\bibnamefont{Ashcroft}} \bibnamefont{and}
\bibinfo{author}{\bibfnamefont{N.~D.}~\bibnamefont{Mermin}},
\emph{\bibinfo{title}{{Solid State Physics}}}
(\bibinfo{publisher}{{Brooks/Cole Thomson Learning, Singapore}}, \bibinfo{year}{2006}).

\bibitem[{\citenamefont{Callen}(1960)}]{Callen1960}
\bibinfo{author}{\bibfnamefont{H.~B.}~\bibnamefont{Callen}},
\emph{\bibinfo{title}{{Thermodynamics}}}
(\bibinfo{publisher}{{Wiley, New York}}, \bibinfo{year}{1960}).

\bibitem[{\citenamefont{Fukushima}(2005)}]{Fukushima2005:JJAP}
\bibinfo{author}{\bibfnamefont{A.}~\bibnamefont{Fukushima}},
\bibinfo{author}{\bibfnamefont{K.}~\bibnamefont{Yagami}},
\bibinfo{author}{\bibfnamefont{A.~A.}~\bibnamefont{Tulapurkar}},
\bibinfo{author}{\bibfnamefont{Y.}~\bibnamefont{Suzuki}},
\bibinfo{author}{\bibfnamefont{H.}~\bibnamefont{Kubota}},
\bibinfo{author}{\bibfnamefont{A.}~\bibnamefont{Yamamoto}}, \bibnamefont{and}
\bibinfo{author}{\bibfnamefont{S.}~\bibnamefont{Yuasa}},
\bibinfo{journal}{{Jap. J. Appl. Phys.}} \textbf{\bibinfo{volume}{44}},
\bibinfo{pages}{L12} (\bibinfo{year}{2005}).

\bibitem[{\citenamefont{Fukushima}(2005)}]{Fukushima2005:IEEE}
\bibinfo{author}{\bibfnamefont{A.}~\bibnamefont{Fukushima}},
\bibinfo{author}{\bibfnamefont{H.}~\bibnamefont{Kubota}},
\bibinfo{author}{\bibfnamefont{A.}~\bibnamefont{Yamamoto}},
\bibinfo{author}{\bibfnamefont{Y.}~\bibnamefont{Suzuki}}, \bibnamefont{and}
\bibinfo{author}{\bibfnamefont{S.}~\bibnamefont{Yuasa}},
\bibinfo{journal}{{IEEE Trans. Magn.}} \textbf{\bibinfo{volume}{41}},
\bibinfo{pages}{2571} (\bibinfo{year}{2005}).

\bibitem[{\citenamefont{Gravier}(2006)}]{Gravier2006:PRB}
\bibinfo{author}{\bibfnamefont{L.}~\bibnamefont{Gravier}},
\bibinfo{author}{\bibfnamefont{S.}~\bibnamefont{Serrano-Guisan}},
\bibinfo{author}{\bibfnamefont{F.}~\bibnamefont{Reuse}}, \bibnamefont{and}
\bibinfo{author}{\bibfnamefont{J-P.}~\bibnamefont{Ansermet}},
\bibinfo{journal}{{Phys. Rev. B}} \textbf{\bibinfo{volume}{73}},
\bibinfo{pages}{024419} (\bibinfo{year}{2006}).

\bibitem[{\citenamefont{Gravier}(2006)}]{Gravier2006:PRB2}
\bibinfo{author}{\bibfnamefont{L.}~\bibnamefont{Gravier}},
\bibinfo{author}{\bibfnamefont{S.}~\bibnamefont{Serrano-Guisan}},
\bibinfo{author}{\bibfnamefont{F.}~\bibnamefont{Reuse}}, \bibnamefont{and}
\bibinfo{author}{\bibfnamefont{J-P.}~\bibnamefont{Ansermet}},
\bibinfo{journal}{{Phys. Rev. B}} \textbf{\bibinfo{volume}{73}},
\bibinfo{pages}{052410} (\bibinfo{year}{2006}).

\bibitem[{\citenamefont{Costache}(2011)}]{Costache2011:NatureMat}
\bibinfo{author}{\bibfnamefont{M.~V.}~\bibnamefont{Costache}},
\bibinfo{author}{\bibfnamefont{G.}~\bibnamefont{Bridoux}},
\bibinfo{author}{\bibfnamefont{I.}~\bibnamefont{Neumann}}, \bibnamefont{and}
\bibinfo{author}{\bibfnamefont{S.~O.}~\bibnamefont{Valenzuela}},
\bibnamefont{to appear in Nature Mater. (DOI: 10.1038/NMAT3201)}

\bibitem[{\citenamefont{Uchida}(2008)}]{Uchida2008:Nature}
\bibinfo{author}{\bibfnamefont{K.}~\bibnamefont{Uchida}},
\bibinfo{author}{\bibfnamefont{S.}~\bibnamefont{Takahashi}},
\bibinfo{author}{\bibfnamefont{K.}~\bibnamefont{Harii}},
\bibinfo{author}{\bibfnamefont{J.}~\bibnamefont{Ieda}},
\bibinfo{author}{\bibfnamefont{W.}~\bibnamefont{Koshibae}},
\bibinfo{author}{\bibfnamefont{K.}~\bibnamefont{Ando}},
\bibinfo{author}{\bibfnamefont{S.}~\bibnamefont{Maekawa}}, \bibnamefont{and}
\bibinfo{author}{\bibfnamefont{E.}~\bibnamefont{Saitoh}},
\bibinfo{journal}{{Nature}} \textbf{\bibinfo{volume}{455}},
\bibinfo{pages}{778} (\bibinfo{year}{2008}).

\bibitem[{\citenamefont{Uchida}(2010)}]{Uchida2010:JoAP}
\bibinfo{author}{\bibfnamefont{K.}~\bibnamefont{Uchida}},
\bibinfo{author}{\bibfnamefont{T.}~\bibnamefont{Ota}},
\bibinfo{author}{\bibfnamefont{K.}~\bibnamefont{Harii}},
\bibinfo{author}{\bibfnamefont{K.}~\bibnamefont{Ando}},
\bibinfo{author}{\bibfnamefont{H.}~\bibnamefont{Nakayama}}, \bibnamefont{and}
\bibinfo{author}{\bibfnamefont{E.}~\bibnamefont{Saitoh}},
\bibinfo{journal}{{J. Appl. Phys.}} \textbf{\bibinfo{volume}{107}},
\bibinfo{pages}{09A951} (\bibinfo{year}{2010}).

\bibitem[{\citenamefont{Uchida}(2010)}]{Uchida2010:NatureMat}
\bibinfo{author}{\bibfnamefont{K.}~\bibnamefont{Uchida}},
\bibinfo{author}{\bibfnamefont{J.}~\bibnamefont{Xiao}},
\bibinfo{author}{\bibfnamefont{H.}~\bibnamefont{Adachi}},
\bibinfo{author}{\bibfnamefont{J.}~\bibnamefont{Ohe}},
\bibinfo{author}{\bibfnamefont{S.}~\bibnamefont{Takahashi}},
\bibinfo{author}{\bibfnamefont{J.}~\bibnamefont{Ieda}},
\bibinfo{author}{\bibfnamefont{T.}~\bibnamefont{Ota}},
\bibinfo{author}{\bibfnamefont{Y.}~\bibnamefont{Kajiwara}},
\bibinfo{author}{\bibfnamefont{H.}~\bibnamefont{Umezawa}},
\bibinfo{author}{\bibfnamefont{H.}~\bibnamefont{Kawai}},
\bibinfo{author}{\bibfnamefont{G.~E.~W.}~\bibnamefont{Bauer}},
\bibinfo{author}{\bibfnamefont{S.}~\bibnamefont{Maekawa}}, \bibnamefont{and}
\bibinfo{author}{\bibfnamefont{E.}~\bibnamefont{Saitoh}},
\bibinfo{journal}{{Nature Mater.}} \textbf{\bibinfo{volume}{9}},
\bibinfo{pages}{894} (\bibinfo{year}{2010}).

\bibitem[{\citenamefont{Uchida}(2009)}]{Uchida2009:JoAP}
\bibinfo{author}{\bibfnamefont{K.}~\bibnamefont{Uchida}},
\bibinfo{author}{\bibfnamefont{S.}~\bibnamefont{Takahashi}},
\bibinfo{author}{\bibfnamefont{J.}~\bibnamefont{Ieda}},
\bibinfo{author}{\bibfnamefont{K.}~\bibnamefont{Harii}},
\bibinfo{author}{\bibfnamefont{K.}~\bibnamefont{Ikeda}},
\bibinfo{author}{\bibfnamefont{W.}~\bibnamefont{Koshibae}},
\bibinfo{author}{\bibfnamefont{S.}~\bibnamefont{Maekawa}}, \bibnamefont{and}
\bibinfo{author}{\bibfnamefont{E.}~\bibnamefont{Saitoh}},
\bibinfo{journal}{{J. Appl. Phys.}} \textbf{\bibinfo{volume}{105}},
\bibinfo{pages}{07C908} (\bibinfo{year}{2009}).

\bibitem[{\citenamefont{Bader}(2010)}]{Bader2010:ARCMP}
\bibinfo{author}{\bibfnamefont{S.~D.}~\bibnamefont{Bader}}, \bibnamefont{and}
\bibinfo{author}{\bibfnamefont{S.~S.~P.}~\bibnamefont{Parkin}},
\bibinfo{journal}{{Annu. Rev. Condensed Matter Phys.}} \textbf{\bibinfo{volume}{1}},
\bibinfo{pages}{71} (\bibinfo{year}{2010}).

\bibitem[{\citenamefont{Jaworski}(2010)}]{Jaworski2010:NatureMat}
\bibinfo{author}{\bibfnamefont{C.~M.}~\bibnamefont{Jaworski}},
\bibinfo{author}{\bibfnamefont{J.}~\bibnamefont{Yang}},
\bibinfo{author}{\bibfnamefont{S.}~\bibnamefont{Mack}},
\bibinfo{author}{\bibfnamefont{D.~D.}~\bibnamefont{Awschalom}},
\bibinfo{author}{\bibfnamefont{J.~P.}~\bibnamefont{Heremans}}, \bibnamefont{and}
\bibinfo{author}{\bibfnamefont{R.~C.}~\bibnamefont{Myers}},
\bibinfo{journal}{{Nature Mater.}} \textbf{\bibinfo{volume}{9}},
\bibinfo{pages}{898} (\bibinfo{year}{2010}).

\bibitem[{\citenamefont{Valenzuela}(2006)}]{Valenzuela2006:Nature}
\bibinfo{author}{\bibfnamefont{S.~O.}~\bibnamefont{Valenzuela}} \bibnamefont{and}
\bibinfo{author}{\bibfnamefont{M.}~\bibnamefont{Tinkham}},
\bibinfo{journal}{{Nature}} \textbf{\bibinfo{volume}{442}},
\bibinfo{pages}{176} (\bibinfo{year}{2006}).

\bibitem[{\citenamefont{Saitoh}(2006)}]{Saitoh2006:APL}
\bibinfo{author}{\bibfnamefont{E.}~\bibnamefont{Saitoh}},
\bibinfo{author}{\bibfnamefont{M.}~\bibnamefont{Ueda}},
\bibinfo{author}{\bibfnamefont{H.}~\bibnamefont{Miyajima}}, \bibnamefont{and}
\bibinfo{author}{\bibfnamefont{G.}~\bibnamefont{Tatara}},
\bibinfo{journal}{{Appl. Phys. Lett.}} \textbf{\bibinfo{volume}{88}},
\bibinfo{pages}{182509} (\bibinfo{year}{2006}).

\bibitem[{\citenamefont{Xiao}(2010)}]{Xiao2010:PRB}
\bibinfo{author}{\bibfnamefont{J.}~\bibnamefont{Xiao}},
\bibinfo{author}{\bibfnamefont{G.~E.~W.}~\bibnamefont{Bauer}},
\bibinfo{author}{\bibfnamefont{K.}~\bibnamefont{Uchida}},
\bibinfo{author}{\bibfnamefont{E.}~\bibnamefont{Saitoh}} \bibnamefont{and}
\bibinfo{author}{\bibfnamefont{S.}~\bibnamefont{Maekawa}},
\bibinfo{journal}{{Phys. Rev. B}} \textbf{\bibinfo{volume}{81}},
\bibinfo{pages}{214418} (\bibinfo{year}{2010}).

\bibitem[{\citenamefont{Adachi}(2010)}]{Adachi2010:APL}
\bibinfo{author}{\bibfnamefont{H.}~\bibnamefont{Adachi}},
\bibinfo{author}{\bibfnamefont{K.}~\bibnamefont{Uchida}},
\bibinfo{author}{\bibfnamefont{E.}~\bibnamefont{Saitoh}},
\bibinfo{author}{\bibfnamefont{J.}~\bibnamefont{Ohe}},
\bibinfo{author}{\bibfnamefont{S.}~\bibnamefont{Takahashi}}, \bibnamefont{and}
\bibinfo{author}{\bibfnamefont{S.}~\bibnamefont{Maekawa}},
\bibinfo{journal}{{Appl. Phys. Lett.}} \textbf{\bibinfo{volume}{97}},
\bibinfo{pages}{252506} (\bibinfo{year}{2010}).

\bibitem[{\citenamefont{Jaworski}(2011)}]{Jaworski2011:PRL}
\bibinfo{author}{\bibfnamefont{C.~M.}~\bibnamefont{Jaworski}},
\bibinfo{author}{\bibfnamefont{J.}~\bibnamefont{Yang}},
\bibinfo{author}{\bibfnamefont{S.}~\bibnamefont{Mack}},
\bibinfo{author}{\bibfnamefont{D.~D.}~\bibnamefont{Awschalom}},
\bibinfo{author}{\bibfnamefont{R.~C.}~\bibnamefont{Myers}}, \bibnamefont{and}
\bibinfo{author}{\bibfnamefont{J.~P.}~\bibnamefont{Heremans}},
\bibinfo{journal}{{Phys. Rev. Lett.}} \textbf{\bibinfo{volume}{106}},
\bibinfo{pages}{186601} (\bibinfo{year}{2011}).

\bibitem[{\citenamefont{Slachter}(2011)}]{Slachter2011:PRB}
\bibinfo{author}{\bibfnamefont{A.}~\bibnamefont{Slachter}},
\bibinfo{author}{\bibfnamefont{F.~L.}~\bibnamefont{Bakker}}, \bibnamefont{and}
\bibinfo{author}{\bibfnamefont{B.~J.}~\bibnamefont{van Wees}},
\bibinfo{journal}{{Phys. Rev. B}} \textbf{\bibinfo{volume}{84}},
\bibinfo{pages}{174408} (\bibinfo{year}{2011}).

\bibitem[{\citenamefont{Flipse}(2011)}]{Flipse2011:preprint}
\bibinfo{author}{\bibfnamefont{J.}~\bibnamefont{Flipse}}, 
\bibinfo{author}{\bibfnamefont{F.~L.}~\bibnamefont{Bakker}},
\bibinfo{author}{\bibfnamefont{A.}~\bibnamefont{Slachter}},
\bibinfo{author}{\bibfnamefont{F.~K.}~\bibnamefont{Dejene}},
\bibnamefont{and}
\bibinfo{author}{\bibfnamefont{B.~J.}~\bibnamefont{{van Wees}}},
\bibinfo{journal}{{arXiv:1109.6898v1}}
(\bibnamefont{unpublished}).

\bibitem[{\citenamefont{Slachter}(2010)}]{Slachter2010:Nature}
\bibinfo{author}{\bibfnamefont{A.}~\bibnamefont{Slachter}},
\bibinfo{author}{\bibfnamefont{F.~L.}~\bibnamefont{Bakker}},
\bibinfo{author}{\bibfnamefont{J-P.}~\bibnamefont{Adam}}, \bibnamefont{and}
\bibinfo{author}{\bibfnamefont{B.~J.}~\bibnamefont{van Wees}},
\bibinfo{journal}{{Nature Phys.}} \textbf{\bibinfo{volume}{6}},
\bibinfo{pages}{879} (\bibinfo{year}{2010}).

\bibitem[{\citenamefont{Bakker}(2010)}]{Bakker2010:PRL}
\bibinfo{author}{\bibfnamefont{F.~L.}~\bibnamefont{Bakker}},
\bibinfo{author}{\bibfnamefont{A.}~\bibnamefont{Slachter}},
\bibinfo{author}{\bibfnamefont{J-P.}~\bibnamefont{Adam}}, \bibnamefont{and}
\bibinfo{author}{\bibfnamefont{B.~J.}~\bibnamefont{van Wees}},
\bibinfo{journal}{{Phys. Rev. Lett.}} \textbf{\bibinfo{volume}{105}},
\bibinfo{pages}{136601} (\bibinfo{year}{2010}).

\bibitem[{\citenamefont{Le Breton}(2011)}]{LeBreton2011:Nature}
\bibinfo{author}{\bibfnamefont{J-P.}~\bibnamefont{Le~Breton}},
\bibinfo{author}{\bibfnamefont{S.}~\bibnamefont{Sharma}},
\bibinfo{author}{\bibfnamefont{H.}~\bibnamefont{Saito}},
\bibinfo{author}{\bibfnamefont{S.}~\bibnamefont{Yuasa}}, \bibnamefont{and}
\bibinfo{author}{\bibfnamefont{R.}~\bibnamefont{Jansen}},
\bibinfo{journal}{{Nature}} \textbf{\bibinfo{volume}{475}},
\bibinfo{pages}{82} (\bibinfo{year}{2011}).

\bibitem[{\citenamefont{Hatami}(2009)}]{Hatami2009:PRB}
\bibinfo{author}{\bibfnamefont{M.}~\bibnamefont{Hatami}},
\bibinfo{author}{\bibfnamefont{G.~E.~W.}~\bibnamefont{Bauer}},
\bibinfo{author}{\bibfnamefont{Q.}~\bibnamefont{Zhang}}, \bibnamefont{and}
\bibinfo{author}{\bibfnamefont{P.~J.}~\bibnamefont{Kelly}},
\bibinfo{journal}{{Phys. Rev. B}} \textbf{\bibinfo{volume}{79}},
\bibinfo{pages}{174426} (\bibinfo{year}{2009}).

\bibitem[{\citenamefont{Ziman}(2007)}]{Ziman2007}
\bibinfo{author}{\bibfnamefont{J.~M.}~\bibnamefont{Ziman}},
\emph{\bibinfo{title}{{Electrons and phonons}}}
(\bibinfo{publisher}{{Clarendon Press, Oxford}}, \bibinfo{year}{2007}).

\bibitem[{\citenamefont{vanSon}(1987)}]{vanSon1987:PRL}
\bibinfo{author}{\bibfnamefont{P.~C.}~\bibnamefont{van~Son}},
\bibinfo{author}{\bibfnamefont{H.}~\bibnamefont{van~Kempen}}, \bibnamefont{and}
\bibinfo{author}{\bibfnamefont{P.}~\bibnamefont{Wyder}},
\bibinfo{journal}{{Phys. Rev. Lett.}} \textbf{\bibinfo{volume}{58}},
\bibinfo{pages}{2271} (\bibinfo{year}{1987}).

\bibitem[{\citenamefont{Valet}(1993)}]{ValetFert1993:PRB}
\bibinfo{author}{\bibfnamefont{T.}~\bibnamefont{Valet}} \bibnamefont{and}
\bibinfo{author}{\bibfnamefont{A.}~\bibnamefont{Fert}},
\bibinfo{journal}{{Phys. Rev. B}} \textbf{\bibinfo{volume}{48}},
\bibinfo{pages}{7099} (\bibinfo{year}{1993}).

\bibitem[{\citenamefont{Hatami}(2010)}]{Hatami2010:SolidStateCommunications}
\bibinfo{author}{\bibfnamefont{M.}~\bibnamefont{Hatami}},
\bibinfo{author}{\bibfnamefont{G.~E.~W.}~\bibnamefont{Bauer}},
\bibinfo{author}{\bibfnamefont{S.}~\bibnamefont{Takahashi}}, \bibnamefont{and}
\bibinfo{author}{\bibfnamefont{S.}~\bibnamefont{Maekawa}},
\bibinfo{journal}{{Solid State Commun.}} \textbf{\bibinfo{volume}{150}},
\bibinfo{pages}{480} (\bibinfo{year}{2010}).

\bibitem[{\citenamefont{Fert and Jaffr\`es}(2001)}]{FertJaffres2001:PRB}
\bibinfo{author}{\bibfnamefont{A.}~\bibnamefont{Fert}} \bibnamefont{and}
\bibinfo{author}{\bibfnamefont{H.}~\bibnamefont{Jaffr\`es}},
\bibinfo{journal}{{Phys. Rev. B}} \textbf{\bibinfo{volume}{64}},
\bibinfo{pages}{184420} (\bibinfo{year}{2001}).

\bibitem[{\citenamefont{Schmidt}(2000)}]{Schmidt2000:PRB}
\bibinfo{author}{\bibfnamefont{G.}~\bibnamefont{Schmidt}}, 
\bibinfo{author}{\bibfnamefont{D.}~\bibnamefont{Ferrand}},
\bibinfo{author}{\bibfnamefont{L.~W. }~\bibnamefont{Molenkamp}},
\bibinfo{author}{\bibfnamefont{A.~T.}~\bibnamefont{Filip}},
\bibnamefont{and}
\bibinfo{author}{\bibfnamefont{B.~J.}~\bibnamefont{{van Wees}}},
\bibinfo{journal}{{Phys. Rev. B}} \textbf{\bibinfo{volume}{62}},
\bibinfo{pages}{(R)4790} (\bibinfo{year}{2000}).

\bibitem[{\citenamefont{Rashba}(2000)}]{Rashba2000:PRB}
\bibinfo{author}{\bibfnamefont{E.~I.}~\bibnamefont{{Rashba}}},
\bibinfo{journal}{{Phys. Rev. B}} \textbf{\bibinfo{volume}{62}},
\bibinfo{pages}{(R)16267} (\bibinfo{year}{2000}).

\bibitem[{\citenamefont{Silsbee}(1980)}]{Silsbee1980:BMR}
\bibinfo{author}{\bibfnamefont{R.~H.}~\bibnamefont{Silsbee}},
\bibinfo{journal}{{Bull. Magn. Reson.}} \textbf{\bibinfo{volume}{2}},
\bibinfo{pages}{284} (\bibinfo{year}{1980}).

\bibitem[{\citenamefont{Jansen}(2011)}]{Jansen2011:preprint}
\bibinfo{author}{\bibfnamefont{R.}~\bibnamefont{Jansen}}, 
\bibinfo{author}{\bibfnamefont{A.~M.}~\bibnamefont{Deac}},
\bibinfo{author}{\bibfnamefont{H.}~\bibnamefont{Saito}}, \bibnamefont{and}
\bibinfo{author}{\bibfnamefont{S.}~\bibnamefont{{Yuasa}}},
\bibinfo{journal}{{arXiv:1112.3430v1}}
(\bibnamefont{unpublished}).

\end{thebibliography}

\end{document}